\documentclass[11pt]{article}

\usepackage{booktabs}
\usepackage{url}
\usepackage{comment}
\usepackage{indentfirst}
\usepackage{setspace}
\usepackage{rotating}
\usepackage{a4wide}
\usepackage{graphicx,subcaption,adjustbox}
\usepackage[margin=1in]{geometry}
\usepackage{gensymb}
\usepackage[multiple]{footmisc}
\usepackage{subcaption}
\usepackage{enumerate}
\usepackage{dcolumn}
\usepackage[flushleft]{threeparttable}   
\usepackage{threeparttablex}   
\usepackage{amsmath,amssymb,mathtools}
\usepackage{fancyhdr}
\usepackage[section]{placeins}
\usepackage{pdflscape}
\usepackage[dvipsnames]{xcolor}
\usepackage{xr-hyper}

\usepackage{hyperref}
\hypersetup{colorlinks,
    linkcolor=RoyalBlue,
    citecolor=RoyalBlue,
    filecolor=RoyalBlue,      
    urlcolor=violet,
    pdfpagemode=FullScreen,
  }
\usepackage{multirow}
\usepackage{lipsum}
\usepackage{CJK}
\usepackage[authoryear]{natbib}
\usepackage[labelsep=period]{caption}
\usepackage{graphicx} 
\usepackage{float} 
\usepackage[toc,page,title,titletoc,header]{appendix}
\usepackage{chngcntr} 
\usepackage{apptools} 
\usepackage{titlesec}
\usepackage[american]{babel} 
\usepackage{csquotes}
\usepackage{xurl} 
\usepackage{siunitx}
\usepackage{environ}

\NewEnviron{mytable}[2][t]{
  \begin{table}[#1]
  \small
  \centering
  \singlespacing
  \begin{minipage}{0.99\textwidth}
  \centering
  \tabcolsep=#2 
  \renewcommand{\tablenotes}[1]{\gdef\mytablenotes{##1}}
  \gdef\mytablenotes{} 
  \BODY 
  
  \ifx\mytablenotes\empty\else
    \vspace{1em}
    \raggedright
    \footnotesize
    \noindent\parbox{\linewidth}{\textit{Notes:} \mytablenotes}
  \fi
  \end{minipage}
  \end{table}
}
\newcolumntype{C}[1]{>{\centering\arraybackslash}p{#1}}
\newcolumntype{R}[1]{>{\raggedleft\arraybackslash}p{#1}}

\newcolumntype{U}[1]{S[table-format=2.2, table-space-text-post={***}, input-symbols={()}, 
    table-column-width = #1]}

\newcolumntype{T}[1]{S[table-format=2.3, table-space-text-post={***}, input-symbols={()}, 
    table-column-width = #1]}

\NewEnviron{myfigure}[1][t]{
  \begin{figure}[#1]
  \centering
  \singlespacing
  \begin{minipage}[c]{0.99\textwidth}
  \centering
  \newcommand{\figurenotes}[1]{\gdef\myfigurenotes{##1}}
  \gdef\myfigurenotes{} 
  \BODY 
  \ifx\mytablenotes\empty\else
    \vspace{1em}
    \raggedright
    \footnotesize
    \noindent\parbox{\linewidth}{\textit{Notes:} \myfigurenotes}
  \fi
  \end{minipage}
  \end{figure}
}

\title{Consumption Stimulus with Digital Coupons: \\ Heterogeneity and Policy Design}
\author{
Ying Chen\thanks{MOE Key Laboratory of Econometrics, School of Economics, and the Paula and Gregory Chow Institute for Studies in Economics, Xiamen University. Email: \href{mailto:ychen@xmu.edu.cn}{ychen@xmu.edu.cn}.}
\and Mingyi Li\thanks{School of Management and Economics, The Chinese University of Hong Kong, Shenzhen. Email: \href{mailto:mingyili@link.cuhk.edu.cn}{mingyili@link.cuhk.edu.cn}.}
\and Jonathan J. Mao\thanks{Corresponding author. Center for Macroeconomic Research and Wang Yanan Institute for Studies in Economics, Xiamen University. Email: \href{mailto:jmao@xmu.edu.cn}{jmao@xmu.edu.cn}.}
\and Jingyi Zhou\thanks{Amazon.com, Inc. Email: \href{mailto:jingyiuohz@gmail.com}{jingyiuohz@gmail.com}.}
\thanks{We thank Ricardo Perez-Truglia, the editor, and three anonymous referees for their valuable guidance and constructive comments. We also thank Nathaniel Baum-Snow, Felipe Carozzi, Shihe Fu, Jessie Handbury, Nathaniel Hendren, Xavier Jaravel, Sophie Calder-Wang, Maisy Wong, and Jianwei Xing for helpful discussions and feedback. Further helpful comments were received from conference and seminar participants at the North American Meeting of the Urban Economics Association, Peking University, Wuhan University, CUHK-Shenzhen, and Xiamen University. Ying Chen acknowledges financial support from the Natural Science Foundation of China [Grants NSFC72203188 and NSFC71988101]. Jonathan Mao acknowledges financial support from the Natural Science Foundation of China [Grants NSFC72133004 and NSFC71988101]. We are grateful to Yue Wang from Rajax Network Technology (Ele.me) for data support, and to the High Performance Computing Platform of the Key Laboratory of Econometrics (Xiamen University), Ministry of Education, for additional data resources.}
}
\date{June 2026}

\begin{document}
\pagenumbering{gobble}
\clearpage
\thispagestyle{empty}
\maketitle

\renewcommand{\abstractname}{\large Abstract \vspace{1cm}}
\begin{abstract}
\large

\noindent We study consumption stimulus using digital coupons, which provide time-limited subsidies contingent on minimum spending. Analyzing a large-scale program in China, we find that the program generates large and heterogeneous short-term effects. Consumption responses vary across both consumers and locations, reflecting both demand-side and supply-side factors. This heterogeneity shapes the incidence of the program: high-response consumers tend to patronize larger businesses, leading to a regressive allocation of stimulus benefits. Through counterfactual analysis, we show that targeting rules can reshape both the size and distribution of stimulus effects. Targeting the most responsive consumers can more than double the aggregate stimulus, while a hybrid design that combines targeted distribution with direct support to small businesses improves both efficiency and equity.
\end{abstract}

\onehalfspacing
\large
\newpage
\clearpage
\pagenumbering{arabic}


\section{Introduction}
\label{sec:intro}

During economic downturns, traditional monetary and fiscal policies often face limitations in their ability to swiftly and effectively boost consumer spending. In response, government-issued digital coupons---a form of mobile-distributed consumption voucher---have emerged as an innovative tool to stimulate demand in targeted sectors. Unlike conventional stimulus measures such as cash transfers or tax rebates, which are often saved rather than spent, digital coupons feature minimum spending thresholds (e.g., spend ¥50 to get ¥15 off) and short expiration periods designed to encourage immediate spending. Their digital distribution enables rapid, large-scale deployment and allows policymakers to direct support toward sectors that are particularly vulnerable during downturns.

In this paper, we study digital coupons as a stimulus tool using comprehensive data from a major program in Beijing, which targeted food delivery services to support the local restaurant industry. We examine both the overall effectiveness of the program in stimulating consumer spending and the heterogeneity in its impact across individuals and businesses. Using detailed transaction records and user attributes from a leading Chinese food delivery platform---augmented with high-resolution housing transaction data and geocoded establishment information---we construct rich profiles of consumer demographic, wealth, and locational characteristics that enable a detailed analysis of the program's impact.

We begin by establishing baseline results using a difference-in-differences design that exploits the program's first-come, first-served allocation rule. We find that, on average, digital coupons increased daily consumption by 12 percent during the program period. For every ¥1 of government subsidy, consumers spent an additional ¥2.38 of their own money, resulting in a spending multiplier of 3.38. Following \citet{Ding2025}, we refer to this multiplier as the coupon MPC---the marginal propensity to consume out of coupon subsidies. Our estimate implies that for every yuan of increased business revenue, approximately one-quarter is financed by government subsidies, while the remaining three-quarters come from consumers' out-of-pocket spending. This result highlights a distinctive feature of digital coupons: unlike conventional stimulus payments where consumers are net recipients of government funds, with digital coupons, both consumers and the government jointly finance the increase in business revenue.\footnote{\citet{Ding2025} refers to this feature as ``consumer-financed stimulus.''}

While average effects are informative, they mask substantial variation in individual responses---variation that is central to understanding both the drivers of stimulus effectiveness and the distribution of benefits. We therefore estimate heterogeneous treatment effects to examine how the program's impact varies across both consumers and businesses. Two key questions guide our heterogeneity analysis. First, we hypothesize that the consumption response to digital coupons is shaped by both demand- and supply-side factors. On the demand side, factors such as personal income and wealth influence how much a person spends after receiving digital coupons. On the supply side, the availability and variety of local businesses determine the opportunity for coupon redemption.\footnote{\label{footnote:2}In this paper, we use the term ``supply-side'' to refer to local consumption opportunities---the set of spending options available to consumers---rather than to firm-level behavioral responses such as price adjustments or changes in operating practices.} For example, two otherwise similar individuals may exhibit different responses if one resides in an area with abundant food delivery options, while the other lives in a neighborhood with limited choices. Consequently, the impact of a digital coupon program varies not only among individuals but across locations. This spatial variation in stimulus effects, which has not been studied in the literature to date, could have important implications for how benefits are distributed among local businesses. Furthermore, wealthier individuals tend to sort into neighborhoods with greater consumption amenities \citep{Couture2020}, leading to an intertwining of demand and supply effects. Although existing research consistently identifies income or wealth as key determinants of consumption responses to fiscal stimulus,\footnote{See, e.g., \citet{Johnson2006,Parker2013,Broda2014,Misra2014,Parker2022a}.} it remains unclear whether these effects are driven by income or wealth itself or are partly attributable to spatial sorting across locations. Disentangling these mechanisms will improve our understanding of how heterogeneous individuals and locations jointly shape program outcomes.

Second, because digital coupons act as a stimulus jointly financed by the government and consumers, they impact businesses and consumers in different ways. This raises the question of which businesses benefit most from the program, who the paying consumers are, and whether these outcomes align with policymakers' objectives. For instance, if the benefits accrue primarily to large businesses or if the financing burden falls mainly on consumers with low income or wealth, the program may fail to achieve its intended policy goals. Understanding these dynamics is essential for assessing the program's distributional impact. Moreover, policymakers may face trade-offs in designing such programs, depending on how consumer spending patterns interact with business characteristics. If large establishments attract wealthier consumers---and these consumers contribute more out-of-pocket in response to digital coupons---then maximizing the total stimulus may require targeting these individuals, potentially concentrating benefits among large businesses. Conversely, prioritizing support for small businesses may require directing more coupons to lower-income consumers, resulting in both a weaker overall stimulus and a program that relies more heavily on their spending as its source of financing.\footnote{\label{fn:intro_welfare} We discuss distributional and policy-design implications in Section~\ref{sec:policy}. Under a standard rational benchmark, receiving digital coupons should weakly increase consumer welfare, so directing more coupons to lower-income consumers would directly benefit them even if their out-of-pocket spending rises. However, if behavioral mechanisms also shape spending responses, then relying more heavily on lower-income consumers' spending to finance the stimulus effect may have ambiguous welfare implications for those consumers, even if the policy raises business revenue.} Recognizing these potential trade-offs can be important, which has so far been overlooked in both academic and policy discussions.

Our empirical analysis builds on modern econometric and machine learning methods. We nonparametrically estimate heterogeneous treatment effects using the causal forest algorithm \citep{Athey2019a}, adapted to a difference-in-differences setting to recover individual-level conditional average treatment effects on the treated based on each consumer's full set of demographic, wealth, and locational characteristics. Following \citet{Chernozhukov2020}, we debias these estimates using augmented inverse-propensity weighting and conduct inference using their best linear projection framework. Finally, we apply the accumulated local effects approach of \citet{AZ2020} to characterize the marginal influence of each characteristic on treatment effect heterogeneity while accounting for correlations among covariates, including spatial sorting between demand- and supply-side variables.

Our analysis yields four main findings. First, we uncover substantial heterogeneity in the consumption response to digital coupons. The distribution of estimated individual treatment effects was highly dispersed: the standard deviation exceeded twice the average effect. This dispersion translated into a wide range of implied coupon MPCs. Over 23 percent of consumers had coupon MPCs above five, indicating very large spending responses. At the other end of the distribution, 19 percent of consumers had coupon MPCs below one and reduced their out-of-pocket expenditure. For these individuals, the coupons generated net savings. Beyond individual heterogeneity, stimulus effects also varied substantially across locations. Aggregating treatment effects within 3km-by-3km neighborhoods, we found that neighborhood-level spending gains ranged from less than 10 percent in some areas to more than 50 percent in others. These patterns highlight the limitations of relying on a single average response when evaluating stimulus policies.

Second, both demand- and supply-side factors contribute to the variation in stimulus effects. On the demand side, we identify a strong positive relationship between individual wealth and consumption response. Consumption habits also matter, with spending increases concentrated among individuals who historically placed the most frequent orders and spent the most per order. On the supply side, we uncover a non-monotonic relationship between consumption response and establishment density. Out-of-pocket spending initially rose with the number of local businesses but declined beyond a threshold. Conditional on total establishment count, neighborhoods with a disproportionate share of large businesses generated significantly higher spending. These findings suggest that digital coupons have limited stimulus effects in both food deserts and areas saturated with small establishments. Thus, both the availability and composition of consumption amenities shape the program's effectiveness. A variance decomposition analysis shows that demand- and supply-side determinants contribute in broadly comparable magnitudes, with demand playing a modestly larger role.

\label{para:third_cont}Third, our findings point to both rational incentives and complementary behavioral mechanisms that may shape consumption responses. Consistent with rational, threshold-based models of coupon use \citep{Xing2023,Ding2025}, we observe clear bunching of expenditures at coupon thresholds, suggesting that some consumers adjusted their spending to qualify for redemption. At the same time, spending responses are not concentrated in narrow regions around the coupon thresholds: treatment effects continue to rise with baseline spending levels well beyond those thresholds, a pattern that is difficult to reconcile with threshold-crossing incentives alone. These patterns are consistent with behavioral mechanisms such as mental accounting \citep{Thaler1999}, under which consumers may treat digital coupons as a distinct budget for discretionary consumption. The design features of digital coupons---minimum spending thresholds and short expiration windows---may further amplify these responses.

Fourth, we examine which types of businesses benefited most from the stimulus. By mapping individual treatment effects onto businesses, we find that larger, higher-priced establishments captured the most additional revenue. We show that this unequal distribution stems from a key form of consumer-business matching: consumers with higher coupon MPCs tended to direct a greater share of their spending to large businesses. Consequently, the incidence of the program---in terms of which businesses ultimately benefited---may not align with policies aimed at supporting small and vulnerable businesses during economic downturns.

Armed with these results, we show that targeting rules can substantially reshape both the size and distribution of stimulus effects. Digital coupons are typically distributed via lotteries or first-come-first-served mechanisms, which do not systematically reach the most responsive consumers and may not align with distributional objectives. Yet, unlike traditional stimulus instruments, digital coupons can leverage mobile platforms to enable real-time targeting based on observable characteristics. This creates scope for improving both the efficiency and distributional consequences of stimulus policy.

We begin by considering the objective of maximizing aggregate stimulus. Leveraging our estimated individual treatment effects, we show that targeting the most responsive consumers could more than double the overall stimulus effect at no additional fiscal cost. These efficiency gains remain substantial even when the government's targeting capacity is limited to a subset of observable characteristics.

However, efficiency-maximizing strategies favor larger businesses and wealthier consumers, potentially undermining objectives to support vulnerable small businesses. We therefore evaluate an alternative targeting strategy that prioritizes consumers more likely to patronize smaller establishments---those with lower wealth, weaker consumption habits, and residing in areas with a higher share of small businesses. We show that doing so effectively shifts revenue gains toward small businesses, but at the cost of lower aggregate stimulus and greater reliance on lower-wealth consumers' out-of-pocket spending as the source of finance.

Recognizing these trade-offs, we evaluate a hybrid policy combining digital coupons with direct small-business support. In this approach, a portion of the government budget is used to fund targeted coupons to the most responsive consumers, while the remainder is allocated as direct subsidies to small businesses. Applying this approach to the Beijing program, we find that the hybrid design can deliver 21 percent more total stimulus while providing substantial direct support to small businesses that received no government assistance under the original program, addressing both efficiency and equity objectives within existing budget constraints.

\subsection{Related Literature}
\label{sec:lit}

Our paper relates to several strands of literature. First, we contribute to the extensive work on consumption responses to fiscal stimulus, which typically estimates the average marginal propensity to consume (MPC) out of tax rebates or cash transfers. Most studies of the 2001 and 2008 U.S. stimulus programs report short-run MPCs between 0.25 and 0.5 for nondurables, though recent work using staggered difference-in-differences designs finds lower estimates around 0.25 for the 2008 payments \citep[e.g.,][]{Johnson2006,Parker2013,Broda2014,Borusyak2024,Orchard2025}.

Our paper is most closely related to the emerging literature on government-issued digital coupons, a novel form of voucher-based fiscal stimulus. Earlier examples of voucher-based stimulus include Japan's 1999 shopping coupon program \citep{Hsieh2010} and Taiwan's 2009 voucher initiative \citep{Kan2017}. Unlike traditional physical vouchers, digital coupons feature minimum spending thresholds and short expiration windows and are distributed via mobile platforms. Studies of digital coupon programs in Hangzhou, Shaoxing, and other Chinese cities estimate coupon MPCs between 3 and 6 \citep{Liu2021,Xing2023,Ding2025}, suggesting substantially larger short-run stimulus effects than those associated with either cash transfers or physical vouchers.\footnote{Compared with the existing literature on digital coupons, our setting is narrower and focuses on food delivery, a largely non-storable consumption category.}

We contribute to this literature by moving beyond average effects to uncover substantial heterogeneity in consumption responses. We analyze both demand- and supply-side drivers of this heterogeneity, and show how it leads to unequal impacts on local businesses and shapes the potential for targeting. Methodologically, we depart from traditional interaction-based approaches to heterogeneity and join a growing literature applying machine learning methods to estimate heterogeneous treatment effects and inform policy design \citep[e.g.,][]{Hino2018,Davis2020,Britto2022,Johnson2023}. 

Our paper also speaks to a literature documenting violations of fungibility in consumption, as evidenced by responses to tax refunds, in-kind transfers, gift cards, and online coupons \citep{Milkman2009,Reinholtz2015,Hastings2018,Baugh2021}. Of particular relevance, \citet{Boehm2025} show that prepaid cards with short expiration windows yield substantially higher short-run MPCs than cash. Our findings document patterns consistent with consumers treating digital coupons as partly nonfungible windfall budgets, with their consumption responses likely amplified by the salience of minimum spending thresholds and expiration constraints.

The rest of the paper is organized as follows. Section \ref{sec:bg-data} provides background on the digital coupon program and describes the data. Section \ref{sec:model} outlines our empirical approach to estimating treatment effects and capturing heterogeneity. Section \ref{sec:result} presents the main results. Section \ref{sec:policy} examines distributional consequences and counterfactual targeting policies. Section \ref{sec:discussion} concludes. References to sections, figures, and tables prefixed by OA refer to the Online Appendix.

\section{Background and Data}
\label{sec:bg-data}

\subsection{Background}
\label{sec:bg}

During the Covid-19 pandemic, China experienced a sharp decline in household consumption and economic activity. Face-to-face service industries such as restaurants and accommodations were hit especially hard, as widespread lockdown measures and limited remote operation options severely restricted their business activities. According to the Beijing Municipal Bureau of Statistics, the value added of the city's accommodation and catering industry fell 13.7 percent in 2022.

In response to these economic challenges, local governments across China implemented a range of fiscal stimulus measures, with digital coupon programs emerging as a key strategy to directly incentivize consumer spending in targeted sectors. By the end of 2022, more than 285 municipal governments had launched digital coupon initiatives, targeting industries such as tourism, retail, and food services. These initiatives were enabled by China's advanced digital infrastructure, which allowed widespread and cost-effective distribution to millions of consumers through mobile platforms.

We study a digital coupon initiative implemented in Beijing during the summer of 2022. In July, the Beijing municipal government launched a restaurant consumption voucher program as part of the 2022 Beijing Consumption Initiatives \citep{Beijing2022}. On the Ele.me food delivery platform, coupons were distributed daily from July 18 to August 27, 2022 on a first-come, first-served basis to individuals with IP addresses located in Beijing. Each day, a limited quota of coupons became available on a dedicated event page within the Ele.me mobile app, where users could attempt to claim that day's coupon bundle. Each eligible individual could claim one bundle of coupons per day, consisting of a ¥15 discount on purchases over ¥50 (the ``50-15'' coupon) and a ¥30 discount on purchases over ¥100 (the ``100-30'' coupon). These coupons expired at midnight on the day of issuance, giving users a brief window to redeem them after claiming.

This quota-constrained, first-come, first-served distribution mechanism generated a daily ``rush'' environment in which users competed to claim coupons before the quota was exhausted. Whether an individual who clicked on the event page successfully obtained a coupon bundle depended on whether their claim attempt occurred before or after the daily quota was reached. We define the treatment group as individuals who successfully obtained at least one coupon bundle during the program period, and the control group as those who participated in the event but never obtained a coupon. To identify the causal impact of coupon receipt on consumer spending, we employ a matched difference-in-differences framework, as detailed in Section \ref{sec:model-did}. We track the expenditures for online delivery orders placed through the partnering platform, where coupons were automatically applied once the purchase met the required threshold. Consumers could only use one coupon per order, providing a clean measure of how coupon availability affected purchasing decisions.

\subsection{Data}
\label{sec:data}
\subsubsection{Sampling and Key Variables}
\label{sec:sampling}

Our main data come from Ele.me (\href{http://ele.me/}{ele.me}), a leading food delivery platform in China and distributor of digital coupons during the Beijing initiative in 2022. As a major player in China's food delivery market, the platform holds over 30 percent market share and employs over four million delivery riders as of 2024 \citep{eleme2024,xinhua2025}.

We implement a stratified random sampling design to select active individuals---defined as those who had placed at least one order on the platform within the past six months---from two strata: individuals who participated in the coupon program and successfully obtained coupons (treatment group), and those who participated but did not obtain coupons (control group). Using a 1:1 stratification ratio between these groups and after excluding invalid accounts, we obtain a sample of 11,765 unique individuals, with 5,980 in the treatment group and 5,785 in the control group.\footnote{The platform's privacy policy limited the number of user-level records we could extract, resulting in an analysis sample that is substantially smaller than the full population of coupon participants. To assess sample representativeness, we conducted an external validation using large-scale mobile-app usage data from a major telecommunications carrier, covering approximately 50 percent of Beijing's population. We identify Ele.me users in this dataset and confirm that their demographic composition closely aligns with our sample. See Section \ref{sec:daas} for details.}

We obtain complete ordering records for each sampled individual covering a 69-day period from July 4 to September 10, 2022. This time frame encompasses the 41-day coupon program (July 18 to August 27), which we refer to as the treatment period, as well as two weeks before program implementation (pre-treatment period) and two weeks after program completion. Each order record provides information on the total order amount and the number of items---referred to as stock keeping units (SKU)---purchased. The order amount reflects both the consumer's out-of-pocket expenditure and any coupon discount subsidized by the government, which together sum to the total payment received by the merchant.

We also collect detailed individual-level characteristics from Ele.me to account for demographic profiles and consumption habits relevant to coupon usage and food delivery demand. These characteristics include age, gender, platform membership status,\footnote{Ele.me offers a paid membership program that provides users with benefits such as waived or reduced delivery fees and exclusive discounts. Users can purchase memberships on a monthly or annual basis.} the geographic coordinates of frequently used delivery addresses, and the price range of their smartphones.\footnote{Age and phone price are reported in bins rather than exact values. We use the median value of each bin in the analysis.} To depict individuals' consumption habits, we also retrieve their average expenditure per order and the total number of orders placed in the six months prior to the coupon event.

Because the platform does not provide direct measures of individual wealth or neighborhood amenities, we supplement the data with additional location-based information. We derive housing price levels from over 0.5 million property transaction records between 2010 and 2020, obtained from Lianjia (\href{http://lianjia.com}{lianjia.com}), the leading property listing platform in China. We assign a local housing price to each individual's frequent delivery address by averaging the prices of the five nearest properties. Since housing prices and smartphone prices are highly correlated, we construct a first principal component from these two variables to create a wealth index that captures individuals' socioeconomic status.

To capture the local opportunities for consumption, we construct a measure of consumption amenities, defined as the number and composition of restaurants in an individual's immediate vicinity. A key hypothesis of our paper is that the supply of nearby restaurants influences how likely individuals are to redeem digital coupons, as it shapes both the ease of access and the attractiveness of dining options. To operationalize this measure, we draw a random sample of 3,000 merchants on Ele.me in Beijing. After excluding grocery stores and new establishments that opened in 2022, we retain a sample of 2,120 restaurants.\footnote{We exclude grocery stores because the coupon program specifically targeted restaurant services. Newly opened establishments from 2022 are omitted because they lack sufficient pre-program sales history to reliably categorize them by size.} We classify restaurants as either small and medium-sized enterprises (SMEs) or large businesses (non-SMEs) based on their median monthly revenue over the six months preceding the coupon event, using the citywide median as the cutoff. For each individual, we create a 3km circular buffer around their frequent delivery address and compute two variables: (1) the total number of restaurants within the buffer, and (2) the share of these restaurants that are classified as large businesses. This 3km buffer corresponds to the typical food delivery service range,\footnote{Over 90 percent of orders in our data fall within 3km of users' delivery addresses.} allowing us to quantify both the density and the composition of consumption amenities available to each individual.

\subsubsection{Sample Definitions and Descriptive Patterns}
\label{sec:sumstats}

We construct our baseline dataset at the individual-day level, aggregating order information each day to compute out-of-pocket expenditure (total expenditure minus coupon subsidies), total expenditure, unsubsidized expenditure (spending on orders without coupon redemption), number of orders, and total items (SKU) per order. On days when a consumer does not make any purchases, these variables are set to zero, ensuring a balanced panel that captures both active and inactive consumption days.

\begin{mytable}[t!]{0pt} 
\caption{Descriptive Statistics}
\label{tab:sumstats}
\begin{tabular}{p{0.5\textwidth} C{.15\textwidth} C{.1\textwidth} C{.15\textwidth} C{.1\textwidth}} 
    \toprule
    \midrule
    & Mean & SD & Mean & SD \\ 
    \cmidrule(lr){2-5}
    & \multicolumn{2}{c}{Treatment} & \multicolumn{2}{c}{Control} \\ 
    \cmidrule(lr){2-3} \cmidrule(lr){4-5}
    \multicolumn{5}{l}{\textbf{Panel A: Individual-day Level}} \\                            
    Out-of-pocket expenditure & 17.584 & 44.390 & 13.989 & 41.245 \\
    Total expenditure         & 18.148 & 45.306 & 13.989 & 41.245 \\
    Unsubsidized expenditure  & 16.235 & 43.186 & 13.989 & 41.245 \\
    Number of orders          & 0.401  & 0.701  & 0.301  & 0.640  \\
    SKU per order             & 0.996  & 2.409  & 0.746  & 2.218  \\
    Observations & \multicolumn{2}{c}{208,285} & \multicolumn{2}{c}{208,285} \\ 
    \midrule
    & \multicolumn{2}{c}{Treatment} & \multicolumn{2}{c}{Control} \\ 
    \cmidrule(lr){2-3} \cmidrule(lr){4-5}
    \multicolumn{5}{l}{\textbf{Panel B: Individual Level}} \\ 
    $\Delta$ Out-of-pocket expenditure & 1.877 & 15.185 & 0.076 & 15.751 \\
    Age                       & 32.250 & 8.160  & 32.281 & 9.047  \\
    Female                    & 0.635  & 0.481  & 0.657  & 0.475  \\
    Platform membership       & 0.382  & 0.486  & 0.395  & 0.489  \\
    Wealth                    & 0.043  & 1.032  & 0.053  & 1.019  \\
    Number of establishments     & 52.445 & 33.987 & 51.284 & 32.638 \\
    Share of non-SME establishments & 0.524 & 0.125 & 0.526 & 0.122 \\
    Number of orders, past 6 months & 55.580 & 55.728 & 54.164 & 56.853 \\
    Spending per order, past 6 months & 45.236 & 26.475 & 44.913 & 27.685 \\
    Observations & \multicolumn{2}{c}{3,787} & \multicolumn{2}{c}{3,787} \\
    \midrule
    & \multicolumn{2}{c}{SME} & \multicolumn{2}{c}{Non-SME} \\ 
    \cmidrule(lr){2-3} \cmidrule(lr){4-5}
    \multicolumn{5}{l}{\textbf{Panel C: Establishment Level}} \\ 
    Average monthly sales, past 6 months (thousand) & 12.965 & 11.816 & 219.903 & 219.060 \\
    Average order price, past 6 months       & 50.845 & 45.119 & 56.552 & 42.698 \\
    Number of establishments (0.5km radius) & 7.012 & 3.973 & 8.228 & 4.399 \\
    Average local housing price (thousand) & 4,562 & 3,791 & 4,975 & 3,921 \\
    Observations & \multicolumn{2}{c}{1,060} & \multicolumn{2}{c}{1,060} \\
    \bottomrule
\end{tabular}
\tablenotes{This table presents descriptive statistics at the individual-day, individual, and establishment (restaurant) levels. Panel A reports individual-day variables related to consumer ordering behavior over the pre-treatment and treatment periods (55 days), where $\Delta$ out-of-pocket expenditure represents consumer spending net of coupon subsidies, total expenditure captures the full payment to sellers (including coupon subsidies), and unsubsidized expenditure measures spending on orders that did not use coupons. All expenditure variables are in yuan (¥). Panel B shows individual-level characteristics. The treatment group includes individuals who  obtained at least one coupon bundle; the control group includes those who participated but did not obtain coupons. References to ``past 6 months'' indicate the six-month period immediately preceding the coupon event (January--July 2022). The wealth index is constructed as the first principal component of smartphone prices and local housing prices, where ``local housing prices'' are calculated using the nearest five transaction records for a given location. Number of establishments and share of non-SME establishments are calculated within a 3km radius of each individual's frequent delivery address. Control group statistics in this panel are weighted to account for matching with replacement. Panel C compares characteristics between SME and non-SME establishments, with SMEs defined as businesses with median monthly revenue below the citywide median over the six months preceding the coupon program.}
\end{mytable}


To further improve comparability between the treatment and control groups, we implement a propensity score matching (PSM) procedure. After dropping individuals with missing values on any key characteristics, we estimate each individual's likelihood of obtaining coupons based on their observed demographic, wealth, and locational attributes. Each treated individual is then matched to a control individual with replacement, based on the estimated propensity scores. This procedure yields a matched sample of 3,787 treated individuals and 3,787 matched control observations, drawn from 1,389 unique control individuals. Altogether, our baseline analysis focusing on the pre-treatment and treatment periods (55 days total) comprises 416,570 individual-day observations (3,787 individuals $\times$ 2 groups $\times$ 55 days). When including the two weeks after program completion for supplementary analyses, our sample expands to 522,606 observations across the full 69-day study period. Further details on the PSM specification and diagnostic checks are provided in Section \ref{sec:psm-det}.

Table \ref{tab:sumstats} presents descriptive statistics for our matched sample. Panel A summarizes individual-day variables related to consumer ordering behavior, including the three expenditure measures, the daily average number of orders and the average number of items per order. For the control group, the three expenditure measures are identical, as they did not receive any subsidies. On average, the treatment group spent substantially more than the control group across all three expenditure measures. Treated individuals also placed slightly more frequent orders and included more items per order. 

Panel B of Table \ref{tab:sumstats} presents summary statistics for the individual-level variables used in our heterogeneity analysis. The data reveal the profile of typical food delivery consumers in Beijing: predominantly young adults (average age around 32), with slightly more females than males. Their wealth index values range from -3 to 3, with our sample centered around the mean (by construction, the index has mean 0 and standard deviation 1). About 39 percent held platform membership status, indicating regular engagement with the delivery service. On average, consumers placed roughly 55 orders in the six months preceding the coupon event, spending about ¥45 per order. Their neighborhoods contained an average of 52 restaurants within delivery range,\footnote{The restaurant count is based on a random sample drawn from the universe of establishments listed on the Ele.me platform. While the absolute number should be interpreted with caution, it provides a valid basis for comparing restaurant availability between treatment and control groups.} with a relatively balanced mix of SME and non-SME establishments. These detailed demographic, wealth, and locational attributes allow us to examine how treatment effects vary systematically across socioeconomic groups and geographic areas.

Panel C of Table \ref{tab:sumstats} presents establishment-level characteristics by SME status. The data reveal notable differences between the large and small establishments in our sample. Large restaurants generated substantially higher sales, with average monthly revenue nearly 17 times greater than that of SMEs. Although they represent 50 percent of restaurants, large establishments accounted for approximately 94 percent of total revenue. On average, they also charged higher order prices (¥56.55 vs. ¥50.85) and were located in areas with greater restaurant density and higher local housing prices. These systematic differences in establishment characteristics and spatial distribution highlight the importance of accounting for neighborhood characteristics in studying the impact of the stimulus program.

\section{Empirical Strategy}
\label{sec:model}

In this section, we present the empirical framework for analyzing the consumption responses to digital coupons. We begin by outlining our approach to estimating the average effect of coupon acquisition on spending behavior. We then describe our methods for estimating heterogeneous treatment effects and examining the factors that drive variation in consumption responses. 

\subsection{Estimating the Average Treatment Effect}
\label{sec:model-did}

We employ a difference-in-differences (DiD) design to identify the average treatment effect of obtaining digital coupons on consumer spending. Our identification strategy exploits the first-come, first-served distribution mechanism described in Section \ref{sec:bg}. All individuals in our sample actively attempted to claim coupons by clicking on the event link, demonstrating comparable intent to obtain and use digital coupons. Their success depended on whether the daily coupon quota had been exhausted at the time of their attempt---a factor plausibly unrelated to unobserved determinants of spending behavior.\footnote{\label{fn:idealrd}An ideal design would restrict comparison to users attempting within very narrow time windows around quota exhaustion, where success is driven primarily by random server load and connection speed. Unfortunately, the platform records timestamps only for successful coupon claims, not for unsuccessful attempts, making such a design infeasible.}

To mitigate potential selection bias arising from individual differences in timing and effort intensity, we first implement propensity score matching, as described in Section \ref{sec:sumstats}, and then conduct our DiD analysis using the matched sample. Matching treated individuals to control units using the full set of demographic, wealth, and locational characteristics yields well-balanced covariates and enhances comparability between the two groups.\footnote{To show that our conclusions do not depend on this particular matching design, Table~\ref{tab:did_app} also reports results from three alternatives: the baseline DiD estimated on the full sample without matching; optimal full matching \citep{Hansen2006}, which implements matched subclasses with varying treated--control ratios rather than one-to-one pair matching; and the doubly robust difference-in-differences estimator of \citet{Sant2020}, which combines propensity-score weighting with outcome regression and remains consistent if either the propensity-score model or the outcome-evolution model is correctly specified. The estimates are quantitatively similar across all three alternatives.} The credibility of this matched DiD design relies on the parallel trends assumption: in the absence of treatment, treated and control individuals would have followed similar spending trajectories. In Section \ref{sec:result-did}, we present graphical and formal evidence supporting this assumption using pre-treatment data.

Formally, we estimate the following two-way fixed effects (TWFE) regression applied to the matched sample at the individual-day level:
\begin{equation}
  \label{eq:block_did}
y_{it} = \alpha \cdot \text{Treat}_i \times \text{Post}_t + \gamma_i + \lambda_t + \epsilon_{it},  
\end{equation}
where $y_{it}$ represents the outcome variables (out-of-pocket expenditure in our baseline model) for individual $i$ on date $t$,  $\text{Treat}_i$ indicates whether individual $i$ ever obtained a coupon bundle during the treatment period, and $\text{Post}_t$ equals one during the treatment period (July 18 to August 27, 2022) and zero during the pre-treatment period. The coefficient $\alpha$ captures the average treatment effect on the treated (ATT), representing the increase in daily out-of-pocket spending by coupon recipients during the treatment period relative to the pre-treatment period, compared to the same temporal difference for the control group. This baseline specification focuses exclusively on the pre-treatment and treatment periods, with the two weeks after the coupon event considered separately in subsequent analyses.

We make several implementation choices to ensure a credible estimation of the treatment effect. First, we define treatment at the period level rather than the daily level: an individual is considered treated if they obtained at least one coupon bundle during any day of the treatment period. This approach avoids endogeneity concerns that would arise if individuals timed their participation in response to anticipated consumption needs.\footnote{In Section \ref{sec:dailydid}, we present an alternative specification that defines treatment at the individual-day level---that is, individuals are considered treated only on days they obtained coupons. This specification does not conform to a standard DiD framework, as treatment status is time-varying and non-absorbing. We report these day-level estimates in the appendix for comparison with our benchmark period-level results presented in the main text.} Second, our estimate reflects an ``intention-to-treat'' effect, measuring the impact of obtaining digital coupons regardless of actual redemption patterns. This distinction is important because not all distributed coupons were redeemed. This approach avoids the selection bias that would arise from conditioning on the endogenous redemption decisions. Third, we adopt a binary treatment definition, classifying individuals as either treated or control rather than modeling treatment intensity based on the number of coupons obtained. While this simplification may understate heterogeneity in exposure, it provides a clean identification of the overall program effect. Finally, since the ``50-15'' and ``100-30'' coupons were always distributed together as a bundle, their individual effects cannot be separately identified. Our estimates therefore represent the combined impact of having obtained both coupon types throughout the program period. 

To alleviate remaining endogeneity concerns, we conduct two robustness exercises that sharply limit the scope for selection. First, individuals differ in the timing of their claim attempts: users who attempt earlier in the day are more likely to succeed and may also differ systematically in their consumption behavior from those who attempt later. To address this concern, we restrict the treatment group to near-cutoff recipients---individuals who obtained at least one coupon bundle within the final decile of a day's successful claim timestamps during the program period.\footnote{One could further restrict the treatment group to users whose entire set of acquired coupons were obtained within the final decile of each day's successful claims. However, such a definition yields too few treated individuals for reliable estimation, so we adopt the broader near-cutoff criterion.} These individuals experienced at least one ``last-minute'' success, securing coupons by only a small margin close to the quota-exhaustion time, making them more comparable to the control group and thereby reducing concerns that our estimates are driven by selection on timing.\footnote{We observe timestamps only for successfully obtained coupons but not for unsuccessful claim attempts. As a result, we cannot isolate control-group individuals who narrowly missed the quota. This prevents us from implementing a formal regression-discontinuity design or a difference-in-differences design centered on daily quota exhaustion.\label{fn:timestamp}} Second, individuals vary in their effort intensity. Users who log in frequently or make repeated attempts are more likely to obtain a coupon bundle and may also differ systematically from casual users in their underlying consumption behavior. To address this concern, we restrict the treatment group to one-time recipients---individuals who obtained exactly one coupon bundle during the program period. This restriction mechanically excludes persistent, high-engagement users, thereby reducing concerns that our estimates are driven by selection on effort intensity. Results for both exercises are reported in Section \ref{sec:did_subsample}. Re-estimating the matched DiD specification on these restricted treatment groups yields effects similar in magnitude to our benchmark estimates, suggesting that timing- and effort-based selection are not driving our main findings.

We further decompose the consumption response along extensive and intensive margins by replacing the dependent variable with measures such as order frequency, expenditure per order, and dishes (SKUs) per order. This decomposition helps identify whether digital coupons primarily stimulate more frequent purchases (extensive margin) or larger purchases per transaction (intensive margin), which has implications for how the program's benefits are distributed across businesses of different sizes and price levels.

We also investigate potential substitution patterns that could limit the net stimulus effect of digital coupons. Using the platform's transaction data, we test whether increased restaurant spending is offset by inter-temporal, inter-category, or intra-household substitution. To test for inter-temporal substitution, we compare spending patterns during the two weeks after the coupon event against those in the two weeks preceding it. If consumers simply shift their planned purchases to coincide with coupon availability and then reduce spending afterward \citep[e.g.,][]{Mian2012}, the net stimulus effect over a longer horizon might be negligible. Specifically, we modify our baseline specification by replacing treatment period observations with data from the two weeks after the coupon event, and redefine $\text{Post}_t$ to indicate this period instead of the treatment period. The coefficient on $\text{Treat}_i \times \text{Post}_t$ then captures the difference in individual daily out-of-pocket expenditure between treatment and control groups during the two weeks after the coupon event relative to the pre-treatment period. This exercise also examines a closely related question: do digital coupons generate sustained increases in consumption that persist after the treatment period ends? Persistent consumption increases would suggest our $\alpha$ estimate in equation \eqref{eq:block_did} underestimates the total impact, while decreased spending would confirm inter-temporal substitution. For inter-category and intra-household substitution, we replace $y_{it}$ in equation \eqref{eq:block_did} with expenditure on grocery orders and with the number of tableware sets requested per order, respectively.

Finally, we examine off-platform substitution, which the platform's transaction data cannot directly capture. Consumers may shift orders from rival delivery platforms to Ele.me (cross-platform substitution) or replace in-person dining with delivery orders (offline-to-online substitution). Since both forms of off-platform substitution would likely appear on Ele.me primarily through additional orders rather than higher spending per existing order, we turn to external data to assess this possibility. Specifically, we use the mobile application usage data introduced in Section \ref{sec:sampling} and compare food-delivery app usage during, and after the coupon event to test whether treated consumers shifted spending from rival platforms to Ele.me during the coupon period. Taken together, these exercises assess whether the estimated consumption response reflects genuine additional spending or redistribution across time, categories, platforms, dining formats, or household members.

\subsection{Estimating Heterogeneous Treatment Effects}
\label{sec:model-hte}

While the average effect estimates provide a measure of the overall effectiveness of the stimulus program, they mask substantial heterogeneity in individual consumption responses. Understanding this heterogeneity is crucial for identifying the micro-drivers of spending behavior, assessing the distributional impact of the program, and designing optimal strategies to support different policy objectives. Furthermore, since individuals live in different locations and shop at different places, heterogeneity in individual consumption responses translates into heterogeneous impact on local businesses. Consequently, the \textit{incidence} of the stimulus program, in terms of which businesses it ultimately benefits, depends fundamentally on consumer heterogeneity.

To analyze how consumption responses vary across individuals, we estimate a heterogeneous treatment effects model. Specifically, let $\Delta y_i$ denote the change in individual $i$'s average out-of-pocket spending, calculated as the difference between their average spending during the treatment period and the pre-treatment period. Let $\text{Treat}_i$ be an indicator for whether individual $i$ belongs to the treatment group. We estimate the following first-difference DiD regression:
\begin{equation}
  \label{eq:HTE}
  \Delta y_i = \alpha(\textbf{X}_i)\cdot \text{Treat}_i + f(\textbf{X}_i) + \varepsilon_i,  
\end{equation}
where $\textbf{X}_i$ includes our full set of observed demographic, wealth, and locational attributes (see Table \ref{tab:sumstats}, Panel B), and $\alpha(.)$ and $f(.)$ are allowed to be any functions. In Section \ref{sec:result-did}, we show that the impact of digital coupons was concentrated within the treatment period and did not extend beyond the program's duration. Accordingly, we focus on analyzing heterogeneity in spending responses during this period.

Since $\text{Treat}_i$ is binary, equation \eqref{eq:HTE} constitutes a complete nonparametric specification.\footnote{Compared to the average effect model \eqref{eq:block_did}, model \eqref{eq:HTE} additionally controls for potential differential trends by $\mathbf{X}_i$, as well as any correlation between treatment status and treatment effect that could arise from residual imbalances in $\mathbf{X}_i$. In Section \ref{sec:hte-driver}, we compare the average treatment effect derived from this heterogeneous effects model to that estimated from the average effect model.} $\alpha(\textbf{X}_i)$ captures the heterogeneous treatment effects of interest. More precisely, it represents the conditional average treatment effect on the treated (CATT) for each individual, given their observed characteristics. In causal inference terminology, the variables $\textbf{X}_i$ are potential \textit{moderators} or \textit{effect modifiers} that predict variation in treatment effects. By modeling the treatment effect as a nonparametric function of $\textbf{X}_i$, we allow for complex patterns to emerge on how different factors jointly shape variation in spending responses. 

To estimate $\alpha(\textbf{X}_i)$, we partial out the influence of $f(\textbf{X}_i)$ by estimating the following residual-on-residual regression:
\begin{equation}
  \label{eq:CF}
    \Delta y_{i}-\mathbb{E}\left[\left.\Delta y_{i}\right|\mathbf{X}_{i}\right]=\alpha(\textbf{X}_{i})\cdot\left(\text{Treat}_{i}-\mathbb{E}\left[\left.\text{Treat}_{i}\right|\mathbf{X}_{i}\right]\right)+\xi_{i}, 
\end{equation}
where the nuisance functions $\mathbb{E}\left[\left.\Delta y_{i}\right|\mathbf{X}_{i}\right]$ and $\mathbb{E}\left[\left.\text{Treat}_{i}\right|\mathbf{X}_{i}\right]$ are estimated using machine learning estimators.\footnote{We implement the residual-on-residual orthogonalization procedure and estimate the causal forest model using the \texttt{grf} package in \texttt{R} \citep{Tibshirani2024}. The \texttt{grf} package allows users to supply pre-estimated nuisance functions, which it uses to perform orthogonalization prior to fitting the causal forest. We estimate these nuisance functions using both the default random forest learners in \texttt{grf} and the Super Learner ensemble method \citep{Laan2007}. Following \citet{Chernozhukov2018}, we apply cross-fitting to reduce overfitting and ensure that Neyman orthogonality holds asymptotically. Additional implementation details are provided in Section \ref{sec:tech}.} This procedure follows the double machine learning (DML) framework of \citet{Chernozhukov2018}.\footnote{Through orthogonalization, this procedure avoids the need to estimate \(f(\mathbf{X}_i)\), whose errors may contaminate the estimation of \(\alpha(\mathbf{X}_i)\). Alternatively, we can estimate the regression model \eqref{eq:HTE} directly, which involves learning both \(\alpha(\mathbf{X}_i)\) and \(f(\mathbf{X}_i)\) jointly. In Section \ref{sec:hteapp}, we implement this direct strategy using alternative machine learning estimators and compare their performance with our benchmark results based on orthogonalization.} The associated moment condition satisfies Neyman orthogonality, which guarantees that the estimation of $\alpha(\textbf{X}_i)$ will be robust to first-order errors in nuisance function estimation.

We estimate equation \eqref{eq:CF} using the causal forest estimator \citep{Wager2018, Athey2019a}, a tree-based machine learning method designed for heterogeneous treatment effect estimation.\footnote{Section \ref{sec:tech} provides details on the implementation, including hyperparameter settings governing forest size, splitting rules, and sample-splitting fractions.} The causal forest constructs an ensemble of causal trees, where each tree recursively partitions the covariate space to maximize variation in estimated treatment effects, thereby identifying subgroups with distinct responses to the treatment. The method employs the ``honest'' approach \citep{Athey2016}, which uses sample-splitting to avoid overfitting and ensure valid inference. By averaging across many such trees, the causal forest produces a nonparametric estimate of the conditional average treatment effect function.\footnote{The causal forest can be interpreted as a locally weighted estimator, with weights determined by the proportion of trees in which an observation shares a leaf with the target point. These weights define an adaptive kernel that places greater mass on regions of the covariate space where treatment effect heterogeneity is most pronounced, unlike classical nearest-neighbor or kernel methods, which apply fixed bandwidths regardless of underlying heterogeneity.} This nonparametric approach offers a systematic way to explore treatment heterogeneity across a rich set of causal effect moderators, while capturing potentially complex and high-order interactions among them.\footnote{For examples of empirical research applying the causal forest method, see \citet{Davis2017}, \citet{Britto2022}, and \citet{Johnson2023}.} 

\subsection{Characterizing Determinants of Treatment Effects}
\label{sec:model-driver}

The estimated heterogeneous treatment effects function allows us to analyze how different factors contribute to variation in consumer responses. A common approach in applied work is to estimate linear interaction models, in which each potential causal effect moderator is interacted with the treatment indicator one at a time. However, this strategy has two important limitations. First, when moderators are correlated with one another, the estimated coefficients may be confounded by omitted variables that are correlated with the interaction term. For instance, both individual wealth and local consumption amenities may influence the consumption response to digital coupons, and these two factors are often correlated due to spatial sorting. In single-interaction models, a positive association between wealth and the treatment effect might therefore reflect either wealth's direct influence---a demand-side effect---or the influence of local consumption amenities---a supply-side effect. This makes the estimated coefficient difficult to interpret and obscures the underlying mechanism. Second, the linear interaction approach is limited to detecting linear relationships between each moderator and the treatment effect. As a result, it may miss important nonlinear or nonmonotonic patterns in treatment effect heterogeneity.

By nonparametrically estimating treatment effects as a function of the full set of observed demographic, wealth, and locational attributes, we can examine the influence of each factor conditional on the others. This multivariate approach mitigates the confounding concerns that arise in single-interaction models and brings our analysis closer to a ``treatment effect on treatment effect'' framework---that is, assessing how individual characteristics shape the direction and magnitude of the treatment's impact. In this section, we pursue two complementary strategies. First, we use a best linear projection to estimate the conditional linear relationship between each variable and the treatment effect, holding other characteristics constant. Second, we employ accumulated local effects curves to uncover potentially nonlinear or nonmonotonic patterns in how individual factors contribute to treatment effect heterogeneity.

\subsubsection{Best Linear Projection}
\label{sec:model-blp}

A simple approach to assess how individual characteristics relate to treatment effect heterogeneity is to regress the estimated treatment effect function $\widehat{\alpha}\left(\mathbf{X}_{i}\right)$ on observed covariates. However, although causal forests yield pointwise consistent estimates of $\alpha(\mathbf{X}_i)$, directly regressing these estimates can lead to biased summaries due to estimation errors that do not cancel out \citep{Athey2019b}.\footnote{Due to regularization, machine learning estimators like causal forests can produce finite-sample biases that are correlated with covariates. In particular, the estimates tend to shrink toward the global mean in regions of the covariate space with sparse support or high variability.} Therefore, following \citet{Chernozhukov2020}, we construct a doubly robust score that corrects for this bias and enables consistent estimation of the best linear approximation to the treatment effect function, also known as the best linear projection (BLP). Specifically, define 
$$
\psi_{i}=\widehat{\alpha}\left(\mathbf{X}_{i}\right)+\frac{\text{Treat}_{i}-\widehat{\mathbb{E}}\left[\left.\text{Treat}_{i}\right|\mathbf{X}_{i}\right]}{\widehat{\mathbb{E}}\left[\left.\text{Treat}_{i}\right|\mathbf{X}_{i}\right]\left(1-\widehat{\mathbb{E}}\left[\left.\text{Treat}_{i}\right|\mathbf{X}_{i}\right]\right)}\cdot\widehat{\varepsilon}_{i},
$$
where $\widehat{\varepsilon}_{i}=\Delta y_{i}-\widehat{\mathbb{E}}\left[\left.\Delta y_{i}\right|\text{Treat}_{i},\mathbf{X}_{i}\right]=\Delta y_{i}-\widehat{\mathbb{E}}\left[\left.\Delta y_{i}\right|\mathbf{X}_{i}\right]-\left(\text{Treat}_{i}-\widehat{\mathbb{E}}\left[\left.\text{Treat}_{i}\right|\mathbf{X}_{i}\right]\right)\cdot \widehat{\alpha}\left(\textbf{X}_{i}\right)$, and $\widehat{\mathbb{E}}\left[\left.\Delta y_{i}\right|\mathbf{X}_{i}\right]$ and $\widehat{\mathbb{E}}\left[\left.\text{Treat}_{i}\right|\mathbf{X}_{i}\right]$ are the estimated nuisance functions from \eqref{eq:CF}. We then estimate:
\begin{equation}
  \label{eq:BLP}
    \psi_{i}=\beta\cdot\widetilde{\mathbf{X}}_{i}+e_{i},
\end{equation}
where $\widetilde{\mathbf{X}}_{i}$ are standardized to have mean zero and unit variance. This approach provides valid standard errors for inference on each coefficient $\beta_{k}$. The standardization allows direct comparison across coefficients as measures of variable importance. In the absence of unobserved moderators correlated with $\widetilde{X}_{ik}$, the coefficient $\beta_{k}$ can be interpreted as the causal effect of $\widetilde{X}_{ik}$ on the treatment effect of digital coupons, allowing us to identify the key drivers of treatment effect heterogeneity.

\subsubsection{Accumulated Local Effect Curves}
\label{sec:model-ale}

While the best linear projection captures linear relationships between treatment effects and observed moderators, it may miss important nonlinear or nonmonotonic patterns. To examine these more complex relationships, we construct accumulated local effects (ALE) curves \citep{AZ2020}, which visualize how each moderator influences the estimated treatment effect while accounting for the joint distribution of the covariates.

Specifically, let $\delta_{k}\left(x\right)=\mathbb{E}\left[\left.\left.\partial\alpha\left(\mathbf{X}_{i}\right)\right/\partial X_{ik}\right|X_{ik}=x\right]$ denote the conditional average partial derivative of the treatment effect function with respect to $X_{ik}$. The ALE curve for $X_{ik}$ is defined as the cumulative integral $h_{k}\left(x\right)=\int_{\underline{x}}^{x}\delta_{k}\left(t\right)dt$, where $\underline{x}$ is the minimum observed value of $X_{ik}$.\footnote{The curve is then centered as $\widetilde{h}_{k}\left(x\right)=h_{k}\left(x\right)-\mathbb{E}\left[h_{k}\left(x\right)\right]$, so that its average over the empirical distribution of $X_{ik}$ is zero.} Intuitively, the ALE curve traces how the treatment effect changes as $X_{ik}$ varies, averaging the marginal effect over the conditional distribution of the remaining moderators at each value of $X_{ik}$. 

ALE curves are particularly useful when covariates are correlated. Unlike partial dependence plots, which average marginal effects over the entire covariate space \citep{Friedman2001}, ALE curves average partial derivatives $\left.\partial\alpha\left(\mathbf{X}_{i}\right)\right/\partial X_{ik}$ only for observations where $X_{ik}=x$ is observed. This local averaging prevents extrapolation into unsupported regions, which can distort inference in the presence of correlated covariates. For example, ALE curves for neighborhood consumption amenities account for spatial sorting between individuals and locations by averaging over wealth levels that are empirically observed within each neighborhood, rather than evaluating all possible---and potentially unrealistic---combinations of wealth and neighborhood characteristics. 

In practice, we estimate the ALE curves using the doubly robust scores $\psi_i$ introduced in Section~\ref{sec:model-blp}, mirroring the bias-correction approach used in the BLP.\footnote{Specifically, we nonparametrically estimate the conditional expectation function $\mathbb{E}[\psi_i | \mathbf{X}_i]$ and compute ALE curves based on the estimated surface. See Section \ref{sec:driverapp} for further details.} This yields a nonparametric visualization of how each moderator influences the treatment effect. Under the assumption that unobserved moderators are uncorrelated with ${X}_{ik}$, the derivative $\delta_{k}\left(x\right)$---and hence the slope of the ALE curve---can be interpreted as the local average treatment effect (LATE) of $X_{ik}$ on the treatment effect of digital coupons for individuals with $X_{ik}=x$. We now turn to the empirical results.

\section{Results}
\label{sec:result}

\subsection{Average Impact of Digital Coupons on Spending}
\label{sec:result-did}

\begin{mytable}{8pt}
    \caption{Average Treatment Effects of Digital Coupons on Different Types of Expenditure}
    \label{tab:did}
\begin{tabular}{p{0.12\textwidth}C{0.25\textwidth}C{0.25\textwidth}C{0.25\textwidth}}
    \toprule
    \midrule
                  &     Out-of-pocket expenditure        &       Total expenditure      &       Unsubsidized expenditure    \\ 
                            &     (1)       &       (2)     &       (3)     \\ \midrule
    Treat$\times$Post       &     1.801***  &    2.558***   &     -0.008     \\
                            &   (0.636)     &     (0.639)   &    (0.635)    \\               
                            &               &               &               \\
    Observations            &  416,570      &    416,570    &   416,570     \\
    \bottomrule
\end{tabular}
\tablenotes{This table presents the average treatment effects of obtaining coupons on daily expenditure among treated individuals, estimated using Equation \ref{eq:block_did} with individual and date fixed effects. Out-of-pocket expenditure represents total expenditure minus government-financed coupon subsidies, total expenditure captures the full payment to sellers (including coupon subsidies), and unsubsidized expenditure measures spending on orders that did not use coupons. We exclude the post-program period when estimating the effects, resulting in 416,570 observations ($3,787$ individuals $\times$ 2 groups $\times$ 55 days). Standard errors are clustered at the individual level. *, **, and *** indicate statistical significance at the 10\%, 5\%, and 1\% levels, respectively.}
\end{mytable}


\begin{myfigure}[t]
  \includegraphics[width=0.75\linewidth]{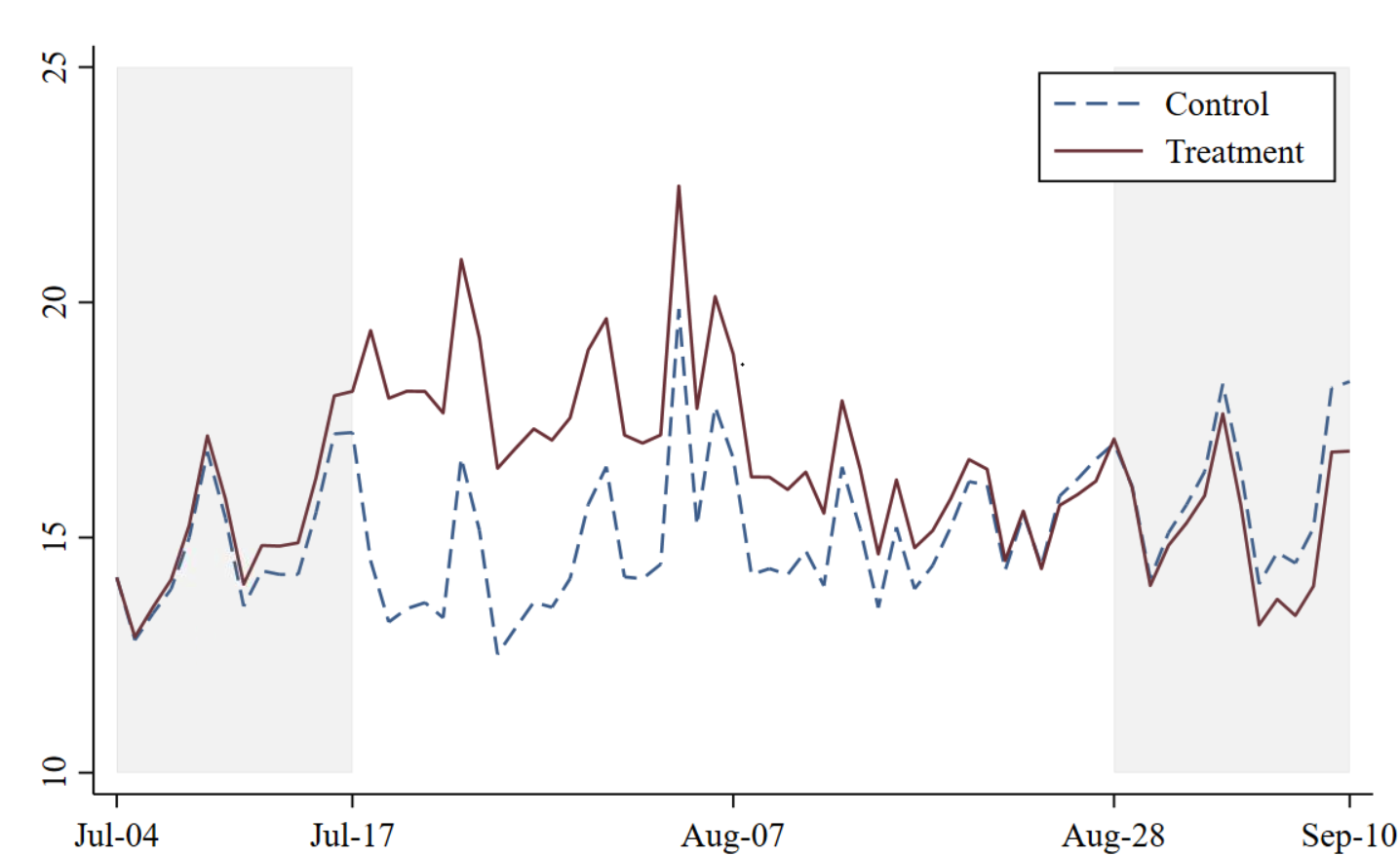}
  \caption{Event Study of Coupon Effects on Out-of-Pocket Expenditure}
  \label{fig:did_ptt}        
        \figurenotes{This figure shows daily average out-of-pocket expenditure for treatment and control groups. The treatment period spans from July 18 to August 27, 2022. Gray shaded areas indicate the two weeks prior to and after the coupon program. The pre-treatment period exhibits parallel trends (F-statistic = 0.31). The treatment-control gap is largest in the first weeks of the program and narrows during the latter half, with full convergence after the coupon event ends.}
\end{myfigure}

\begin{myfigure}[t]
  \centering
  \includegraphics[width=0.8\linewidth]{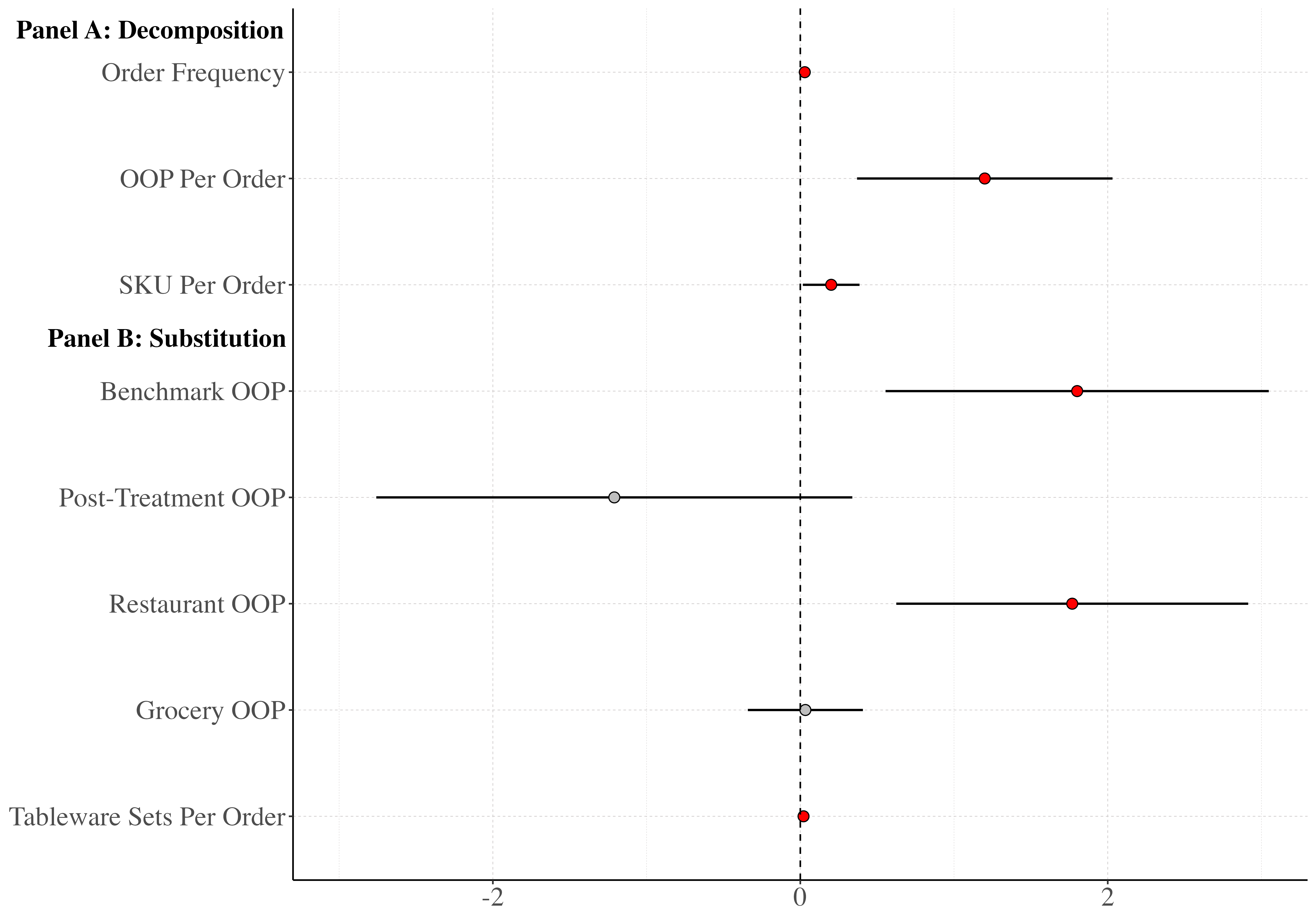}
  \caption{Treatment Effect Decomposition and Substitution Patterns}
  \figurenotes{This figure presents average treatment effects of digital coupons on alternative dependent variables. Panel A decomposes the consumption response along extensive and intensive margins, showing effects on order frequency, out-of-pocket (OOP) expenditure per order, and dishes (SKUs) per order. Panel B examines potential substitution effects. The benchmark OOP shows the baseline average treatment effect on daily OOP expenditure. Post-treatment OOP tests for inter-temporal substitution by comparing spending in the two weeks after the coupon event to the pre-treatment period. Restaurant OOP and grocery OOP show effects on restaurant and grocery spending, respectively, with the latter testing for inter-category substitution. Tableware sets per order proxies for the number of diners and tests for intra-household or workplace sharing. Dots represent point estimates with 95 percent confidence intervals, clustering standard errors at the individual level. See Table \ref{tab:decomp_sub} for the full set of estimates and standard errors.}
  \label{fig:decomp_sub}
\end{myfigure}

Table \ref{tab:did} presents the estimated average treatment effects of coupon receipt following Equation \eqref{eq:block_did}. We focus on three key outcomes: out-of-pocket expenditure, total expenditure, and unsubsidized expenditure. Examining these outcomes provides a comprehensive view of how coupon receipt influences both consumer spending and whether coupon-induced changes spill over to non-discounted transactions.

On average, obtaining coupons raised consumer daily out-of-pocket expenditure by ¥1.80---a roughly 12 percent increase relative to the average daily unsubsidized expenditure of ¥15.04 during the treatment period. Total spending rose even more, reflecting the additional government subsidy. Meanwhile, expenditure on unsubsidized orders showed no statistically significant change, suggesting that the treatment effects of digital coupons do not spill over to non-discounted transactions.

We next examine whether the estimated effects are supported by the parallel trends assumption. Figure \ref{fig:did_ptt} illustrates the group-level time trends for the treatment and control groups. Prior to the coupon event, the two groups followed parallel trends, supported by an F-statistic of 0.31. During the treatment period, the treatment group's out-of-pocket expenditure outpaced that of the control group, with the gap largest in the first weeks of the program and tapering visibly during the latter half. After the program ended, the two groups converged fully. The figure also shows that the control group consumed slightly more than the treatment group during the two weeks after the coupon event, though this visual pattern requires formal statistical testing, which we conduct in the following analysis.

Our estimated effect translates into a spending multiplier of 3.38.\footnote{We define the spending multiplier as the ratio of the increase in total expenditure (out-of-pocket expenditure plus coupon subsidies) to the government subsidy. The estimated daily increase in out-of-pocket expenditure is ¥1.80 per treated individual. The corresponding government subsidy, defined as the daily average coupon discount redeemed per treated individual (see Section \ref{sec:cost}), is ¥0.76. Total expenditure therefore increased by ¥2.56 per treated individual per day, implying a spending multiplier of \(2.56 / 0.76 \approx 3.38\). See Section \ref{sec:acc_iden}  for the accounting identities linking coupon MPC to out-of-pocket spending changes.} While this cannot be given a conventional MPC interpretation---since coupons do not directly increase recipients' income---we follow \citet{Ding2025} and refer to this multiplier as the ``coupon MPC'', the marginal propensity to consume out of coupon subsidies. In other words, for every yuan of government expenditure on coupons, consumers increased their total spending by more than triple that amount. Such a pronounced response aligns with recent studies \citep{Liu2021, Xing2023, Ding2025} documenting similar magnitudes of coupon MPCs. This result exemplifies the distinctive joint-financing feature of digital coupon programs: unlike conventional stimulus payments where consumers are net recipients of government funds, digital coupons generate earnings for local businesses by amplifying government expenditure through consumer contributions, effectively creating a public-private partnership in supporting local businesses.

To understand the primary channels driving this increase in spending, Figure \ref{fig:decomp_sub} decomposes the coupon-induced out-of-pocket expenditure effect. Panel A shows that while the effect of obtaining coupons on order frequency is statistically significant, its magnitude is near zero, indicating minimal extensive margin effects. Moreover, we observe a significant increase in out-of-pocket expenditure per order alongside an increase in dishes per order, suggesting that discount thresholds motivate some consumers to bundle additional items to qualify for the discounts. This response to minimum spending thresholds is further evidenced by clear ``bunching'' behavior in the distribution of order amounts. Figure \ref{fig:bunch} presents histograms of order frequencies by amount, contrasting distributions across different day types. Redemption days---defined at the individual-day level as days during the treatment period when consumers placed at least one order using a coupon---show substantial clustering of orders just above the ¥50 and ¥100 discount thresholds. No such clustering appears on non-redemption days.

Panel B of Figure \ref{fig:decomp_sub} examines the potential substitution patterns discussed in Section \ref{sec:model-did}. We find minimal evidence of inter-temporal substitution, consistent with the perishable nature of delivered meals that limits consumers' ability to shift consumption across time periods. This result also aligns with the visual pattern in Figure \ref{fig:did_ptt}, where the treatment-control gap converges once the treatment period ends, indicating that digital coupons generate immediate but temporary consumption responses.

We also detect no evidence of inter-category substitution within the platform. Replacing the outcome variable with out-of-pocket expenditure on grocery orders yields an estimated effect that is essentially zero, indicating that the increase in restaurant expenditure does not come at the expense of other food-related purchases on Ele.me. Panel B further shows that the number of tableware sets requested per order is statistically significant but economically negligible in magnitude, pointing to no meaningful intra-household or workplace sharing to meet the spending thresholds.

Turning to off-platform substitution, the near-zero extensive-margin effect in Panel A makes substantial reallocation across platforms or dining formats unlikely, since both cross-platform switching and offline-to-online substitution would primarily manifest through additional orders (Section \ref{sec:model-did}). The mobile application usage data are consistent with this interpretation: Ele.me's share of total food-delivery app usage time in Beijing remained stable across the coupon month and subsequent comparison months, with no temporary increase during the event, implying little reallocation of app usage from rival platforms to Ele.me.\footnote{Section \ref{sec:daas-sub} presents details of this exercise. Because coupon recipients constitute a nontrivial share of Ele.me users during the coupon month, any substantial coupon-induced shift of activity onto Ele.me should generate a detectable increase in Ele.me's share of total food-delivery app usage time relative to the non-coupon comparison months.} Taken together, these results indicate that the estimated consumption response reflects genuine additional spending rather than redistribution across time, categories, platforms, dining formats, or household members.

Our analysis reveals several key features of digital coupon effectiveness. The program successfully increased consumer expenditure primarily through the intensive margin, raising spending per order rather than order frequency. Consistent with \citet{Liu2021}, the consumption boost in the Beijing initiative is short-lived. This temporary nature, combined with threshold-induced bunching, suggests that the design features---minimum spending requirements and short expiration periods---are important to its effectiveness. However, the program's impact varies substantially across merchants. Figure~\ref{fig:redem} further reveals that coupon redemptions disproportionately occur at establishments that had higher sales revenue and higher average order prices in the six months prior to the coupon event. This concentration of redemptions at larger, higher-priced establishments reflects the distribution of consumer spending patterns when using coupons, potentially amplifying revenue gains for these businesses. These patterns suggest complex interactions between consumer characteristics and business attributes that we investigate in the following sections.

\subsection{Heterogeneous Consumption Responses}
\label{sec:result-hte}

\begin{myfigure}
  \begin{subfigure}{0.4\textwidth}
    \includegraphics[width=\linewidth]{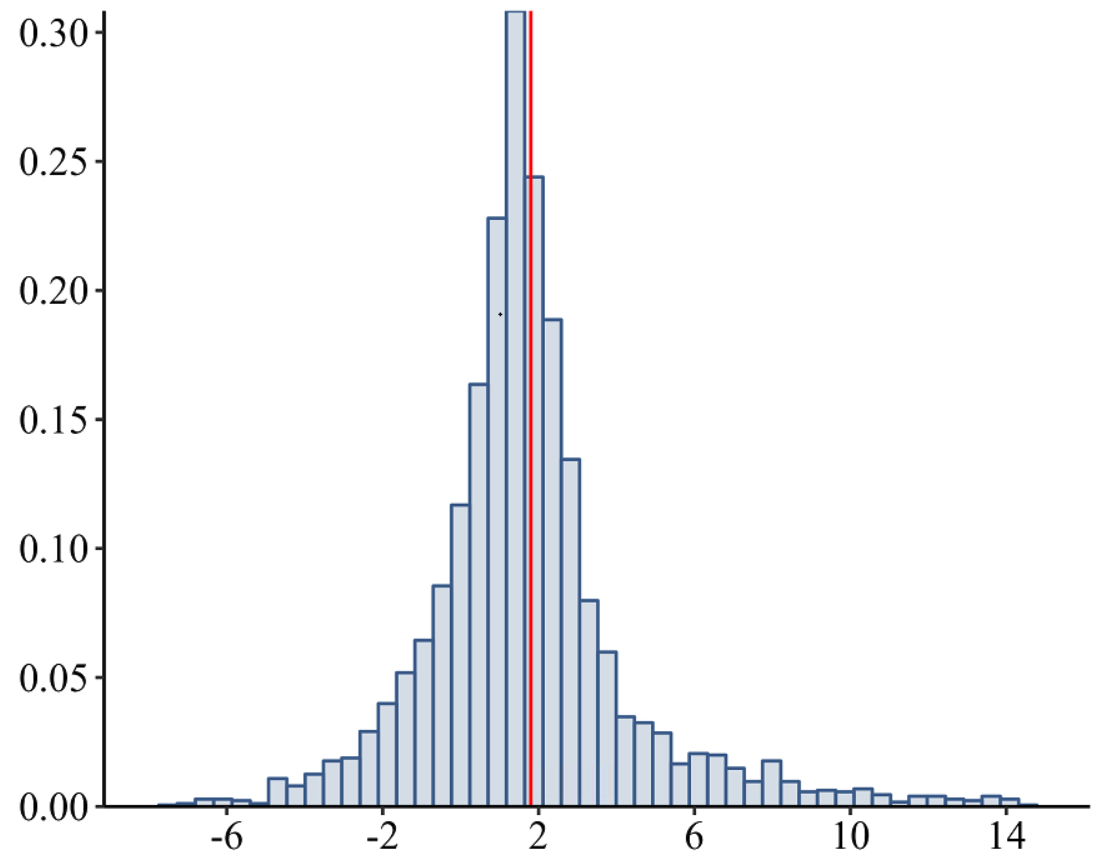}
    \caption{Conditional Average Treatment Effects}
    \label{fig:catt_hist}
  \end{subfigure} \quad{}
  \begin{subfigure}{0.4\textwidth}
    \includegraphics[width=\linewidth]{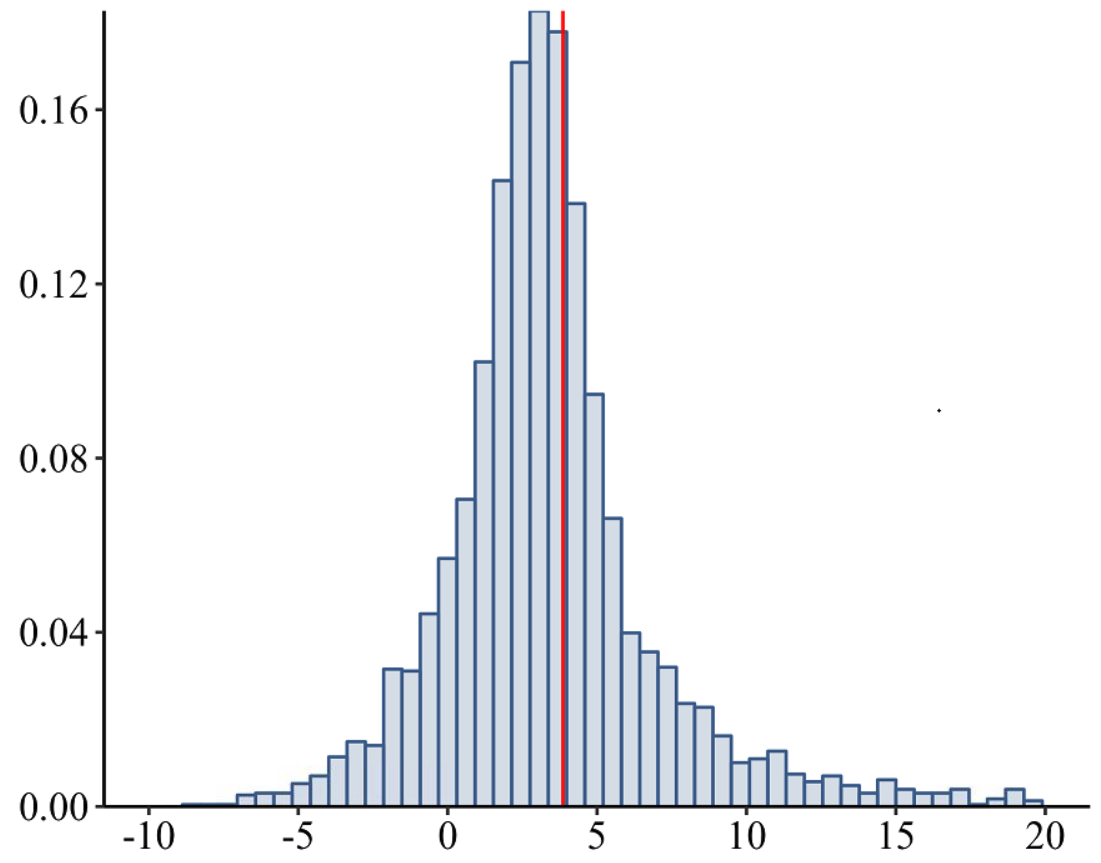}
    \caption{Conditional Coupon MPCs}
    \label{fig:mpc_hist}
  \end{subfigure}
  \caption{Heterogeneity in Individual Responses to Digital Coupons}
  \figurenotes{Panel (a) displays the distribution of estimated conditional average treatment effects (CATTs) on daily out-of-pocket spending among treated individuals. The red vertical line indicates the average treatment effect on the treated (ATT), obtained from the average effects model \eqref{eq:block_did}. Panel (b) shows the distribution of conditional marginal propensities to consume out of coupon subsidies (coupon MPCs), calculated as the ratio of each individual's estimated increase in total spending to the expected coupon subsidy they redeem. The red vertical line marks the benchmark coupon MPC, constructed using the ATT and the average coupon subsidy redeemed across treated individuals.}
  \label{fig:hist}
\end{myfigure} 

\begin{myfigure}  \includegraphics[width=0.8\linewidth]{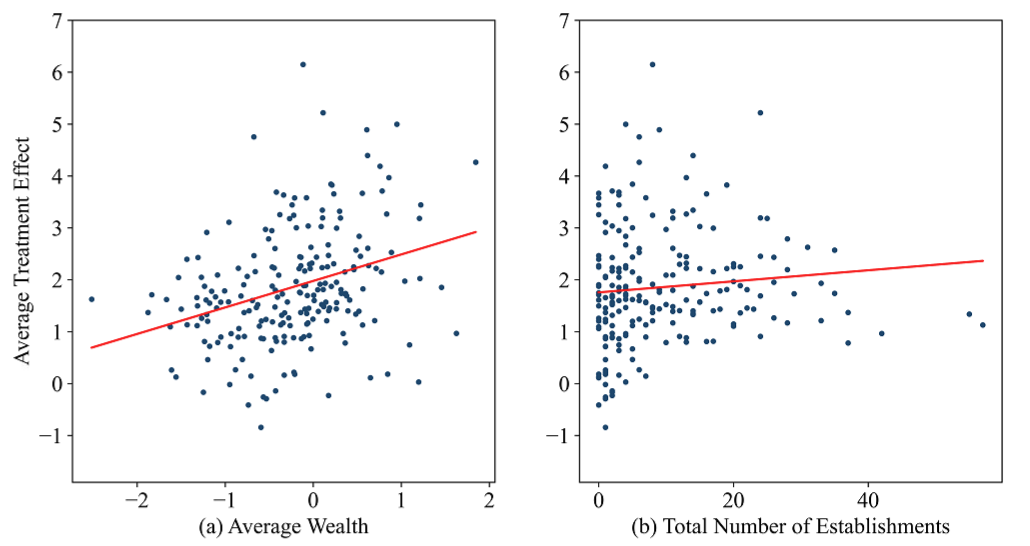}
    \caption{Spatial Distribution of Treatment Effects}
  \figurenotes{This figure plots neighborhood-level average treatment effects against two neighborhood characteristics. Each point represents a neighborhood, defined as a 3km $\times$ 3km grid cell. For each grid, the neighborhood-level average treatment effect is defined as the mean estimated treatment effect on daily out-of-pocket spending among treated individuals residing in that grid. In panel (a), the horizontal axis reports neighborhood average wealth, measured as the mean wealth of treated individuals residing in the grid. In panel (b), it reports the total number of establishments located in the grid. The red line in each panel shows the fitted linear relationship.}
  \label{fig:nbhd}
\end{myfigure} 

The average effects presented in the previous section mask substantial variation in individual responses to the program. Figure~\ref{fig:catt_hist} plots the distribution of estimated heterogeneous treatment effects on daily out-of-pocket spending. While the average effect is ¥1.80 according to the average effect model (marked by the vertical line), the standard deviation of the estimated individual treatment effects is ¥4.42, revealing wide dispersion in consumer responses. Approximately 13 percent of individuals exhibited responses at least twice as large as the average, while 9 percent of individuals accounted for nearly half of the total aggregate effect---indicating that a small group of high spenders drove most of the increased revenue for local businesses. On the other hand, 19 percent of consumers reduced their out-of-pocket spending after receiving digital coupons. For these individuals, coupon redemption translated into net savings.

We can compute conditional coupon MPCs for each individual using their estimated treatment effect on out-of-pocket expenditure combined with their expected government subsidy.\footnote{The expected subsidy reflects variation in both the number of coupon bundles obtained and the probability of redemption across treated individuals. We estimate it using a nonparametric regression of realized subsidies on individual attributes. For details, see Section \ref{sec:cost}.} These individual-level coupon MPCs measure how much additional total spending could be generated per yuan of government subsidy if coupons were allocated to individuals with specific characteristics. Figure~\ref{fig:mpc_hist} presents the resulting distribution. As with treatment effects, the estimated MPCs show considerable heterogeneity: more than 23 percent of individuals had MPCs exceeding 5, while the 19 percent of individuals who reduced their out-of-pocket spending after receiving digital coupons had MPCs below 1. These patterns suggest that governments could significantly boost the stimulus effects of digital coupon programs through targeted distribution. In Section \ref{sec:policy}, we explore targeting rules designed to maximize different policy objectives, including total spending and support for small businesses.

The response to the digital coupon program also exhibited significant spatial variation across neighborhoods. Figure \ref{fig:nbhd_space} maps average per-consumer treatment effects across Beijing's neighborhoods, defined as 3km-by-3km grids. The results show pronounced geographic disparities in program impact. Ranking neighborhoods by their total contribution to aggregate consumer spending, we find that just 11 percent of neighborhoods accounted for half of the total citywide increase in spending. Figure~\ref{fig:nbhd} explores how these neighborhood-level effects relate to local characteristics. On average, wealthier neighborhoods with a higher density of restaurants experienced larger average treatment effects. These findings highlight heterogeneity in stimulus effects at both the individual and spatial levels. In the next section, we examine the underlying drivers of this heterogeneity, focusing on both demand- and supply-side mechanisms.

\subsection{Drivers of Treatment Effect Heterogeneity}
\label{sec:hte-driver}

\begin{myfigure}
  \includegraphics[width=0.7\linewidth]{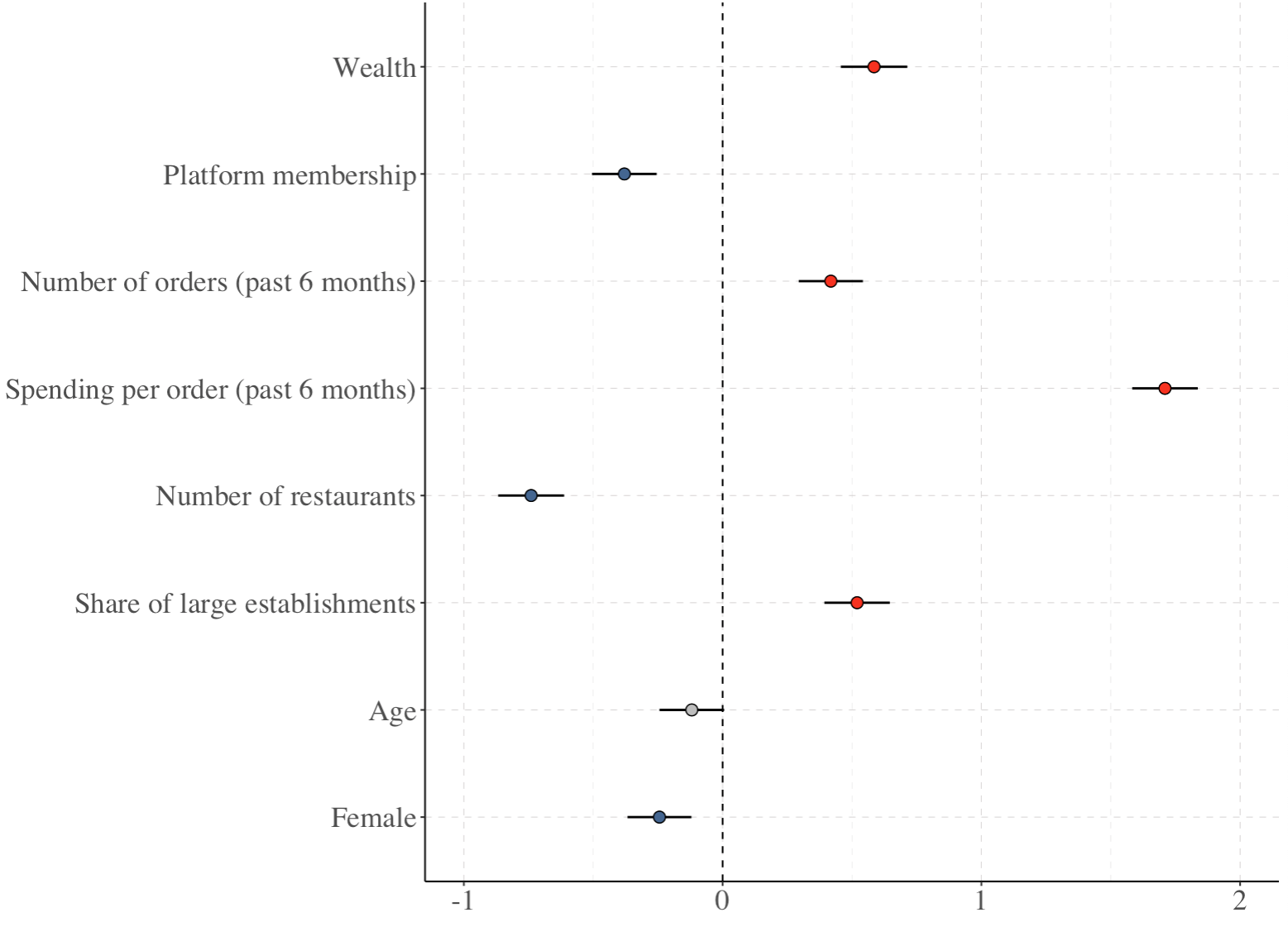}
  \caption{Best Linear Projection of Heterogeneous Treatment Effects}
  \figurenotes{This figure plots the estimated coefficients from the best linear projection (BLP) of individual treatment effects on standardized covariates. All continuous covariates are standardized to have a mean of zero and unit variance, allowing for direct comparison across coefficients as measures of variable importance. The dots represent point estimates; horizontal lines denote 95 percent confidence intervals. See Table \ref{tab:cf_blp} for the full set of estimates and standard errors.}
  \label{fig:cf_blp}
\end{myfigure}

\begin{myfigure}
  \begin{subfigure}{0.32\textwidth}
    \includegraphics[width=\linewidth]{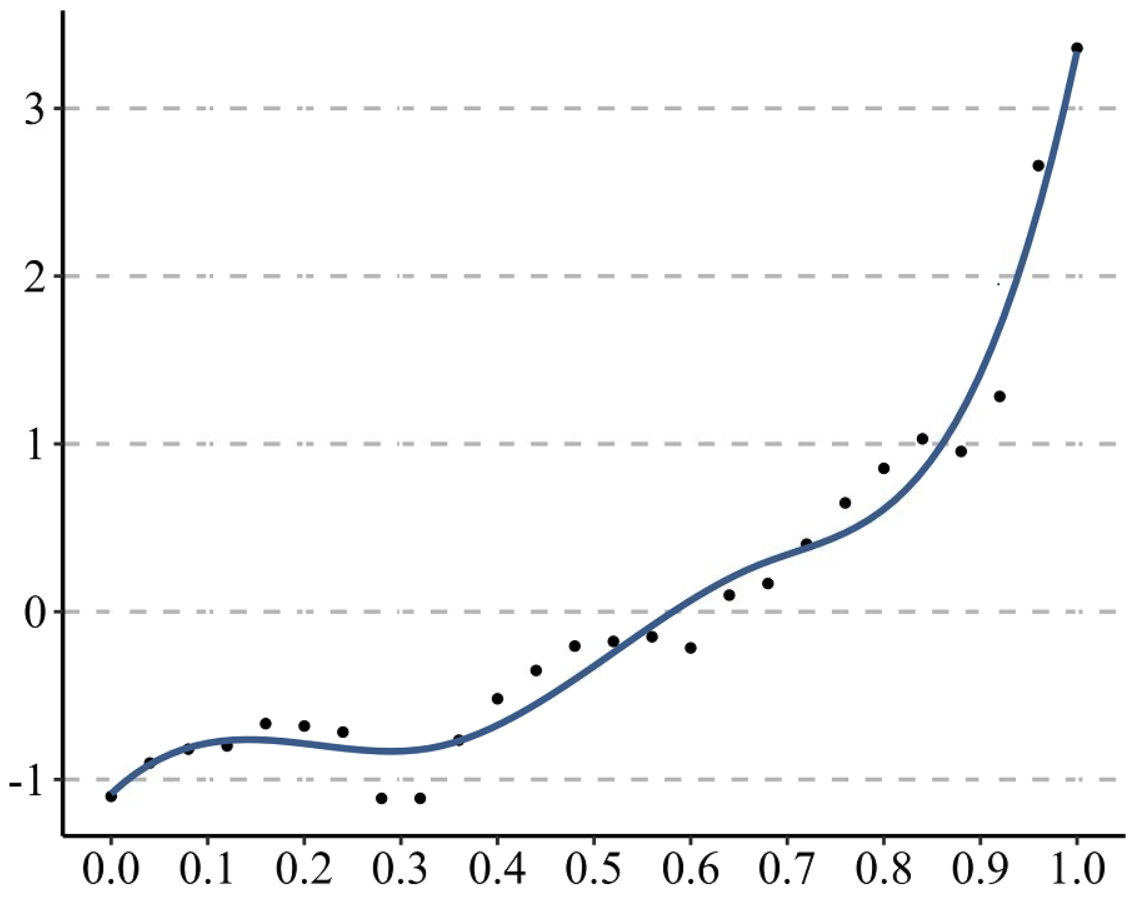}
    \caption{Wealth}
    \label{fig:ale_wealth}
  \end{subfigure}
  \begin{subfigure}{0.32\textwidth}
    \includegraphics[width=\linewidth]{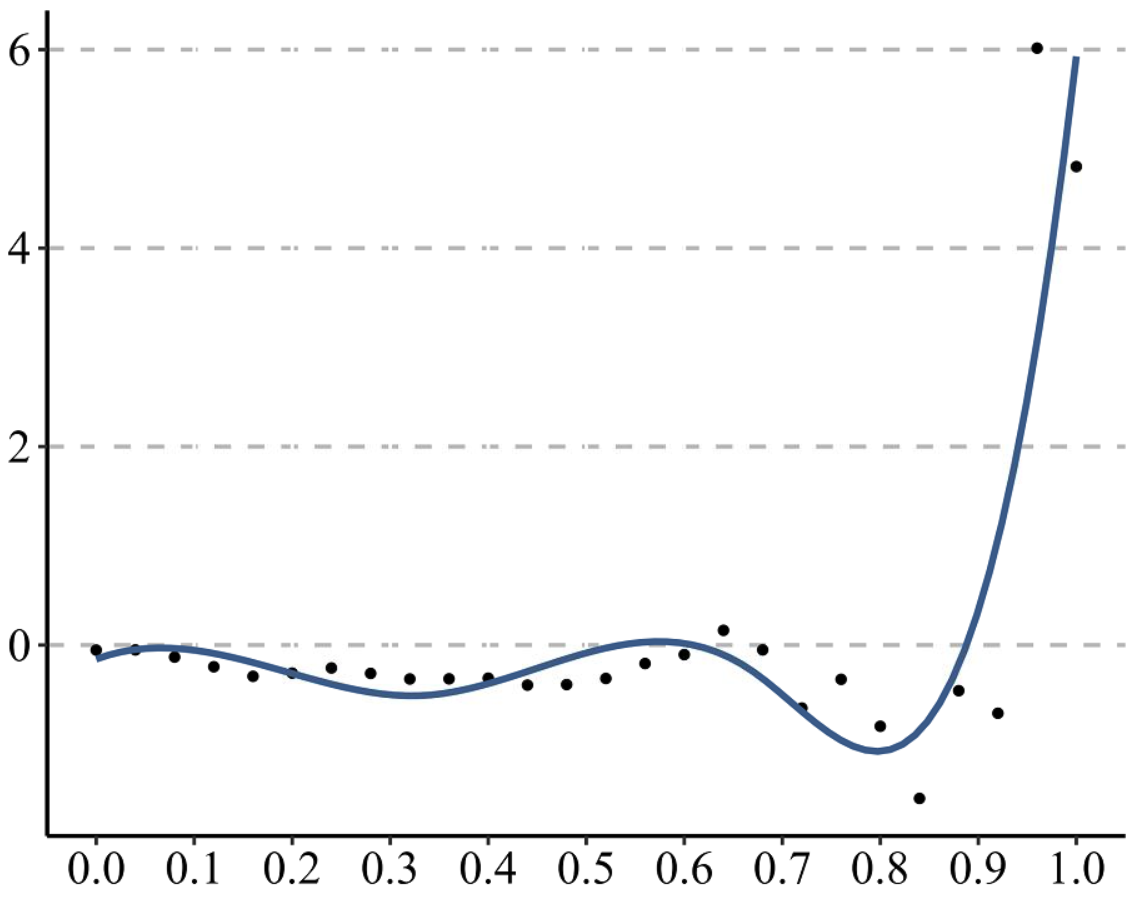}
    \caption{Order Frequency}
    \label{fig:ale_freq}
  \end{subfigure}
    \begin{subfigure}{0.32\textwidth}
    \includegraphics[width=\linewidth]{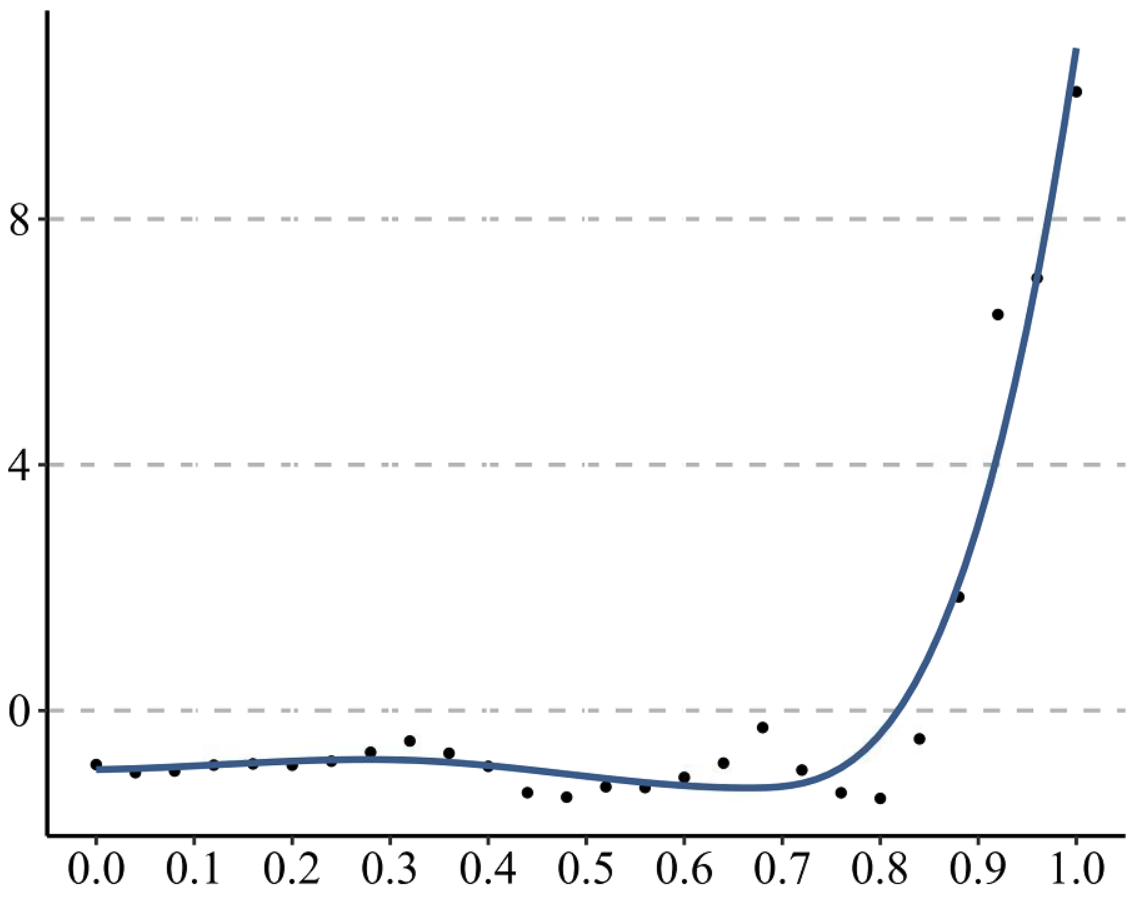}
    \caption{Spending per Order}
    \label{fig:ale_threshold}
  \end{subfigure}
  \caption{Accumulated Local Effects of Demand-Side Factors on Consumption Response}
  \label{fig:ale_d}
  \figurenotes{This figure displays accumulated local effects (ALE) curves for wealth, order frequency (total number of orders placed in the six months prior to the coupon event), and average spending per order (average out-of-pocket expenditure per order over the same period). The horizontal axis represents empirical quantiles of each variable, and the vertical axis shows the variable's marginal influence on the estimated treatment effect, averaging over the observed distribution of the remaining covariates. Dots denote the estimated ALE values, and the solid line is a spline-smoothed fit. All curves are centered to have mean zero.}
  \end{myfigure} 

\begin{myfigure}
  \begin{subfigure}{0.36\textwidth}
    \includegraphics[width=\linewidth]{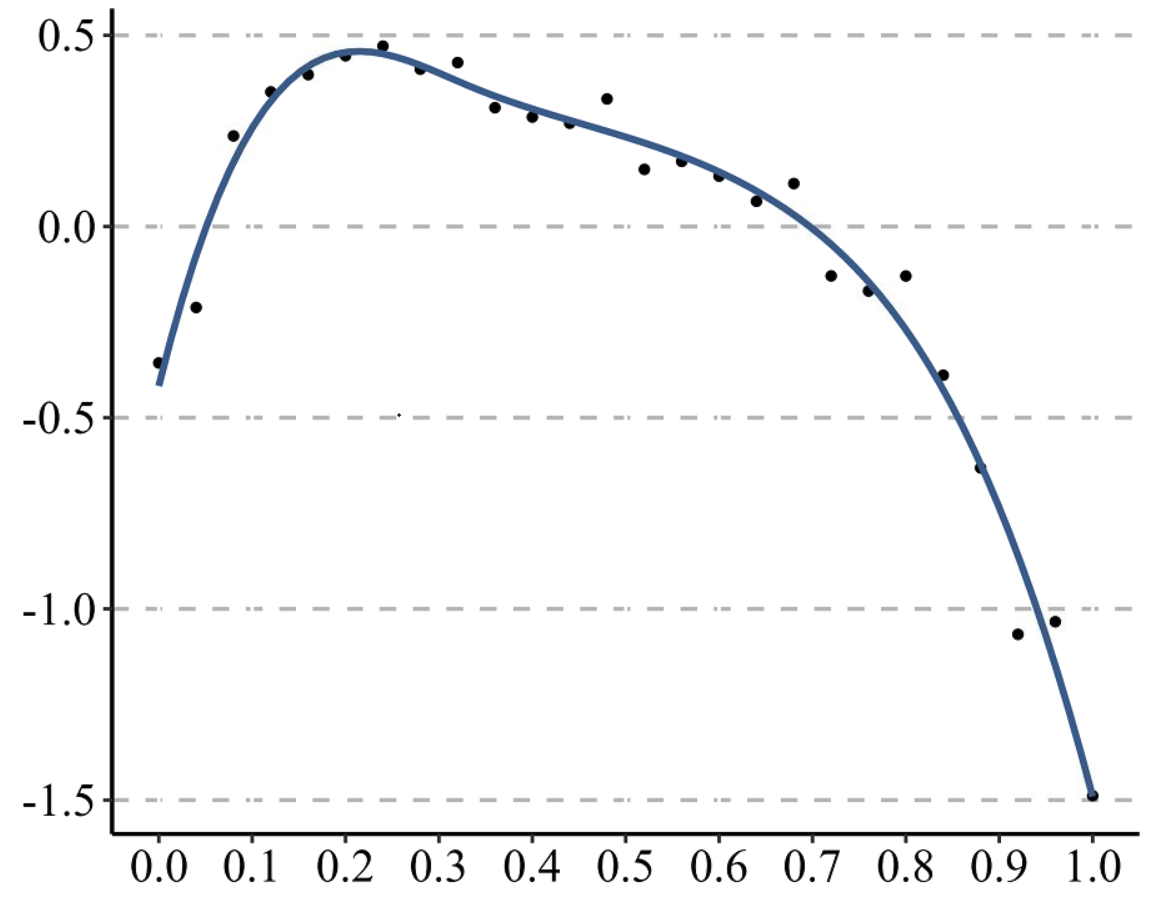}
    \caption{Number of Establishments}
    \label{fig:ale_res}
  \end{subfigure} \quad{}
  \begin{subfigure}{0.36\textwidth}
    \includegraphics[width=\linewidth]{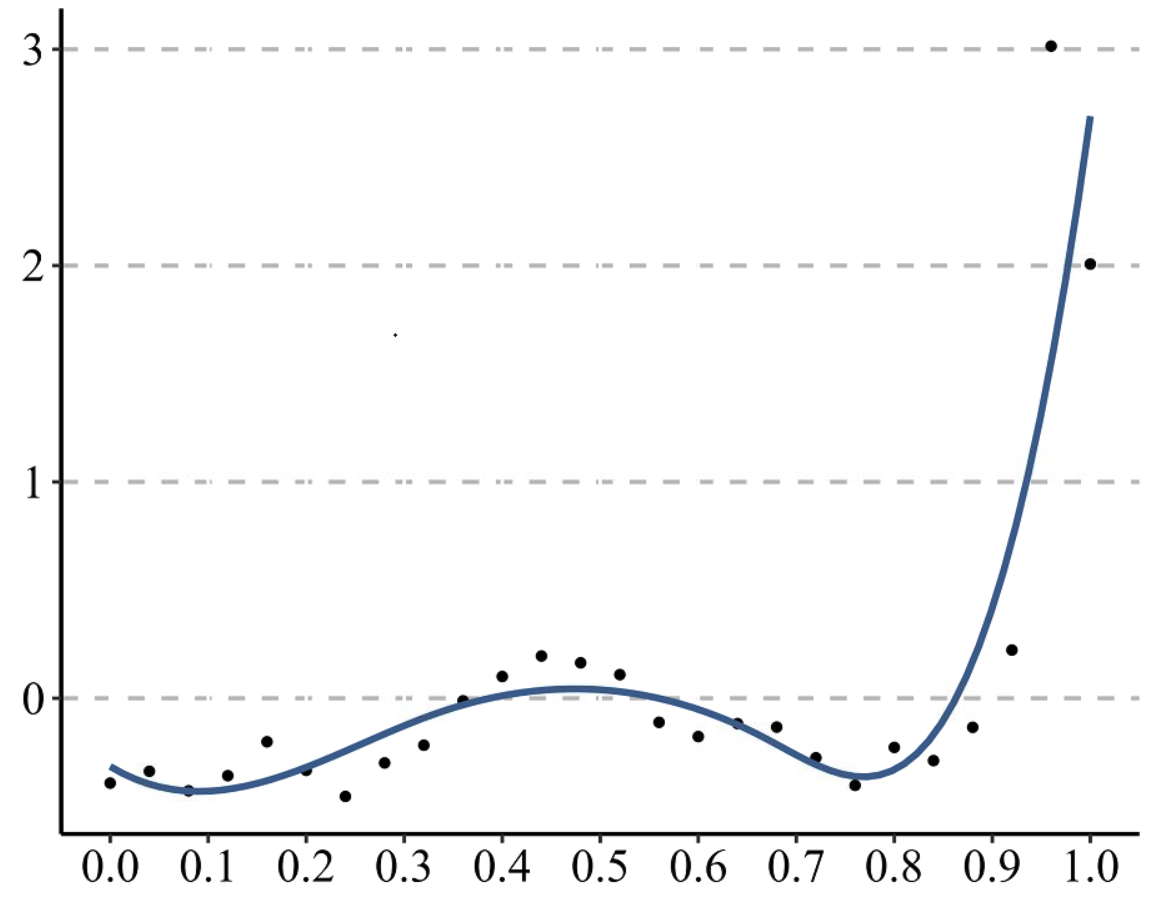}
    \caption{Share of Non-SME Establishments}
    \label{fig:ale_bigshare}
  \end{subfigure}
  \caption{Accumulated Local Effects of Supply-Side Factors on Consumption Response}
  \label{fig:ale_s}
  \figurenotes{This figure displays accumulated local effects (ALE) curves for the number of establishments and the share of non-SME establishments in an individual's neighborhood. The horizontal axis represents empirical quantiles of each variable, and the vertical axis shows the variable's marginal influence on the estimated treatment effect, averaging over the observed distribution of the remaining covariates. Dots denote the estimated ALE values, and the solid line is a spline-smoothed fit. All curves are centered to have mean zero.}
\end{myfigure} 

Figure \ref{fig:cf_blp} displays the estimated coefficients from the best linear projection (BLP) regression \eqref{eq:BLP}, along with standard errors.\footnote{Table \ref{tab:cf_blp} reports the full set of coefficient estimates and standard errors in tabular form. The intercept in the regression is 1.96 and represents the average of the bias-corrected individual treatment effects, i.e., the ATT implied by the heterogeneous effects model.} All potential moderators, except age, are statistically significant predictors of treatment effect heterogeneity. In particular, individuals with higher wealth (as proxied by our wealth index), greater consumption habits (measured by order frequency and average spending per order over the prior six months), and those living in neighborhoods with a larger share of non-SME establishments exhibited larger consumption responses, while being female, having platform membership, and living in areas with more total establishments were associated with smaller treatment effects, holding other factors constant.\footnote{Since the covariates in the best linear projection are standardized, the coefficients in Figure~\ref{fig:cf_blp} directly measure variable importance. An alternative measure is the frequency and magnitude with which each covariate contributes to splitting across trees in causal forest estimation \citep{Athey2019b}. Using this metric, the most important predictors of treatment effect heterogeneity are consumption habits, wealth, and neighborhood consumption amenities (see Figure \ref{fig:vip}). This aligns with the findings from Figure~\ref{fig:cf_blp}.} 

Among the identified drivers of consumption responses, wealth, platform membership, and consumption habits can be considered demand-side factors, reflecting an individual's willingness and capacity to spend after receiving digital coupons,\footnote{We classify consumption habits as demand-side factors in this section because they primarily reflect an individual's baseline propensity to consume on the platform. However, since consumption habits may also partly reflect local consumption opportunities, the Shapley decomposition reported later considers alternative allocations of these variables between demand- and supply-side factors.} while the availability of consumption amenities---captured by the number of establishments and the share of non-SME establishments in the neighborhood---represents supply-side factors that determine the local consumption opportunity set. A central insight of our analysis is that the consumption response to digital coupons---as well as to traditional forms of fiscal stimulus such as cash payments and tax rebates---is jointly shaped by both demand- and supply-side factors. These factors are often correlated, as wealthy individuals tend to reside in neighborhoods with greater consumption amenities, making it difficult to disentangle their respective contributions. While the literature on tax rebates has consistently identified income or wealth as key determinants of consumption responses \citep{Shapiro2003,Johnson2006,Sahm2010,Parker2013},\footnote{The empirical literature evaluating the MPC out of cash-based fiscal stimulus programs has consistently identified wealth and income as key predictors of consumption responses, but presents mixed findings regarding the direction of the relationship. \citet{Sahm2010} and \citet{Shapiro2003,Shapiro2009} find that MPC increases with wealth or income, while \citet{Johnson2006} and \citet{Parker2013} report a negative association. Theoretical models of consumption responses to unanticipated income shocks emphasize the role of liquidity constraints: individuals with limited liquid assets---regardless of total wealth---may exhibit hand-to-mouth behavior and respond more strongly to fiscal transfers \citep{Kaplan2014}. Recent evidence from \citet{Boehm2025}, however, based on randomized experiments designed to mimic realistic cash transfers, suggests that liquid wealth may play only a limited role in explaining variation in consumption responses.} it remains unclear whether these effects partly stem from spatial sorting into neighborhoods with unequal access to consumption amenities.

\label{para:hte-driver-causal}In this paper, through nonparametric estimation of treatment effects based on all observed demand- and supply-side factors, we can disentangle their influences by examining how each factor is related to treatment effect heterogeneity while controlling for the others. For example, the wealth coefficient of 0.59 implies that, conditional on all other observed factors---including local consumption amenities---a one-unit increase in the wealth index is associated with a \yen 0.59 increase in the treatment effect of digital coupons. While the BLP captures the linear influence of each factor, we next examine their potential nonlinear and nonmonotonic effects using ALE curves.

Figure~\ref{fig:ale_d} presents the ALE curves for wealth, order frequency, and average spending per order. Each curve illustrates how the estimated treatment effect varies with the corresponding variable, averaging over the conditional distribution of other covariates. To aid interpretation, the horizontal axis in each plot is scaled to the empirical quantiles of the variable, and each curve is centered to have mean zero. As the figure shows, the effects of consumption habit variables---order frequency and average spending per order---are concentrated among individuals in the top deciles of the distribution. This nonlinearity qualifies the average effects reported in Figure~\ref{fig:cf_blp}, suggesting that the positive association between consumption habits and treatment effects is largely driven by a relatively small group of heavy users. On the other hand, wealth emerges as a consistently strong predictor of consumption responses: the ALE curve for wealth rises steadily across the distribution. On average, a 10 percent increase in the wealth index corresponds to an additional ¥0.45 in daily out-of-pocket expenditure. Importantly, the ALE curve accounts for the correlation between individual wealth and neighborhood consumption amenities by showing how a marginal increase in wealth influences an individual's consumption response within the context of the neighborhoods where individuals of similar wealth reside.

Turning to the supply side, Figure \ref{fig:ale_s} presents the ALE curves for local consumption amenities---the number of establishments and the share of non-SME establishments in the neighborhood. Two conclusions emerge from these plots. First, while Figure~\ref{fig:nbhd} shows that neighborhoods with more restaurants tend to exhibit larger average treatment effects, Figure~\ref{fig:ale_s} reveals a non-monotonic relationship once other factors are controlled for: treatment effects initially rise with establishment density up to the 25th percentile of the distribution, and decline thereafter. This pattern suggests that digital coupons generate the largest consumption response in neighborhoods with a moderate number of establishments, all else equal. On average, as shown in Figure \ref{fig:cf_blp}, establishment density is negatively associated with treatment effects, indicating diminishing returns to consumption amenities in already dense areas.\footnote{This pattern is also consistent with consumers in restaurant-dense areas having stronger offline dining alternatives.} Second, conditional on the number of establishments in a neighborhood, a higher share of non-SME establishments leads to larger treatment effects. Much like the patterns observed for consumption habits, these effects are concentrated in neighborhoods with the highest proportion of large businesses. As shown in Section \ref{sec:result-did}, consumers are more likely to redeem their coupons at larger, higher-priced restaurants. The high concentration of large establishments thus facilitates greater consumption through digital coupons, which helps explain the patterns observed in the ALE plot. Together, these findings demonstrate how both the quantity and composition of neighborhood consumption amenities influence individual responses to digital coupon programs.

We can quantify the relative importance of demand-side and supply-side factors using a Shapley-based variance decomposition of the estimated treatment effects. Section~\ref{sec:shapley} provides the details. Because consumption habits may partly reflect local consumption opportunities and thus the interplay of both demand and supply forces, we first consider the demand side to include only wealth and platform membership. Under this comparison, the relative contribution of the demand side is 58 percent, while that of the supply side is 42 percent.\footnote{The reported shares are normalized to sum to one to reflect their relative contributions.} Alternatively, we allocate the contribution of consumption habits equally between the demand and supply sides. Under this comparison, the demand side accounts for 52 percent of the explained variation, while the supply side accounts for 48 percent. Taken together, these results show that demand- and supply-side factors contribute in broadly comparable magnitudes to shaping the consumption response to digital coupons, with the demand side playing a modestly larger role.

\subsection{Implications for Models of Consumption Behavior}
\label{sec:result-implication}

We now discuss the implications of our findings for models of consumption behavior. \citet{Xing2023} and \citet{Ding2025} develop rational choice frameworks to explain consumption responses to coupons with minimum spending thresholds. In these models, consumer responses depend on their optimal spending absent the coupon. Marginal consumers, whose baseline spending lies just below the threshold, increase their spending to qualify for redemption, while inframarginal consumers, whose baseline spending already exceeds the threshold, treat the coupon subsidy as cash equivalent and respond only through the income effect of the discount. Both models therefore predict bunching at the threshold, with coupon-induced spending arising from threshold-driven reallocation.

Our empirical results provide partial support for this threshold-induced rational optimization mechanism. We observe clear bunching at coupon thresholds (Figure~\ref{fig:bunch}), consistent with marginal consumers adjusting spending to qualify for redemption. At the same time, we also document substantial spending responses among consumers whose baseline spending per order exceeds one or both coupon thresholds and who are therefore likely to be inframarginal with respect to at least part of the coupon bundle. As shown in Figure~\ref{fig:ale_threshold}, treatment effects rise smoothly and sharply in the upper tail of the baseline spending-per-order distribution.\footnote{Figure~\ref{fig:inframarginal} overlays the ¥50 and ¥100 thresholds on the same ALE curve and shows that treatment effects begin to rise around the ¥50 threshold and continue increasing further above ¥100. Table~\ref{tab:interaction} reports corresponding subgroup DiD estimates for three baseline spending-per-order groups---below ¥50, ¥50--100, and above ¥100---providing a simple grouped summary that complements the nonparametric evidence in Figure~\ref{fig:inframarginal}.} Under a simple threshold-crossing account, spending responses would be expected to be concentrated in relatively narrow regions around the relevant coupon thresholds, rather than increasing smoothly throughout the upper tail of the spending distribution. In particular, one would expect relatively small responses both among consumers whose baseline spending lies well below the thresholds and among those whose baseline spending already exceeds them. The pattern we observe is therefore difficult to reconcile with the rational threshold models alone. It is also not readily explained by consumers simply placing additional orders to use the second coupon in the bundle. As shown in Section~\ref{sec:result-did}, the treatment effect operates primarily through the intensive margin, with economically small effects on order frequency.\footnote{\label{fn:infra_limit}A rigorous test of the rational threshold models in our setting would require identifying which consumers are marginal or inframarginal with respect to the coupon bundle in the absence of treatment. That distinction cannot be inferred sharply from baseline spending per order alone. Because the coupons are bundled and only one coupon can be used per order, bundle-level inframarginality depends on the consumer's counterfactual same-day ordering pattern---that is, whether the consumer would have placed no order, one qualifying order, or multiple qualifying orders on that day even without the coupon bundle. Because we do not observe such same-day order histories over the six months prior to the coupon event, baseline spending per order can provide only an imperfect guide to bundle-level marginality or inframarginality. The evidence presented here should therefore be interpreted as suggestive rather than conclusive.}

These patterns motivate consideration of complementary behavioral mechanisms. One candidate is mental accounting \citep{Thaler1999}, under which consumers may treat digital coupons as a distinct budget for extra or hedonic consumption.\footnote{\citet{Boehm2025} formalize this mechanism in a model where recipients of windfall transfers perceive the funds as ``special money'' and experience disutility from allocating them to routine expenditures. Because households face search costs in identifying non-routine uses, the salience of a short expiration window induces earlier spending and raises marginal propensities to consume.} This interpretation is consistent with the finding that the largest treatment effects are observed among individuals who are wealthy, have high spending levels, and reside in areas with a high share of large establishments---those who are more capable of discretionary consumption. The design features of digital coupons may further amplify such responses. Minimum spending thresholds create salient spending targets, while short expiration windows heighten the perceived cost of non-redemption. Consistent with salience models \citep{Bordalo2012,Bordalo2013} and loss aversion \citep{Tversky1991}, these features may elevate spending beyond what standard price incentives alone would predict. Although our empirical design does not separately identify these channels, these behavioral mechanisms may complement rational threshold-based incentives in explaining the heterogeneous consumption responses documented in this section. We now turn to the distributional consequences of those responses.


\section{Distributional Consequences and Policy Design}
\label{sec:policy}

Digital coupons stimulate consumption through heterogeneous consumer responses. Because consumers differ systematically in where they spend, this heterogeneity necessarily translates into heterogeneous impacts across businesses. Understanding this mapping from consumer-side stimulus to firm-level incidence is central to evaluating digital coupon programs as a policy tool. In this section, we document the mapping from consumer responses to business gains and examine how alternative targeting rules affect both overall stimulus and the distribution of benefits across firms.

\subsection{Distributional Impact on Local Businesses}
\label{subsec:welfarebiz}

\begin{myfigure}
  \begin{subfigure}{0.48\textwidth}
    \includegraphics[width=\linewidth]{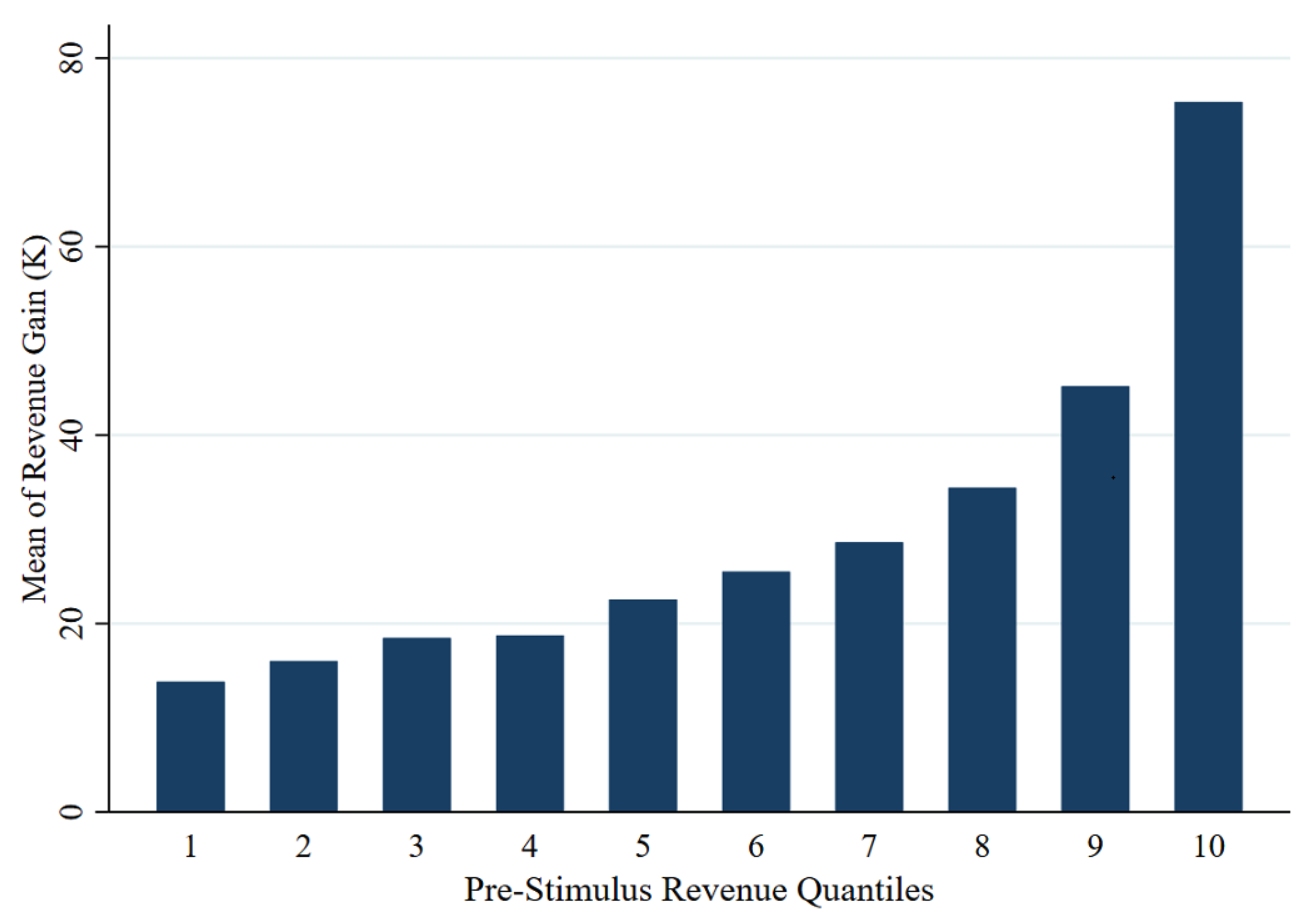}
    \caption{By Establishment Size}
    \label{fig:gains_size}
  \end{subfigure}
  \begin{subfigure}{0.48\textwidth}
    \includegraphics[width=\linewidth]{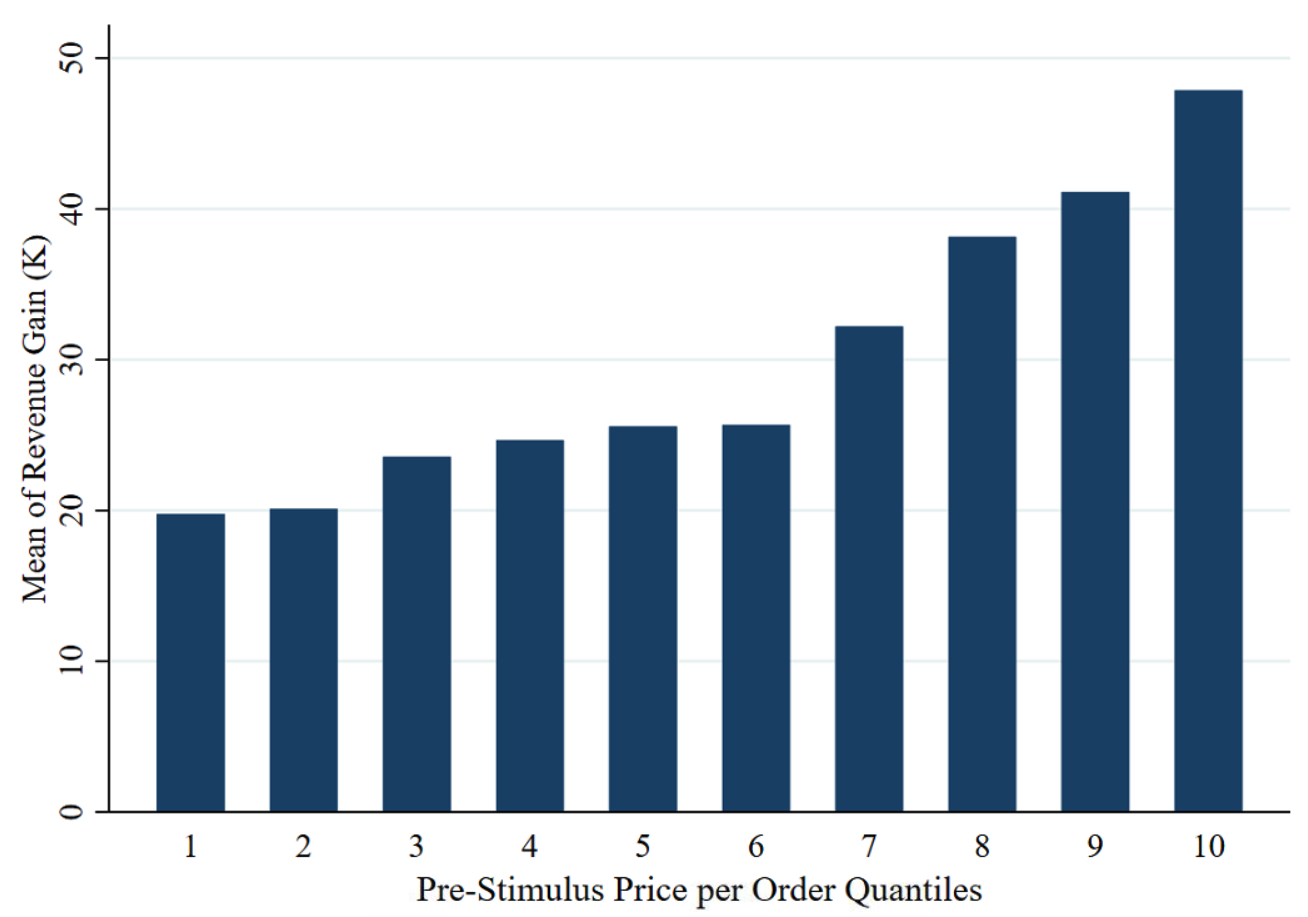}
    \caption{By Price Level}
    \label{fig:gains_price}
  \end{subfigure}
    \caption{Coupon-Induced Revenue Gains Across Establishments}
  \label{fig:gains_business}      
  \figurenotes{This figure presents estimated coupon-induced revenue gains across establishments. Revenue gains are computed by allocating each consumer's estimated increase in total spending---including both out-of-pocket expenditure and the coupon subsidy---across establishments they patronized during the treatment period in proportion to their observed spending shares. Panel (a) groups establishments by sales revenue (six months prior to the coupon event) and reports average gains within each size decile. Panel (b) groups establishments by average order price (six months prior to the coupon event) and reports average gains within each price decile.}
\end{myfigure}

Because digital coupons operate through consumers rather than as direct transfers to businesses, the distribution of gains across firms depends jointly on which consumers respond most strongly and where they allocate their spending. 

To quantify business-level gains, we link individual treatment effects to establishments using transaction data. For each consumer, we allocate their estimated increase in total spending---including both out-of-pocket expenditure and the coupon subsidy---across the establishments they patronized during the treatment period in proportion to their observed spending shares. Section \ref{sec:mapping} provides details.\footnote{\label{footnote:37}Our procedure assumes that each consumer's additional spending is distributed proportionally across the businesses they patronized during the treatment period. This isolates the role of consumer–business matching and abstracts from within-consumer reallocation across establishments. To the extent that coupons can shift incremental spending toward larger or higher-priced establishments, our results represent an underestimate of both the dispersion of business gains and the concentration of benefits among large establishments.} Our results reveal pronounced inequality in revenue gains across businesses, with larger and higher-priced establishments capturing the greatest increases. As shown in Figure~\ref{fig:gains_business}, average gains rise steadily with establishment size, while higher-priced establishments experience systematically larger increases in revenue. These patterns align with consumer redemption behavior documented in Figure~\ref{fig:redem}, which shows more frequent coupon redemptions at businesses with higher historical sales and order prices. 

A key empirical correlation underlies these unequal impacts: consumers with higher coupon MPCs tended to allocate a larger share of their spending to large establishments (Figure~\ref{fig:mpc_pct}). Table~\ref{tabA:btmean} provides additional evidence. Consumers who mainly patronized SMEs differed markedly from those who frequented larger businesses. The former had lower wealth, placed smaller orders, and lived in areas with fewer restaurants overall but a higher share of SMEs. We can quantify the importance of this form of consumer-business matching with a simple counterfactual: suppose all individuals had identical spending allocations equal to overall market shares. Under this scenario, the variance in revenue gains across businesses would decline by 52 percent. Together, these results show that firm-level incidence was shaped by the interaction of heterogeneous consumer responses and consumer-business matching: larger establishments experienced larger revenue gains because they disproportionately served consumers with larger consumption responses to the coupon program.\footnote{This observed matching between high-response consumers and larger establishments should be interpreted as conditional on the threshold structure of the Beijing program. Under alternative coupon designs, a different set of consumers may respond most strongly, potentially altering the observed matching between consumer responsiveness and establishment type.}

These findings have direct implications for policy design. If the goal of a stimulus program is to provide broad-based support to all businesses, particularly smaller or more vulnerable ones, then the existing design of digital coupon programs may not achieve that objective.

\subsection{Policy Design}
\label{subsec:design}

Consumer heterogeneity creates scope for targeting and makes coupon allocation a central policy design choice. Directing coupons toward consumers with the largest expected spending responses has the potential to significantly increase the total stimulus generated by a given budget. However, as shown in the previous subsection, those consumers also disproportionately patronize larger businesses. Efficiency-maximizing targeting may thus concentrate the program's gains among large firms, potentially conflicting with objectives such as supporting vulnerable small businesses. More generally, allocation rules affect not only total stimulus but also business incidence. Coupon policy must therefore be tailored to the policymaker's objective. In this subsection, we evaluate alternative allocation rules under different policy goals.

\subsubsection{Maximizing Stimulus Effect}

\begin{myfigure}
  \begin{subfigure}{0.45\textwidth}
    \includegraphics[width=0.8\linewidth]{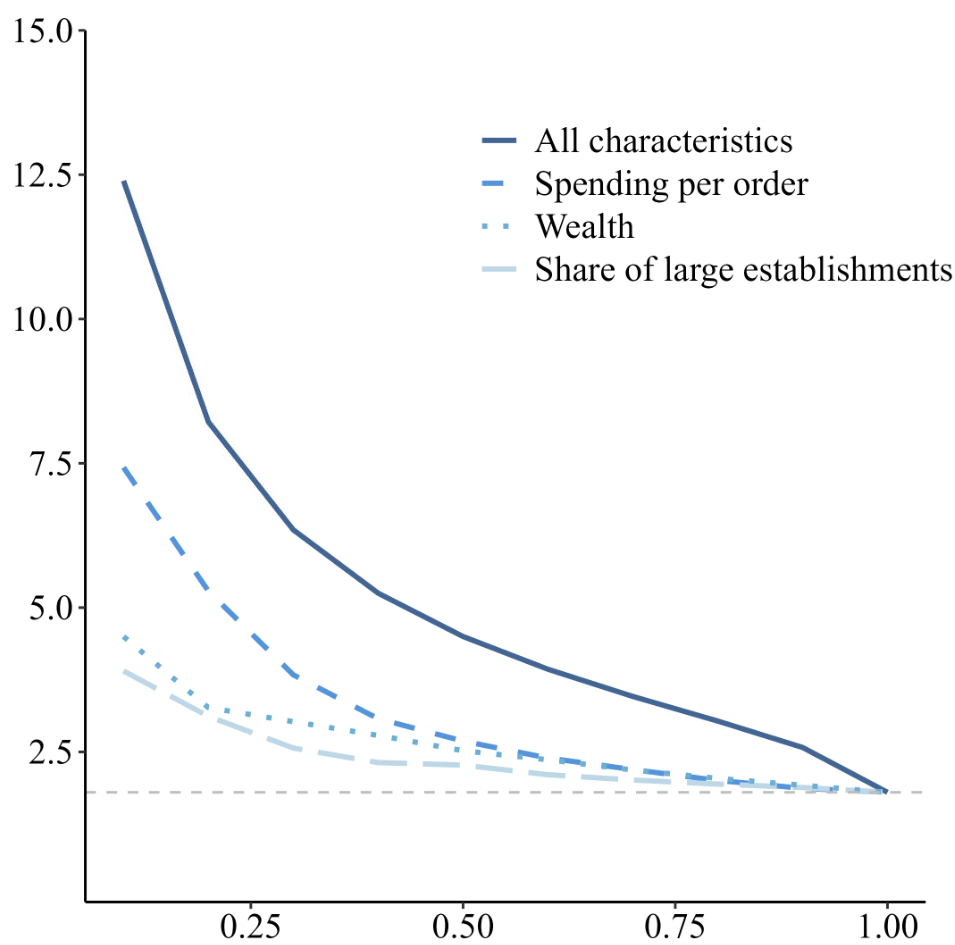}
    \caption{Targeting Efficiency}
    \label{fig:toc}
  \end{subfigure}\quad{}
  \begin{subfigure}{0.45\textwidth}
    \includegraphics[width=0.8\linewidth]{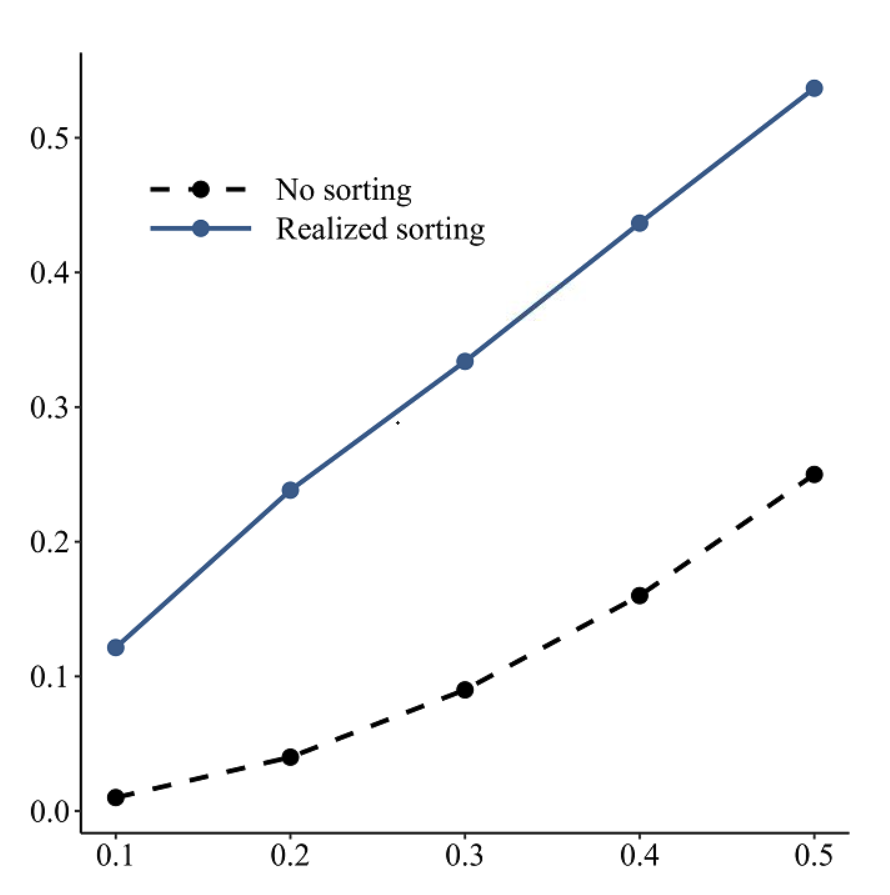}
    \caption{Correlation among Characteristics}
    \label{fig:sorting}
  \end{subfigure}
  \caption{Efficiency Gains from Targeted Coupon Distribution}
  \figurenotes{This figure evaluates the potential efficiency gains from targeting coupon allocation based on consumer characteristics. Panel (a) shows how the average treatment effect varies with the fraction of consumers targeted. For each targeting dimension, consumers are ranked by their predicted treatment effect and each curve plots the average effect when coupons are allocated to the most responsive consumers up to that fraction. The solid navy curve targets using the full set of observable characteristics; the remaining curves target using a single characteristic: spending per order, wealth, and the share of large establishments, respectively. The horizontal gray dashed line indicates the benchmark ATT under the actual coupon allocation. Panel (b) illustrates residential sorting between individual wealth and the share of non-SME establishments within 3km of the individual's home address. For each percentile cutoff (from the top 10\% to 50\%), the vertical axis shows the probability that an individual is simultaneously in the top of both the wealth distribution and the neighborhood non-SME share distribution. The solid blue line shows the actual joint probability, while the dashed black line shows the probability expected under independence. Because targeting on one characteristic also selects consumers who score highly on correlated characteristics, single-characteristic targeting can approximate multidimensional targeting more closely than independent variation would suggest.}

\end{myfigure}

We first examine efficiency gains from targeting coupons based on consumers' predicted responsiveness to the stimulus. Following the rank-weighted average treatment effect (RATE) framework of \citet{Tibshirani2024}, we rank consumers by their estimated treatment effects and compute the average effect when coupons are allocated to the most responsive consumers up to a given fraction of the population.

Figure \ref{fig:toc} presents the results. The vertical axis shows the predicted average treatment effect among targeted consumers, while the horizontal axis indicates the fraction of the population receiving coupons. The solid curve represents targeting based on the full set of individual characteristics, while the remaining curves target based on a single characteristic: spending per order, wealth, and the share of large establishments, respectively.\footnote{Section \ref{subsec:policy-design} presents the full set of targeting strategies and shows the predicted average treatment effects in numeric values by targeting decile.} The horizontal dashed line indicates the benchmark OOP effect of ¥1.80 under the actual coupon allocation. The results demonstrate substantial efficiency gains from targeting: for example, allocating coupons to the top 10 percent of consumers based on their predicted treatment effects increases daily consumption by ¥12.39---more than six times the benchmark.

In practice, governments may not have the capacity to target based on a large range of consumer characteristics. We therefore consider targeting strategies based on single characteristics that are more realistic to implement. The single-characteristic targeting curves in Figure \ref{fig:toc} show that even targeting on one dimension can yield substantial gains: treating the top 10 percent of consumers can achieve daily consumption increases of ¥7.43, ¥4.50, and ¥3.90 for spending per order, wealth, and the share of large establishments, respectively. Expanding to the top 20 percent, spending-per-order targeting more than doubles the stimulus effect compared to the benchmark.

While single-characteristic targeting is less effective than using comprehensive consumer data, spatial sorting patterns partially mitigate efficiency losses. Targeting based on one attribute can identify a population similar to one selected through more complex criteria due to correlations among consumer characteristics. Figure \ref{fig:sorting} illustrates this sorting effect by examining the correlation between wealth and local consumption amenities. The horizontal axis represents different targeting thresholds based on the most responsive consumers (from the top 10 percent to the top 50 percent), while the vertical axis shows the probability that an individual falls within these top percentiles for both wealth and neighborhood share of large establishments. The dashed line shows the expected overlap under independence between these two dimensions (e.g., if 10 percent of individuals are in the top 10 percent of both distributions, the probability would be 0.01 under independence). The solid line shows the actual realized overlap, which is substantially higher. For example, among the top 10 percent wealthiest consumers, approximately 12 percent also live in areas with the top 10 percent highest share of large establishments---twelve times the 1 percent expected under independence. This gap demonstrates that targeting one dimension can capture many of the same consumers as multidimensional targeting, partially compensating for limited information in practical policy implementation.\footnote{While such correlations among consumer characteristics create challenges for disentangling the individual drivers of treatment effects---which we address in this paper---they are helpful for targeting purposes by allowing simple rules to approximate more sophisticated selection criteria.}

\subsubsection{Supporting Small Businesses}
\label{sec:sme_support}
\begin{myfigure}
  \begin{subfigure}{0.32\textwidth}
    \includegraphics[width=\linewidth]{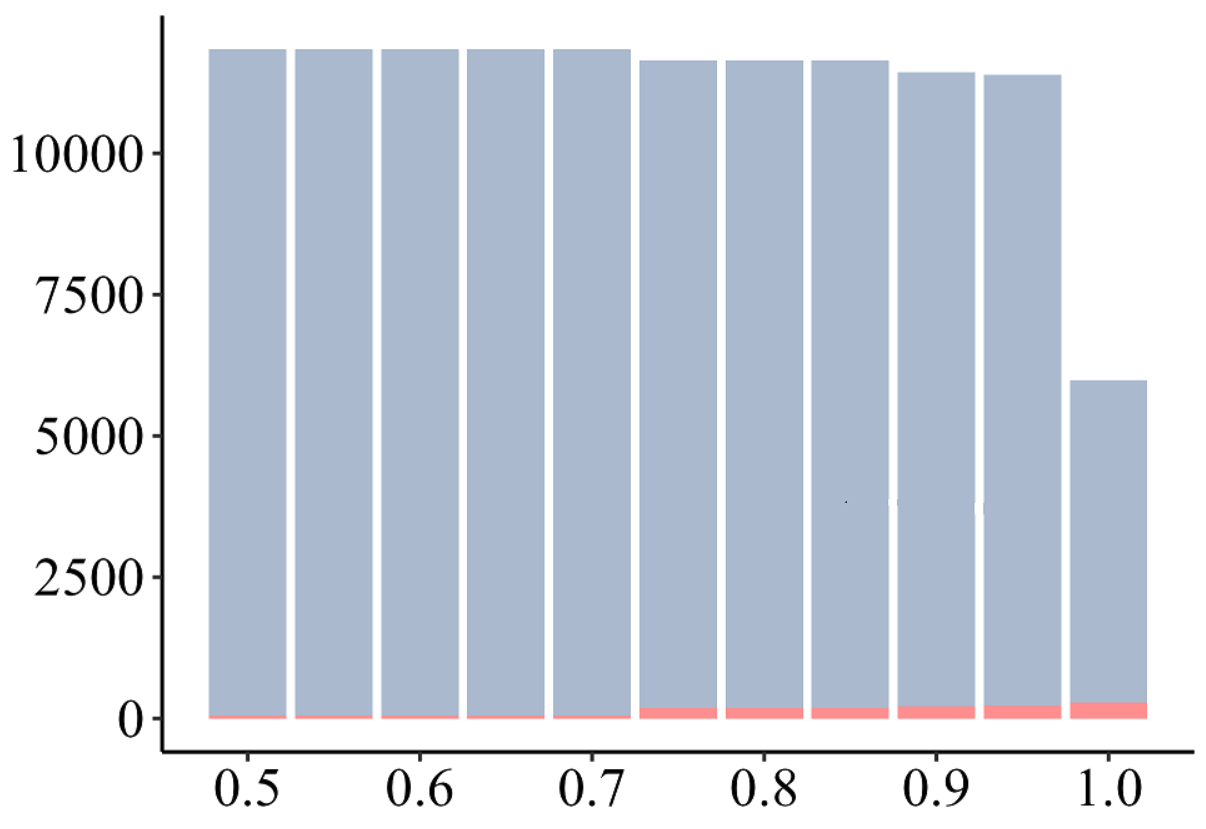}
    \caption{SME at 20th percentile}
    \label{fig:sme20}
  \end{subfigure}
  \begin{subfigure}{0.32\textwidth}
    \includegraphics[width=\linewidth]{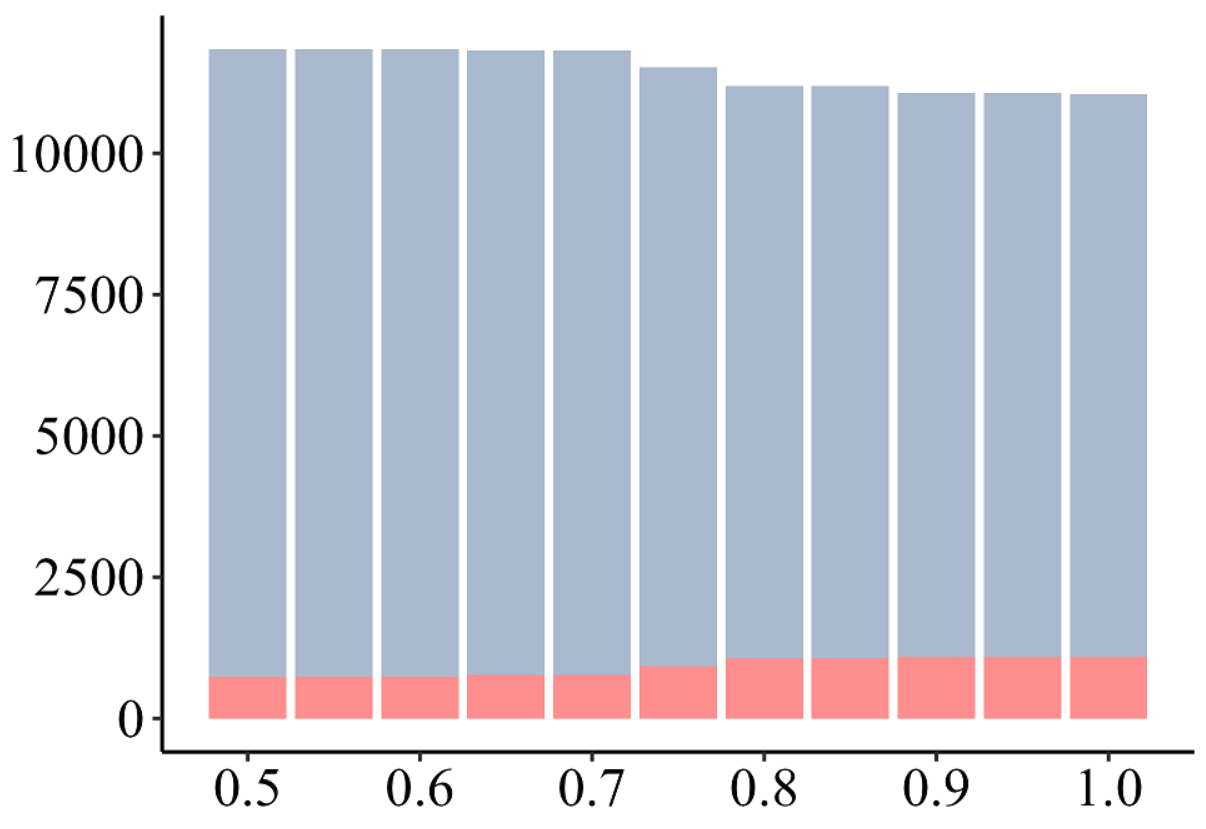}
    \caption{SME at 50th percentile}
    \label{fig:sme50}
  \end{subfigure}
  \begin{subfigure}{0.32\textwidth}
    \includegraphics[width=\linewidth]{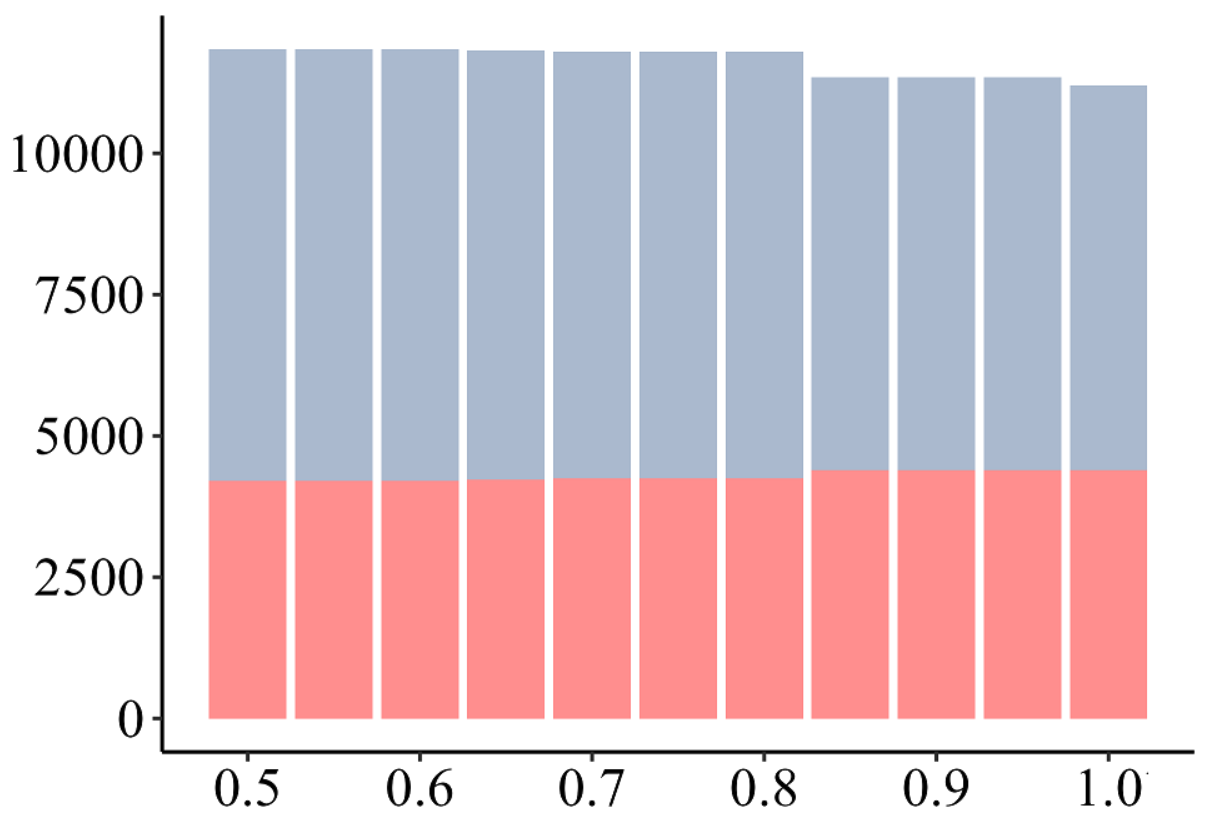}
    \caption{SME at 86th percentile}
    \label{fig:sme80}
  \end{subfigure}
  \caption{Tradeoff Between Overall Stimulus and SME Support}
  \label{fig:sme}
 
  \figurenotes{This figure illustrates the tradeoff between maximizing total coupon-induced revenue and increasing SME revenue share under alternative targeting policies. We define a weighted objective function that places weight $\lambda$ on SME revenue and $1-\lambda$ on non-SME revenue, and construct targeting rules using the optimal policy tree algorithm \citep{Athey2021} to identify the coupon allocation that maximizes this objective. Panel (a) defines SMEs as establishments below the 20th percentile of sales revenue in the six months prior to the coupon event, Panel (b) as those below the 50th percentile, and Panel (c) as those below the 86th percentile. Each bar represents total coupon-induced revenue, decomposed into SME revenue (red) and non-SME revenue (blue). As weight on SMEs increases (moving right along the horizontal axis), the SME share rises while total revenue declines. The corresponding details are presented in Section~\ref{sec:oa-policytree}.}

\end{myfigure}

Supporting SMEs is often a central policy objective for local governments during economic downturns.\footnote{For example, the Beijing municipal government's policy document ``Specific Measures to Help Enterprises Overcome Difficulties and Promote Rapid Recovery of Consumption'' explicitly stated that its digital coupon program aimed to maximize support for restaurant business recovery and development, with specific provisions to support small and medium enterprises in pandemic-affected sectors \citep{Beijing2022}.} SMEs face greater vulnerability to demand shocks due to limited access to credit markets and lower liquidity buffers, and their closure can generate persistent losses in local employment and neighborhood amenities \citep{Chodorow2014, Buera2015, Bartik2020b}. However, targeting based solely on consumer responsiveness may direct spending toward larger businesses, as discussed in Section \ref{subsec:welfarebiz}. These larger businesses may also employ more workers and generate greater tax revenue. This creates a fundamental tradeoff between maximizing total stimulus (which may favor large businesses) and achieving distributional objectives that support vulnerable SMEs. This tradeoff is rooted in the distinct characteristics of consumers who patronize different types of establishments: since higher-MPC consumers generate larger stimulus effects but tend to frequent large establishments, prioritizing SME patronage necessarily involves targeting consumers with lower average responsiveness.

To systematically evaluate policies that balance overall stimulus effectiveness with SME support, we apply optimal policy trees \citep{Athey2021} to search for the best allocation rule by defining a weighted objective. Specifically, we search for policies that maximize $\lambda \cdot R_{\text{SME}} + (1-\lambda) \cdot R_{\text{large}}$, where $R_{\text{SME}}$ and $R_{\text{large}}$ represent coupon-induced consumer spending at SMEs and large establishments, respectively, and $\lambda \in [0,1]$ represents the weight placed on SME outcomes in the policy objective.\footnote{We implement the policy tree algorithm using the \texttt{policytree} package in \textit{R}, restricting the policy class to depth-2 decision trees. Section~\ref{sec:oa-policytree} provides details.} When $\lambda = 0.5$, this objective places equal weight on SME and non-SME revenue and is therefore equivalent to maximizing total stimulus. As $\lambda$ increases from 0.5 to 1.0, the objective places progressively greater value on revenue generated at SMEs relative to large establishments. When $\lambda = 1$, it places weight exclusively on SME revenue. By varying $\lambda$, we trace out the efficient frontier of policies that balance these competing objectives.

We present results across three different classification thresholds of SMEs based on percentiles of sales revenue in the six months prior to the coupon event: establishments in the lowest 20th, 50th, and 86th percentiles. The 50th percentile represents a commonly used threshold and is the definition used throughout our analysis, while the 86th percentile aligns with definitions used by the U.S. Small Business Administration \citep{Census2021} for small business classification.\footnote{The U.S. Small Business Administration defines ``small'' establishments as those accounting for 50 percent of total industry revenue. Applied to our Beijing restaurant data, this criterion corresponds to the 86th percentile of six-month pre-event sales revenue.} These thresholds show how the tradeoff varies with the breadth of the SME definition.

Figures \ref{fig:sme20} to \ref{fig:sme80} present results across these three SME definitions. The analysis reveals a clear tradeoff: increasing the weight on SMEs boosts their share of induced revenue but reduces overall stimulus effectiveness. For example, in Figure \ref{fig:sme50}, prioritizing SMEs increases their revenue share from 6 percent to 10 percent, but at the cost of a 7 percent reduction in total revenue generated. This pattern persists across all SME definitions, illustrating the tension between maximizing overall economic impact and ensuring equitable distribution of benefits.\footnote{\label{fn:sme_welfare}In addition to the efficiency-equity tradeoff quantified here, there is also a consumer-side normative consideration. Because consumers who patronize SMEs tend to have lower wealth (Table~\ref{tabA:btmean}), policies that shift more revenue toward SMEs also rely more heavily on lower-wealth consumers' out-of-pocket spending to generate the stimulus effect. Under a standard rational benchmark, receiving digital coupons should always yield non-negative utility gains, so targeting lower-wealth consumers would directly benefit them. However, the empirical findings in Section~\ref{sec:result-implication} suggest that the observed responses are not fully explained by simple rational threshold-based models alone. If stimulus relies on consumer co-financing and behavioral mechanisms also shape spending responses, then shifting more of the financing burden toward lower-wealth consumers may have ambiguous welfare implications for those consumers, even if such targeting better supports SMEs. Recognizing this consideration can be important in the policy design of digital coupons, even though our paper does not attempt a behavioral welfare analysis.}

\subsubsection{Reconciling Efficiency and Equity}
\label{sec:eff-equ}

The previous section shows that targeting can be used to navigate the efficiency–equity tradeoff inherent in digital coupon programs. By reallocating coupons across consumers, policymakers can shift the balance between maximizing total stimulus and directing benefits toward SMEs. Yet demand-side targeting has an inherent limitation: policymakers can decide which consumers receive coupons, but not which firms those consumers ultimately patronize. Because business incidence is mediated by consumer spending choices, even an optimally targeted coupon policy cannot precisely direct support to particular establishments. Supply-side instruments such as direct transfers have the opposite property: they can target firms precisely, but they do not harness consumer co-financing and therefore cannot generate the same stimulus multiplier as digital coupons. These complementary strengths and limitations motivate a hybrid policy that combines targeted coupon distribution with direct transfers to SMEs. Specifically, we propose allocating coupons to a targeted subset of high-response consumers and using the resulting budget savings to provide direct transfers to SMEs. Because these consumers generate more spending per yuan of government subsidy, a given level of aggregate stimulus can be achieved with lower coupon expenditure, freeing fiscal resources for firm-side support.

To illustrate, Table \ref{tab:policy} compares three allocation strategies under a fixed government budget of ¥117,108, which corresponds to the observed coupon subsidy expenditure in our sample. We compare the total stimulus achieved by each strategy, defined as the total monetary support delivered to local businesses. The ``Random allocation'' strategy randomly assigns coupons to half of the sample.\footnote{In practice, the resulting allocation corresponds to our matched estimation sample, where treated and control units are balanced on observed characteristics and half of the sample is treated.} Under this allocation, consumer out-of-pocket spending totals ¥315,773. Together with the ¥117,108 in coupon subsidies, this yields a total stimulus of ¥432,881. The ``Full targeting'' strategy ranks all consumers in our sample by their predicted treatment effects and allocates coupons to the most responsive individuals until the full government budget is exhausted. Despite treating a comparable share of consumers (51 percent), this approach more than doubles consumer out-of-pocket spending to ¥725,960 and raises total stimulus to ¥843,068, but provides no targeted support to SMEs.

The ``Hybrid policy'' targets only the most responsive 11 percent of consumers. Under this allocation, consumer out-of-pocket spending totals ¥405,220, while requiring only ¥27,551 in coupon subsidies. The coupon component of the policy therefore generates a total stimulus of ¥432,771, approximately matching the ¥432,881 achieved under random allocation, while using less than one quarter of the original budget. The remaining ¥89,557 of the government budget can then be transferred directly to SMEs. With these direct transfers counted one-for-one as support delivered to businesses, total stimulus reaches ¥522,328, exceeding the random allocation benchmark by approximately 21 percent while also providing targeted support for vulnerable businesses.
\begin{mytable}[t]{0pt}
    \centering
    \caption{Comparison of Alternative Coupon Distribution Policies}
    \label{tab:policy}
      \begin{tabular}{p{0.23\textwidth}C{0.15\textwidth}C{0.15\textwidth}C{0.15\textwidth}C{0.15\textwidth}C{0.15\textwidth}}
      \toprule
      \midrule
      \textbf{Policy Design} & \textbf{Consumers Treated} & \textbf{Coupon Subsidies} & \textbf{Consumer Out-of-pocket Spending} & \textbf{Direct Transfers to SMEs} & \textbf{Total Stimulus} \\
      \midrule
Random allocation & 50\% & 117,108 & 315,773 & 0 & 432,881 \\
Full targeting & 51\% & 117,108 & 725,960 & 0 & 843,068 \\
Hybrid policy & 11\% & 27,551 & 405,220 & 89,557 & 522,328 \\
      \bottomrule
      \end{tabular}
      
\tablenotes{This table compares three coupon allocation strategies under a fixed total policy budget of ¥117,108. The ``Random allocation'' strategy randomly assigns coupons to half of the sample. The ``Full targeting'' strategy ranks all consumers by their predicted treatment effects and allocates coupons to the most responsive individuals until the full budget is exhausted on coupon subsidies. Under the ``Hybrid policy,'' the algorithm targets only enough high-response individuals to approximately match the total stimulus achieved under random allocation. This requires only ¥27,551 in coupon subsidies, allowing the remaining ¥89,557 of the fixed budget to be redirected as direct transfers to SMEs. All monetary values are aggregates over the entire coupon period. Total stimulus is defined as the sum of consumer out-of-pocket spending, coupon subsidies, and any direct transfers to SMEs.}
  \end{mytable}

In summary, targeting coupons toward the most responsive consumers can substantially improve program efficiency but concentrates benefits among large establishments. The hybrid approach addresses this limitation by combining the multiplier advantage of targeted coupons with the targeting precision of direct transfers: coupons generate aggregate stimulus, while direct transfers allow policymakers to channel support to SMEs without relying on consumer spending choices. More broadly, our results demonstrate that when policy objectives include both overall stimulus and the distribution of gains across firms, combining demand-side and supply-side instruments can outperform either instrument used in isolation.

\section{Conclusion and Policy Implications}
\label{sec:discussion}

Digital coupon programs have emerged as a popular stimulus tool, yet we do not fully understand their distributional consequences. Our analysis of Beijing's 2022 initiative reveals that digital coupons amplify government subsidies through additional consumer spending, so that both their stimulus effect and firm-level incidence depend critically on heterogeneous consumer responses. While the program generates significant increases in short-term spending, the resulting revenue gains are distributed unevenly across businesses. These findings offer actionable insights for the design of digital coupon programs, with broader implications for demand-side stimulus policies.

Governments typically employ two broad categories of stimulus tools during economic downturns. Demand-side instruments seek to raise consumer spending through transfers or consumption incentives directed at households, such as the U.S. Economic Stimulus Payments in 2001, 2008, and 2020, as well as more recent policies using prepaid cards or time-limited consumption vouchers \citep{Boehm2025}. Supply-side instruments, by contrast, provide direct support to businesses. During the COVID-19 pandemic, many governments implemented business-assistance programs, such as the U.S. Paycheck Protection Program and related schemes in other countries, to help vulnerable enterprises survive temporary demand shortfalls.

Digital coupons represent a demand-side instrument in the form of consumption vouchers with minimum spending thresholds and short expiration windows. Our study finds that they generate significant and highly heterogeneous consumption responses. This heterogeneity, in turn, shapes the incidence of the program: high-response consumers tend to patronize larger businesses, leading to an uneven allocation of stimulus benefits. More broadly, this mechanism extends beyond digital coupons. Whenever a demand-side stimulus policy elicits heterogeneous responses from consumers, those differences will translate into unequal firm-side gains so long as consumers are systematically matched to firms---that is, so long as different types of consumers frequent different types of businesses. Consequently, the incidence of demand-side stimulus policies, in terms of which businesses ultimately benefit, cannot be understood from consumer responses alone.

At the same time, our setting is more specific than many traditional coupon or stimulus environments. The program we study targets food delivery, a largely non-storable consumption category, which likely reduces the scope for intertemporal substitution through stockpiling, although cross-category substitution remains possible. Our results are therefore most directly informative about digital coupon stimulus in an immediately consumable service category. The heterogeneity we document generalizes beyond this particular consumption category and has immediate implications for how digital coupon programs should be designed.

The substantial heterogeneity in consumer responses implies that a uniform allocation of coupons is unlikely to be optimal and creates scope for more deliberate policy design. Through targeted distribution, policymakers can significantly increase aggregate stimulus by directing coupons toward consumers with the strongest predicted responses. Alternatively, targeting can be used to steer spending toward businesses that policymakers wish to support, such as SMEs. Doing so entails an efficiency-equity tradeoff: shifting revenue gains toward small businesses comes at the cost of lower aggregate stimulus. Importantly, we show that such targeting can be implemented using only a few observable characteristics, making it feasible even in settings with limited data infrastructure.

Hybrid policy provides a way to pursue the dual objectives of raising aggregate stimulus and supporting desired businesses by combining the strengths of demand-side and supply-side instruments. Demand-side targeting can generate stimulus efficiently, but it influences business outcomes only indirectly through the consumers who receive coupons and the businesses they happen to patronize. Supply-side instruments, by contrast, can channel support directly to specific businesses, but they do not leverage additional consumer spending and therefore do not produce the same multiplier effects as digital coupons. A hybrid approach thus uses each instrument for what it does best: targeted coupons maximize consumption stimulus, while direct business subsidies ensure that support reaches vulnerable firms such as SMEs.

Taken together, our analysis points to several broader lessons for stimulus design. First, policymakers should recognize explicitly that many stimulus instruments involve efficiency-equity tradeoffs: policies that appear effective in aggregate may distribute gains unevenly. Second, heterogeneity in consumer responses can be a valuable input into policy design, allowing governments to improve the effectiveness of stimulus policies through targeting. Third, when policy objectives include both aggregate demand support and support for vulnerable businesses, combining demand-side and supply-side instruments may be more effective than relying on either margin alone.

Finally, while digital coupons appear especially well suited to generating immediate, targeted stimulus, cash-transfer programs may be better suited to providing medium-term, broad-based consumption support. The choice between instruments should depend on specific policy objectives: immediate targeted stimulus versus broader consumption support, and short-term stimulus versus longer-term economic stabilization.

One important margin of policy optimization lies beyond the scope of our analysis. Digital coupon programs can be optimized along two distinct margins: targeting, which determines who receives coupons, and coupon design, which determines features such as minimum spending thresholds and expiration windows. Because the Beijing program does not provide the independent variation needed to separately identify the causal effects of alternative coupon designs, our analysis focuses only on the first margin. We therefore do not draw conclusions about how changes in thresholds or expiration windows would affect consumption responses or business-side incidence. Studying this second margin, and characterizing optimal design parameters and targeting rules under alternative policy objectives, is an important direction for future research.

\clearpage 
\setstretch{1} 
\singlespacing 


\clearpage
\begin{center}
{\LARGE Online Appendix for ``Consumption Stimulus with Digital
Coupons: Heterogeneity and Policy Design'' \par}
\vspace{1.5em}
{\large by Ying Chen, Mingyi Li, Jiaming Mao and Jingyi Zhou \par}
\end{center}
\vspace{1em}

\addcontentsline{toc}{section}{Online Appendix}
\renewcommand{\thefigure}{OA.\arabic{figure}}
\renewcommand{\thetable}{OA.\arabic{table}}
\renewcommand{\theequation}{OA.\arabic{equation}}
\renewcommand{\thesection}{OA\arabic{section}}
\renewcommand{\thesubsection}{\thesection.\arabic{subsection}} 

\setcounter{figure}{0}
\setcounter{table}{0}
\setcounter{equation}{0}
\setcounter{section}{0}
\onehalfspacing

\section{Data and Sample Construction}

\subsection{Mobile Application Usage Data}
\label{sec:daas}
Our analysis sample comprises users who participated in the coupon event on the Ele.me platform. To assess whether these participants are representative of the broader Ele.me user base, we draw on a large-scale mobile application usage dataset from a major telecommunications provider in China. The dataset tracks monthly usage time for all mobile applications with at least 500 users among all subscribers to the provider's service in Beijing during August 2022, August 2023, and November 2023. The telecommunications provider collaborates with the Bureau of Statistics to calculate weights using population census data, enabling representative estimates of app usage across Beijing's population. The sample covers 10.87 million subscribers in August 2022 (43.5 percent of Beijing's residential population) and 11.20 million in August 2023 (45.7 percent), representing approximately 21.8 million weighted residents in both periods. We also use this data to assess potential platform substitution in Section \ref{sec:daas-sub}.

We assess the representativeness of our baseline sample by comparing it to the population of all Ele.me users identified in the mobile application usage data. Among coupon event participants in our sample, the average age is 32 years and the female share is 64 percent. Among all Ele.me users in the mobile application usage data, the average age is 35 years and the female share is 53 percent. While these samples are not directly comparable (the analysis sample is drawn through stratified random sampling from coupon program participants, whereas the mobile application usage data captures all Ele.me users in Beijing), the demographic profiles suggest that coupon participants skew somewhat younger and more female than the broader Ele.me user base. The modest differences indicate that our analysis sample, while focused on the event participants, reflects a reasonably representative subset of the platform's active user population.

\subsection{Matching Procedure}
\label{sec:psm-det}
We employ a Logit model to estimate propensity scores using 4,237 treatment and 2,225 control users with complete individual characteristics data. The model includes our full set of observed characteristics, including age, gender, platform membership, wealth, consumption habits (order frequency and expenditure per order over the six months prior to the coupon event) and local consumption amenities (the number of establishments and the share of non-SME establishments in the neighborhood). Table \ref{tab:psm} compares individual characteristics between groups before and after matching. Panel A shows significant differences across all variables before matching. Panel B demonstrates successful balancing post-matching, with no significant differences and small magnitudes relative to mean values.


\begin{mytable}[h]{0pt}
    \caption{Comparison Between the Treatment Group and Control Group}
    \label{tab:psm}
    \begin{tabular}{p{0.4\textwidth} R{0.15\textwidth} R{0.15\textwidth} R{0.15\textwidth} U{0.15\textwidth}}
        \toprule
        \midrule
        \multicolumn{5}{c}{Panel A: Before Matching}      \\ 
        \midrule
        & \multicolumn{1}{R{0.15\textwidth}}{Treatment} & \multicolumn{1}{R{0.15\textwidth}}{Control} & \multicolumn{1}{R{0.15\textwidth}}{Difference} & \multicolumn{1}{C{0.15\textwidth}}{$t$-statistics} \\
        \midrule
        Age & 32.301 & 31.716 & 0.585 & 2.45{**} \\
        Female & 0.631 & 0.549 & 0.081 & 6.18{***} \\     
        Platform membership & 0.361 & 0.236 & 0.125 & 10.71{***} \\     
        Wealth & 0.032 & -0.061 & 0.093 & 3.31{***} \\     
        Number of orders (past 6 months) & 54.582 & 38.840 & 15.743 & 12.15{***} \\
        Spending per order (past 6 months) & 45.059 & 45.581 & -0.522 & -0.52 \\
        Number of establishments & 51.831 & 48.056 & 3.775 & 4.41{***} \\   
        Share of non-SME establishments & 0.523 & 0.528 & -0.004 & -1.29 \\
        \midrule
        \multicolumn{5}{c}{Panel B: After Matching}      \\ 
        \midrule
        & \multicolumn{1}{R{0.15\textwidth}}{Treatment} & \multicolumn{1}{R{0.15\textwidth}}{Control} & \multicolumn{1}{R{0.15\textwidth}}{Difference} & \multicolumn{1}{C{0.15\textwidth}}{$t$-statistics} \\  
        \midrule
        Age & 32.250 & 32.281 & -0.030 & -0.09 \\
        Female & 0.635 & 0.657 & -0.022 & -1.25 \\     
        Platform membership & 0.382 & 0.395 & -0.013 & -0.68 \\     
        Wealth & 0.043 & 0.053 & -0.010 & -0.25 \\     
        Number of orders (past 6 months) & 55.580 & 54.164 & 1.416 & 0.59 \\
        Spending per order (past 6 months) & 45.236 & 44.913 & 0.324 & 0.34 \\
        Number of establishments & 52.445 & 51.284 & 1.161 & 0.90 \\   
        Share of non-SME establishments & 0.524 & 0.526 & -0.002 & -0.36 \\
        \bottomrule
    \end{tabular}
\tablenotes{This table compares user characteristics between the treatment and control groups before and after propensity score matching. Propensity scores are estimated using a Logit model with the covariates listed above, and each treated individual is matched to one control individual with replacement. The pre-matching sample includes 4,237 treated and 2,225 control individuals with non-missing values on all covariates. The post-matching sample includes 3,787 matched pairs. The difference column reports the difference in group means, and $t$-statistics test the null of equal means. *, **, and *** indicate statistical significance at the 10\%, 5\%, and 1\% levels, respectively.}
\end{mytable}

\section{Coupon MPC and Out-of-Pocket Spending: Accounting Identities}
\label{sec:acc_iden}

Let $E_i$ denote total expenditure (merchant revenue) and $S_i$ the coupon subsidy (discount)
redeemed by individual $i$ during the treatment period. Out-of-pocket spending is defined as
\[
\text{OOP}_i = E_i - S_i.
\]

We define the coupon marginal propensity to consume as
\[
\text{MPC}_i^{\text{coupon}} \equiv \frac{\Delta E_i}{S_i},
\]
where $\Delta E_i$ denotes the change in total expenditure over the treatment period relative to the no-coupon counterfactual. Since individuals receive no coupon subsidy in the absence of treatment, we have $\Delta S_i = S_i$.

Using this normalization, the change in out-of-pocket spending satisfies the accounting identity:
\begin{equation}
\Delta \text{OOP}_i
= \Delta(E_i - S_i)
= (\text{MPC}_i^{\text{coupon}} - 1)\, S_i .
\end{equation}

Therefore, $\text{MPC}_i^{\text{coupon}} < 1$ implies $\Delta \text{OOP}_i < 0$, indicating that the coupon user reduces out-of-pocket spending (i.e., generates net savings for the consumer), while $\text{MPC}_i^{\text{coupon}} > 1$ implies an increase in out-of-pocket spending.

\section{Average Treatment Effects: Estimation and Results}
\label{sec:oa-att}

\subsection{Same-Day Effects of Coupon Receipt}
\label{sec:dailydid}

We estimate a daily entry-exit DiD specification that captures the immediate impact of coupon receipt on consumer spending. Unlike our baseline model, which defines treatment at the individual level for the entire coupon event period, this alternative specification defines treatment at the individual-day level, identifying the effect only on days when consumers actually obtained coupons.
We estimate the following two-way fixed effects (TWFE) regression:

\begin{equation*}
y_{it} = \gamma_i + \lambda_t + \alpha \cdot \text{Coupon}_{it} + \epsilon_{it},    
\end{equation*}

\noindent where $\gamma_i$ and $\lambda_t$ represent individual and date fixed effects, respectively; and Coupon$_{it}$ is a binary indicator equal to 1 if individual $i$ obtained a coupon on date $t$, and 0 otherwise. The coefficient $\alpha$ is intended to capture the same-day spending response.

\begin{mytable}[h]{8pt}
    \caption{Same-Day Coupon Receipt: Spending Impacts by Expenditure Type}
    \label{tabA:dailydid}
\begin{tabular}{p{0.31\textwidth}C{0.18\textwidth}C{0.18\textwidth}C{0.18\textwidth}}
    \toprule
    \midrule
                            &     Out-of-pocket        &       Total      &       Unsubsidized     \\ 
                            &     (1)       &       (2)     &       (3)     \\ \midrule
    Coupon   &     20.747***  &    30.45***   &     -2.928*** \\
                            &   (0.563)     &     (0.703)   &    (0.441)    \\               
                            &               &               &               \\
    Observations            &  416,570      &    416,570    &   416,570     \\
    \bottomrule
\end{tabular}    

\tablenotes{This table presents estimates from a daily entry-exit DiD specification with individual and date fixed effects. The treatment variable is an indicator equal to one on days when individual $i$ obtained a coupon, and zero otherwise. Out-of-pocket expenditure represents total expenditure minus government-financed coupon subsidies, total expenditure captures the full payment to sellers (including coupon subsidies), and unsubsidized expenditure measures spending on orders that did not use coupons. Observations from the post-program period are excluded. Standard errors are clustered at the individual level. *, **, and *** indicate statistical significance at the 10\%, 5\%, and 1\% levels, respectively.}
\end{mytable}


Table \ref{tabA:dailydid} presents the results of this daily entry-exit DiD specification. On days when consumers received coupons, their out-of-pocket expenditure increased by ¥20.75 on average (Column 1), while total expenditure (including subsidies) increased by ¥30.45 (Column 2). Interestingly, unsubsidized expenditure---spending on orders that did not use coupons—decreased by ¥2.93 (Column 3), suggesting some substitution from unsubsidized to subsidized orders on coupon receipt days.

These results reveal a substantially larger same-day effect compared to our baseline estimate of ¥1.80 daily average effect over the entire coupon period. This difference is expected since this daily specification captures the concentrated impact on specific days when coupons were obtained, rather than averaging the effect across all days during the coupon period.

While this specification provides insight into the immediate spending response, we prefer our baseline approach for several reasons discussed in the main text. First, the daily specification may be subject to endogeneity concerns if individuals strategically time their participation based on anticipated consumption needs on specific days. Second, this entry-exit treatment definition raises additional concerns beyond those in our baseline specification. Recent literature on staggered DiD designs highlights potential biases when treatment timing varies across units and when previously treated units can revert to untreated status \citep{deChaisemartin2020, Goodman2021}. Such designs may produce biased estimates when treatment effects are heterogeneous across time or units, as earlier-treated observations can serve as controls for later-treated ones. Nevertheless, these daily results corroborate our main finding that coupon receipt increases consumer spending, while providing additional granularity on the timing of these effects.

\subsection{Robustness to Timing- and Effort-Based Selection}
\label{sec:did_subsample}

We conduct two robustness checks to ensure our results are not driven by selection bias. We apply more conservative criteria to define treated users. First, to address selection on timing, we use the timestamps of successful coupon claims to approximate the daily coupon quota-exhaustion point and restrict the treatment group to near-cutoff recipients---individuals who obtained at least one coupon bundle within the final decile of a day's successful claim timestamps during the program period. These users experienced at least one ``last-minute'' success and therefore relied less heavily on systematically early claim attempts. Second, to address selection on effort intensity, we restrict the treated group to one-time recipients---individuals who obtained exactly one coupon bundle during the program period. This definition excludes persistent, high-engagement users who claimed coupons repeatedly across multiple days. 

Table \ref{tab:did_subsample} reports the results. Both alternative treatment definitions yield estimates similar to our baseline specification, with OOP treatment effects of ¥1.97 and ¥2.53, respectively, compared to our baseline estimate of ¥1.80. The consistency across these subsamples suggests that our main findings are not driven by selection on timing or effort intensity and remain robust to substantially more restrictive definitions of treatment status.


\begin{mytable}[H]{6pt}
    \centering
    \caption{Robustness to Timing- and Effort-Based Selection}
    \label{tab:did_subsample}
\begin{tabular}{p{0.3\textwidth}C{0.2\textwidth}C{0.2\textwidth}C{0.2\textwidth}}
    \toprule
    \midrule
            \multicolumn{4}{c}{Panel A: Near-Cutoff Recipients} \\ \midrule
                            &     Out-of-pocket        &       Total      &       Unsubsidized     \\ 
                            &     (1)       &       (2)     &       (3)     \\ \midrule
    Treat$\times$Post      &     1.972**  &    2.897***   &     -0.342     \\
                            &   (0.770)     &     (0.784)   &    (0.760)    \\               
                            &               &               &               \\
    Observations            &  85,360      &    85,360    &   85,360     \\
    \midrule
        \multicolumn{4}{c}{Panel B: One-Time Recipients} \\ \midrule
    Treat$\times$Post      &     2.531***  &    3.032***   &     1.030     \\
                            &   (0.847)     &     (0.847)   &    (0.846)    \\               
                            &               &               &               \\
    Observations            &  98,670      &    98,670    &   98,670     \\
    \bottomrule
    \end{tabular}
    \tablenotes{This table re-estimates the baseline DiD specification (Equation \ref{eq:block_did}) with individual and date fixed effects, using more conservative definitions of the treatment group. In Panel A, the treated group is restricted to near-cutoff recipients, defined as individuals who obtained at least one coupon bundle within the final decile of a day’s successful claim timestamps during the program period. In Panel B, the treated group consists of one-time recipients, defined as individuals who obtained exactly one coupon bundle during the program period. Both panels implement propensity score matching prior to estimation, following the procedure described in Section \ref{sec:psm-det}. Out-of-pocket expenditure represents consumer spending excluding coupon subsidies, total expenditure captures the full payment to sellers (including government-financed coupon subsidies), and unsubsidized expenditure measures spending on orders that did not use coupons. Standard errors are clustered at the individual level. *, **, and *** indicate statistical significance at the 10\%, 5\%, and 1\% levels, respectively.}
    \end{mytable}
    

\subsection{Alternative DiD Estimators}
\label{sec:alt_did}

We show that our conclusions do not depend on the matching design used in the baseline specification by considering three alternative estimators. First, we re-estimate the baseline DiD model on the full sample without matching. Second, we implement optimal full matching following \citet{Hansen2006}. Unlike one-to-one pair matching, this approach forms matched subclasses with varying treated--control ratios and thus makes fuller use of the sample. Third, we implement the doubly robust DiD estimator of \citet{Sant2020}. This estimator combines inverse-probability weighting with outcome regression and remains consistent if either the treatment-assignment model or the outcome-evolution model is correctly specified.

Table \ref{tab:did_app} reports the results of these alternative estimators. Across all three panels, the estimates are highly consistent with our baseline in Table \ref{tab:did}. The estimated effects on out-of-pocket and total expenditure remain similar in magnitude to the baseline, whereas the effect on unsubsidized expenditure remains statistically insignificant. These exercises suggest that neither the decision to match nor the particular matching procedure drives our findings.


\begin{mytable}[H]{6pt}
    \centering
    \caption{Alternative DiD Estimates}
    \label{tab:did_app}
\begin{tabular}{p{0.3\textwidth}C{0.2\textwidth}C{0.2\textwidth}C{0.2\textwidth}}
    \toprule
    \midrule
            \multicolumn{4}{c}{Panel A: No Matching DiD} \\ \midrule
                            &     Out-of-pocket        &       Total      &       Unsubsidized     \\ 
                            &     (1)       &       (2)     &       (3)     \\ \midrule
    Treat$\times$Post      &     1.587***  &    2.339***   &     -0.207     \\
                            &   (0.383)     &     (0.387)   &    (0.382)    \\               
                            &               &               &               \\
    Observations            &  355,410      &    355,410    &   355,410     \\
    \midrule
        \multicolumn{4}{c}{Panel B: Full Matching DiD} \\ \midrule
    Treat$\times$Post      &     2.003***  &    2.771***   &     -0.097     \\
                            &   (0.494)     &     (0.497)   &    (0.493)    \\               
                            &               &               &               \\
    Observations            &  313,775      &  313,775      &  313,775      \\
    \midrule

            \multicolumn{4}{c}{Panel C: Doubly Robust DiD} \\ \midrule
    Treat$\times$Post      &     1.992***  &    2.756***   &     0.166     \\
                            &   (0.621)     &     (0.628)   &    (0.604)    \\               
                            &               &               &               \\
    Observations            &  11,490       &  11,490       &  11,490       \\
    \bottomrule
    \end{tabular}
    \tablenotes{This table reports alternative DiD estimates for the same three expenditure outcomes as in Table~\ref{tab:did}. Panel A re-estimates Equation~\eqref{eq:block_did} on the underlying unmatched user-day sample (N=355,410). Panel B applies optimal full matching using the full set of observed covariates. Since complete covariate information is required, the sample size is reduced to 313,775. Panel C implements the doubly robust DiD estimator of Sant’Anna and Zhao (2020), which uses a 2×2 block design at the period level and therefore reports 11,490 period-level observations rather than daily user-day observations. The reported observation counts are not directly comparable to those in Table~\ref{tab:did}, because the baseline specification in Table~\ref{tab:did} is estimated on a matched sample constructed with replacement, so matched control users appear multiple times according to their matching weights. Standard errors are clustered at the individual level. *, **, and *** indicate statistical significance at the 10\%, 5\%, and 1\% levels, respectively.}
\end{mytable}


\subsection{Decomposition of Treatment Effects and Substitution Patterns}
\label{sec:substitute}
Table \ref{tab:decomp_sub} reports estimates underlying Figure \ref{fig:decomp_sub}. Panel A decomposes the aggregate spending response along extensive and intensive margins, while Panel B tests for substitution across time, categories, and household members. Panel A separates the total treatment effect into extensive and intensive margins. While the extensive margin (order frequency) exhibits a statistically significant increase, the magnitude of 0.029 additional orders per day (Column 2) is economically negligible. Instead, we observe a significant increase in out-of-pocket expenditure per order alongside an increase in dishes per order. 

Panel B evaluates the potential substitution patterns discussed in Section \ref{sec:model-did}. Column 1 tests for inter-temporal substitution by estimating the spending response in the post-coupon event period. The estimate is statistically indistinguishable from zero, showing that the coupon event generated an immediate but temporary stimulus. Column 2 reports the baseline effect on restaurant spending, while Column 3 tests for inter-category substitution using grocery spending as the outcome. Restaurant out-of-pocket expenditure increases significantly, whereas the estimated effect on grocery spending is near zero. This indicates that the increase in restaurant spending did not come at the expense of other food-related purchases on Ele.me. Column 4 evaluates intra-household or workplace substitution using the number of tableware sets requested per order, which proxies for the number of diners. The estimated effect is statistically significant but economically negligible, pointing to no meaningful intra-household or workplace sharing to meet the minimum thresholds.

    \begin{mytable}[H]{6pt}
    \centering
    \caption{Decomposition and Substitution Effects}
    \label{tab:decomp_sub}

    \begin{tabular}{lcccc}
        \toprule
        \midrule
        \multicolumn{5}{c}{Panel A: Decomposition Effects}\\
        \midrule
                                &  OOP per order    &  Order frequency     &    SKU per order  &  OOP per SKU   \\ 
                                &      (1)         &         (2)           &      (3)          &    (4)          \\ \midrule
        Treat$\times$Post      &      1.200***     &        0.029***          &     0.201**      &  -1.047**  \\
                                &      (0.424)     &        (0.010)        &     (0.094)       &  (0.506)  \\                   
        Observations            &       416,570    &       416,570         &      109,640      &    109,640 \\ 
        \midrule
        \multicolumn{5}{c}{Panel B: Substitution Effects} \\ \midrule
                                & Post-treatment OOP  &     Restaurant OOP          &          Grocery OOP         & Tableware sets per order \\ 
                                &      (1)         &         (2)         &      (3)      &      (4)\\ \midrule
        Treat$\times$Post      &      -1.210      &      1.769***    &       0.033        &      0.021***   \\
                                &     (0.790)      &      (0.584)     &       (0.191)       &     (0.007)   \\           
        Observations            &     212,072      &       416,570    &       416,570       &    416,570    \\ 
        \bottomrule
     
    \end{tabular}

    \tablenotes{This table presents estimates from Equation \ref{eq:block_did} using alternative dependent variables. Panel B Column 1 additionally redefines $\text{Post}_t$ to indicate the two weeks after the coupon event and restricts the sample accordingly (see Section \ref{sec:model-did}) Panel A decomposes the consumption response along extensive and intensive margins. Column 1 shows the effect on out-of-pocket (OOP) expenditure per order, Column 2 on order frequency, Column 3 on the number of dishes (SKUs) per order, and Column 4 on OOP per dish (SKU). Columns 3 and 4 condition on days with at least one order, resulting in fewer observations. Panel B examines potential substitution patterns. Column 1 tests for inter-temporal substitution by comparing OOP spending in the two weeks after the coupon event to the pre-treatment period. Column 2 shows the baseline effect on restaurant OOP spending. Column 3 tests for inter-category substitution using grocery OOP as the outcome. Column 4 tests for intra-household substitution using tableware sets per order, which proxies for the number of diners. Standard errors are clustered at the individual level. *, **, and *** indicate statistical significance at the 10\%, 5\%, and 1\% levels, respectively.}
\end{mytable}

\subsection{Supplementary Evidence on Platform Substitution}
\label{sec:daas-sub}

The main text presents within-platform evidence that the coupon program's spending response reflects genuine consumption rather than reallocation across platforms or dining formats. We provide supplementary evidence using mobile application usage data from a major telecommunications provider, which tracks monthly app usage time for all subscribers in Beijing (see Section \ref{sec:daas} for details).

To assess cross-platform substitution, we compute Ele.me's share of food delivery app usage time in Beijing, with the denominator comprising usage time on Ele.me and Meituan, the two major food delivery platforms. If the coupon program attracted users from the rival platform, Ele.me's share should be elevated during the coupon month relative to comparable periods without coupon programs. We use August 2023 and November 2023 as comparison periods, neither of which featured coupon programs. Ele.me's share was 14.5 percent during the coupon month (August 2022), compared to 14.7 percent in both comparison periods.

To gauge whether this comparison can detect substitution at the relevant scale, we estimate coupon recipients' share of Ele.me's user base. Our baseline estimates imply a per-person, per-day average government subsidy of \textyen 0.76 (Section \ref{sec:cost}). Dividing total government expenditure on coupon subsidies distributed through the platform (approximately \textyen 20 million, based on information provided by the platform) by this per-person subsidy over the 41-day program period yields an estimated 642,000 coupon recipients. The weighted number of active Ele.me users in Beijing during August 2022 is approximately 2.8 million, so coupon recipients constitute roughly 23 percent of the platform's active user base. Under full substitution, in which all coupon recipients were users who switched to Ele.me from rival platforms specifically to obtain coupons, their usage time would be absent from Ele.me in months without the program. Removing their contribution from Ele.me's coupon-month total, while holding total food delivery usage constant, would imply a non-coupon usage share of approximately 11.2 percent ($14.5 \times 0.77$, where $0.77 = 1 - 0.23$ is the non-recipient share). If one in four coupon recipients were platform switchers, the predicted non-coupon share would be approximately 13.7 percent. The observed shares of 14.7 percent in both August and November 2023 suggest limited scope for cross-platform substitution.

We interpret this evidence as a consistency check rather than a definitive test, since other forces affecting Ele.me's market share could mask smaller substitution effects. This test also relies on the assumption that any coupon-induced platform switching is transitory: if switchers remained on Ele.me after the program ended, the platform's share would stay elevated in non-coupon months and the comparison would not detect substitution.

Beyond cross-platform switching, the coupon program could also have induced offline-to-online substitution by shifting consumers from in-person dining to food delivery. As a proxy for interest in offline dining, we examine usage of DaZhongDianPing (DZDP), an offline restaurant review platform. If such substitution occurred, we would expect a decline in DZDP engagement during the month with the coupon program. We compute DZDP's share of active users across the major food-related apps (Ele.me, Meituan, and DZDP). This share was 31.6 percent during the month with the coupon program (August 2022), 30.7 percent one year later (August 2023), and 30.9 percent in November 2023. This stability is consistent with continued interest in offline dining throughout the coupon period. As with the cross-platform test, we interpret this as a consistency check subject to the same caveats about confounding trends; together, the two exercises complement the within-platform intensive-margin evidence in the main text.

\section{Heterogeneous Consumer Behavior}
\subsection{Bunching Pattern}

\begin{myfigure}[H]
\includegraphics[width=0.8\linewidth]{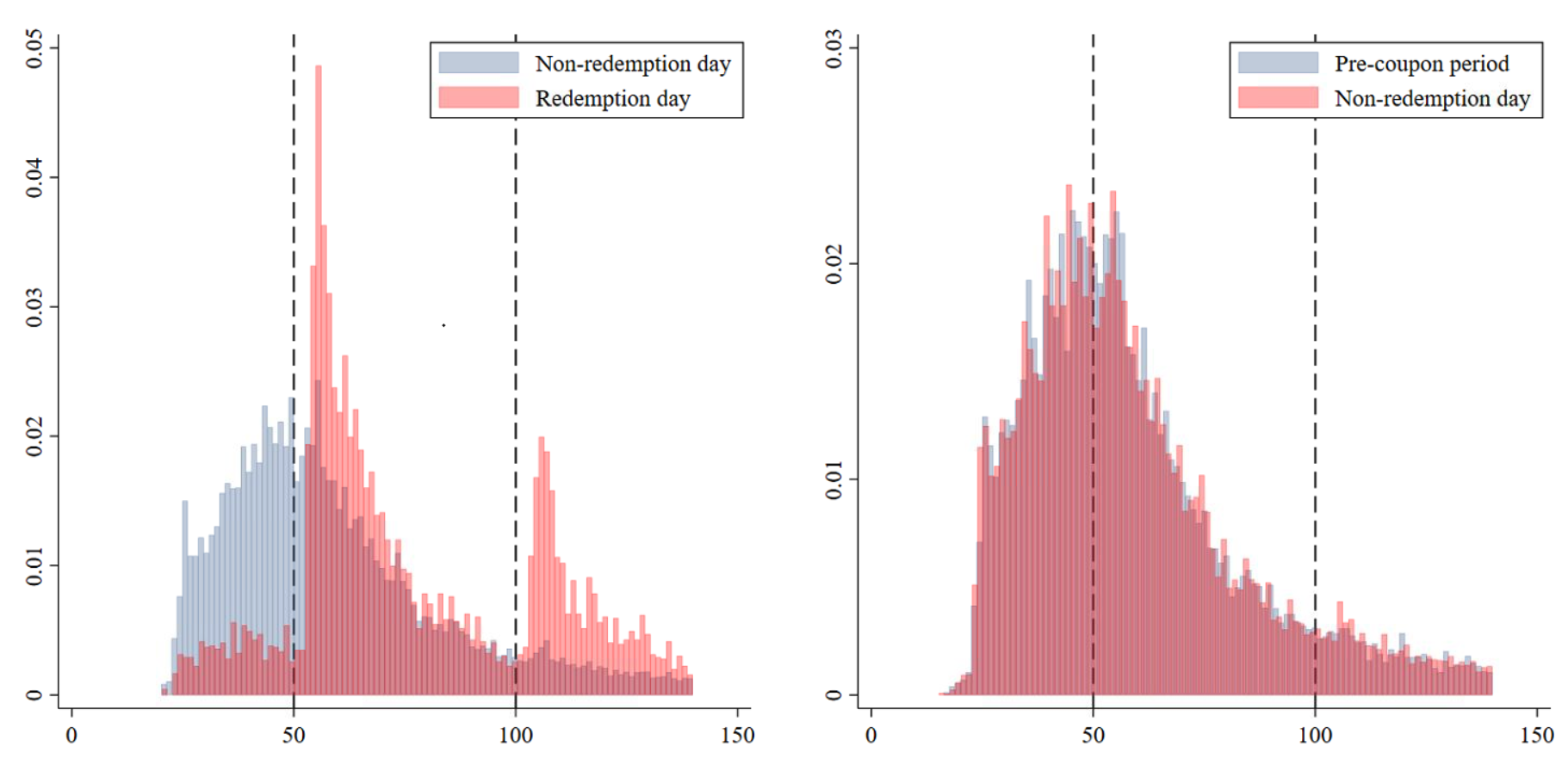}
\caption{Discount Threshold and Bunching}
\label{fig:bunch}
  \figurenotes{This figure plots the transaction-level probability density of order amounts to illustrate strategic consumer bunching around the program's minimum spending thresholds. The horizontal axis represents the gross order amount, and the vertical axis represents the empirical density. A redemption day is defined at the individual-day level as any day during the program period when a consumer placed at least one subsidized order. The left panel compares the distribution of order amounts on redemption days (red) against non-redemption days (blue) during the coupon event period. The right panel serves as a placebo test, comparing the distribution of order amounts on non-redemption days during the program period (red) against the baseline distribution from the pre-coupon event period (blue). Vertical dashed lines mark the ¥50 and ¥100 discount thresholds. The spikes just above these thresholds on redemption days demonstrate strategic consumer behavior to qualify for discounts. In contrast, no such bunching appears in the pre-coupon event period or on non-redemption days.}
\end{myfigure} 

\subsection{Coupon Redemption Patterns by Establishment Type}
\begin{myfigure}[H]
  \begin{subfigure}{0.48\textwidth}
    \includegraphics[width=\linewidth]{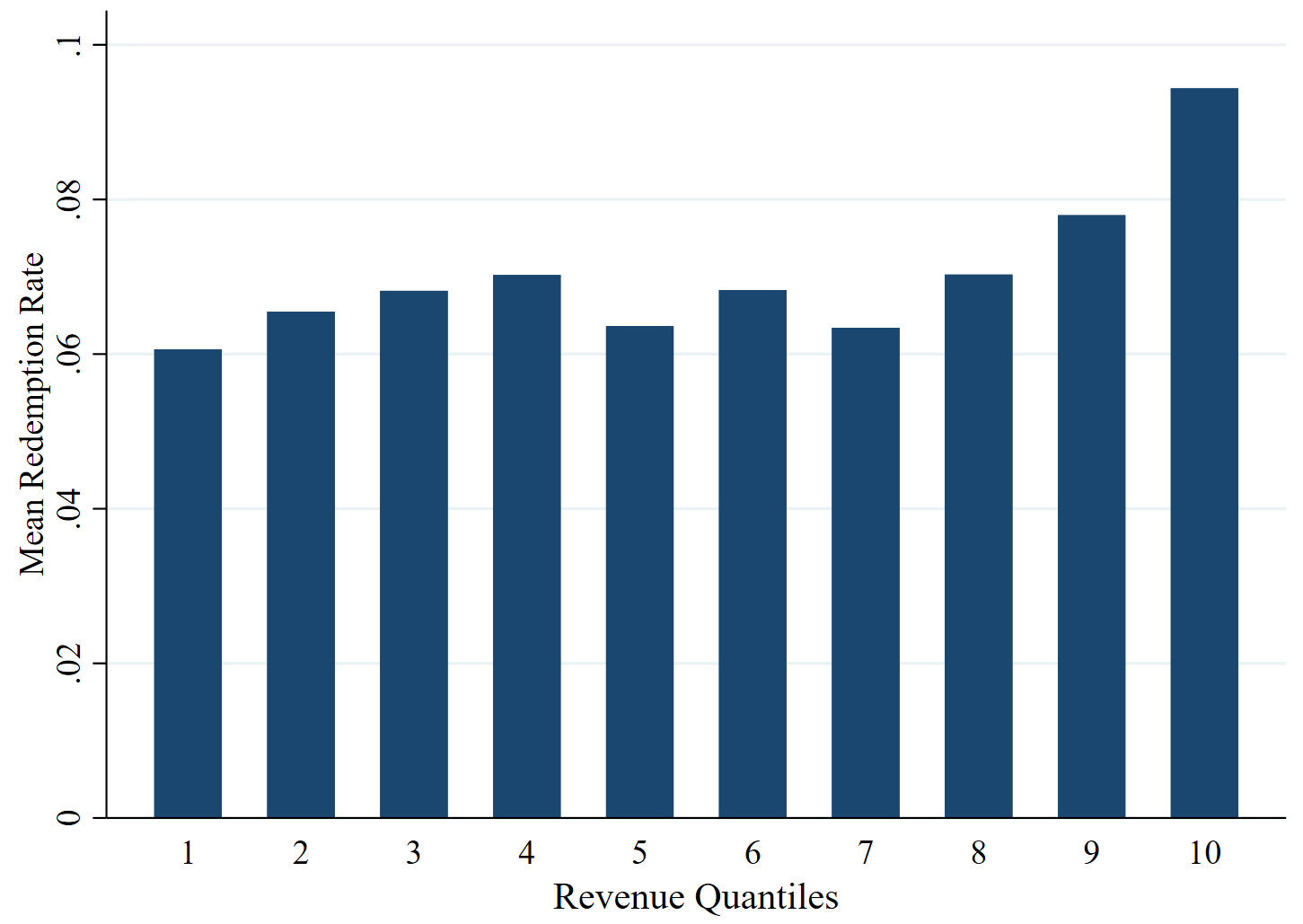}
    \caption{By Establishment Size}
    \label{fig:redemption_size}
  \end{subfigure} 
  \begin{subfigure}{0.48\textwidth}
    \includegraphics[width=\linewidth]{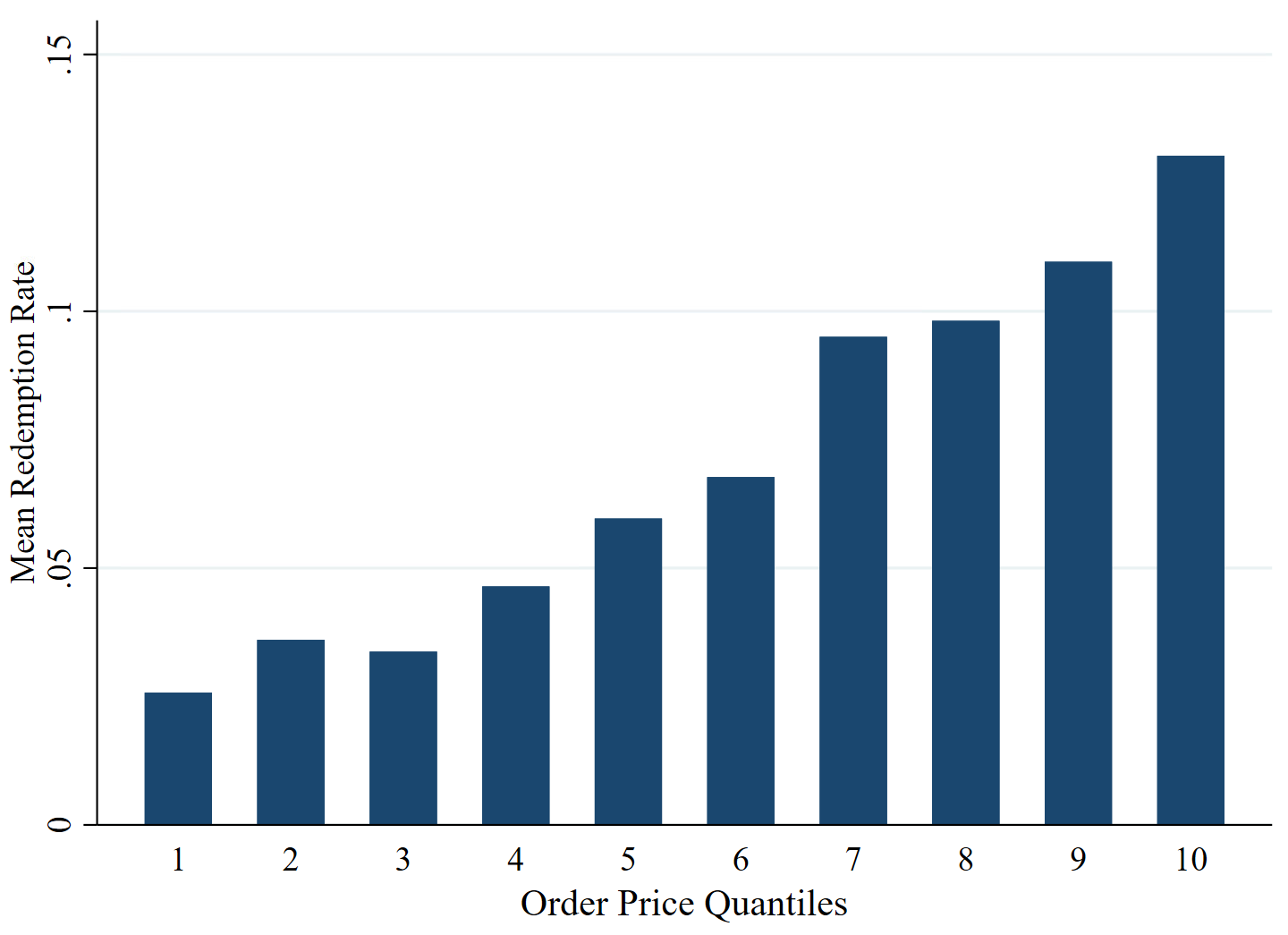}
    \caption{By Price Level}
    \label{fig:redemption_price}
  \end{subfigure} 
\caption{Coupon Redemption Rates Across Establishment Types}
  \label{fig:redem}
    \figurenotes{This figure presents the heterogeneity in coupon redemption rates across establishments. Panel (a) groups establishments by sales revenue (six months prior to the coupon event) and reports the mean coupon redemption rate within each size decile. Panel (b) groups establishments by average order price (six months prior to the coupon event) and reports the mean coupon redemption rate within each price decile.}
\end{myfigure} 

\subsection{Consumer Characteristics by Non-SME Expenditure Share}

\begin{myfigure}[H]
    \includegraphics[width=0.55\linewidth]{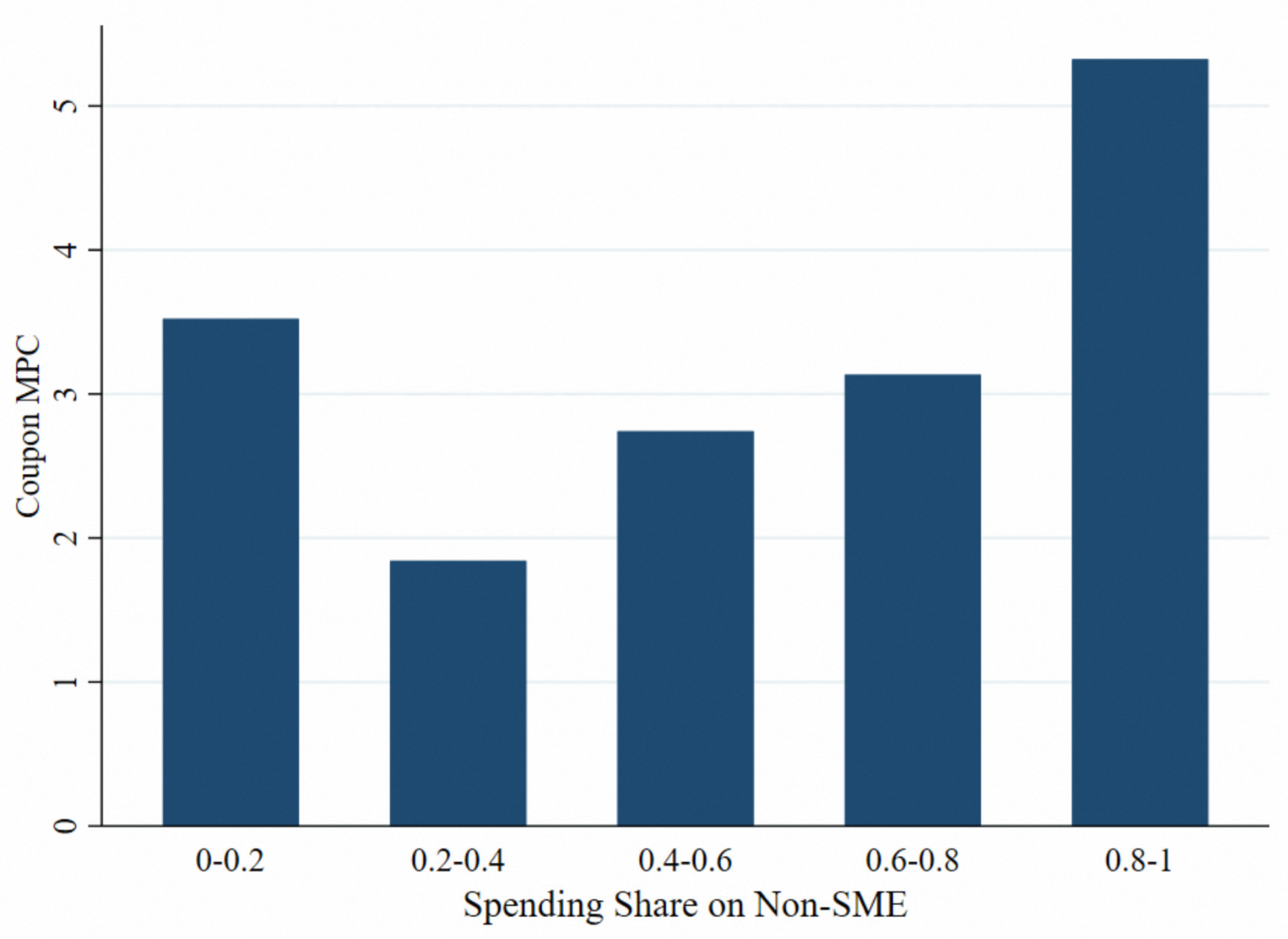}
    \caption{Coupon MPC by Non-SME Expenditure Share}
    \label{fig:mpc_pct}
    \figurenotes{This figure presents the average marginal propensity to consume out of coupon subsidies (Coupon MPC) across quintiles of consumers' non-SME expenditure share. The horizontal axis groups consumers by their pre-treatment share of spending allocated to non-SME establishments. The vertical axis reports mean estimated Coupon MPC within each quintile.}
\end{myfigure}

\section{Heterogeneous Treatment Effects: Estimation and Results}

\subsection{Causal Forest Estimation}
\label{sec:tech}

We implement the causal forest framework of \citet{Athey2019a} using the \texttt{grf} package in \textit{R}. Following the residual-on-residual orthogonalization procedure outlined in Section \ref{sec:model-hte}, we first estimate the nuisance functions $\mathbb{E}[\Delta y_i | \mathbf{X}_i]$ and $\mathbb{E}[\text{Treat}_i | \mathbf{X}_i]$. In \texttt{grf}, these nuisance components are estimated automatically using regression forests. As an independent check, we also estimate them using the Super Learner \citep{Laan2007} (see Section~\ref{sec:hteapp}). The resulting treatment effect estimates are very similar and available upon request. 

We then estimate the causal forest using the package's ``honest'' procedure (\texttt{honesty = TRUE}). Under honesty, each tree uses separate subsamples for split selection and for treatment effect estimation within the resulting leaves, which helps reduce overfitting and supports valid statistical inference under the conditions in \citet{Athey2019a}. Hyperparameters are selected using the automated \texttt{tune.parameters} routine, focusing on the number of variables considered at each split (\texttt{mtry}) and the subsampling fraction (\texttt{sample.fraction}). We set \texttt{tune.num.trees = 1{,}000} and \texttt{tune.num.reps = 200} for tuning, and grow the final forest with \texttt{num.trees = 5{,}000} to ensure stable treatment effect estimates.

\subsection{Alternative Machine Learning Estimators}
\label{sec:hteapp}

Our benchmark strategy relies on Neyman orthogonalization to partial out the influence of covariates from both the outcome and the treatment indicator, allowing us to estimate the treatment effect directly. Alternatively, we can estimate heterogeneous treatment effects using a direct approach that jointly learns the relationship between the covariates, the treatment indicator, and the outcome by modeling a single conditional response surface:

\begin{equation}\label{eq:direct_surface}
    \Delta y_i = \mu(\mathbf{X}_i, \text{Treat}_i) + \epsilon_i,
\end{equation}
where $\mu(\mathbf{X}_i, \text{Treat}_i) = \mathbb{E}[\Delta y_i | \mathbf{X}_i, \text{Treat}_i]$. Once the joint response surface is estimated, the conditional average treatment effect, $\alpha(\mathbf{X}_i)$, is extracted by taking the difference in the predicted potential outcomes for the treated and untreated states:
\begin{equation}\label{eq:direct_HTE}
    \hat{\alpha}(\mathbf{X}_i) = \hat{\mu}(\mathbf{X}_i, 1) - \hat{\mu}(\mathbf{X}_i, 0).
\end{equation}

To estimate the joint response surface, we implement the Super Learner framework of \citet{Laan2007} using the \texttt{SuperLearner} package in \textit{R}, which is an ensemble machine learning algorithm that optimally combines predictions from a set of candidate models. By utilizing cross-validated weighting, it ensures that the ensemble prediction minimizes overall cross-validated risk. 

Specifically, our candidate models include a standard linear model (\texttt{SL.lm}), a generalized additive model (\texttt{SL.gam}), a kernel-based support vector machine (\texttt{SL.ksvm}), and a random forest (\texttt{SL.randomForest}). We retain the default hyperparameter configurations for each candidate model and leverage this direct estimation strategy to calculate $\hat{\alpha}(\mathbf{X}_i)$.

Figure \ref{figA:sl_cf} presents a scatterplot of the predicted conditional average treatment effects from the direct approach against those from the causal forest benchmark. The clustering of estimates around the 45-degree reference line demonstrates that the predicted treatment effects remain consistent across both approaches.

\begin{myfigure}[H]
    \centering
    \includegraphics[width=0.6\linewidth]{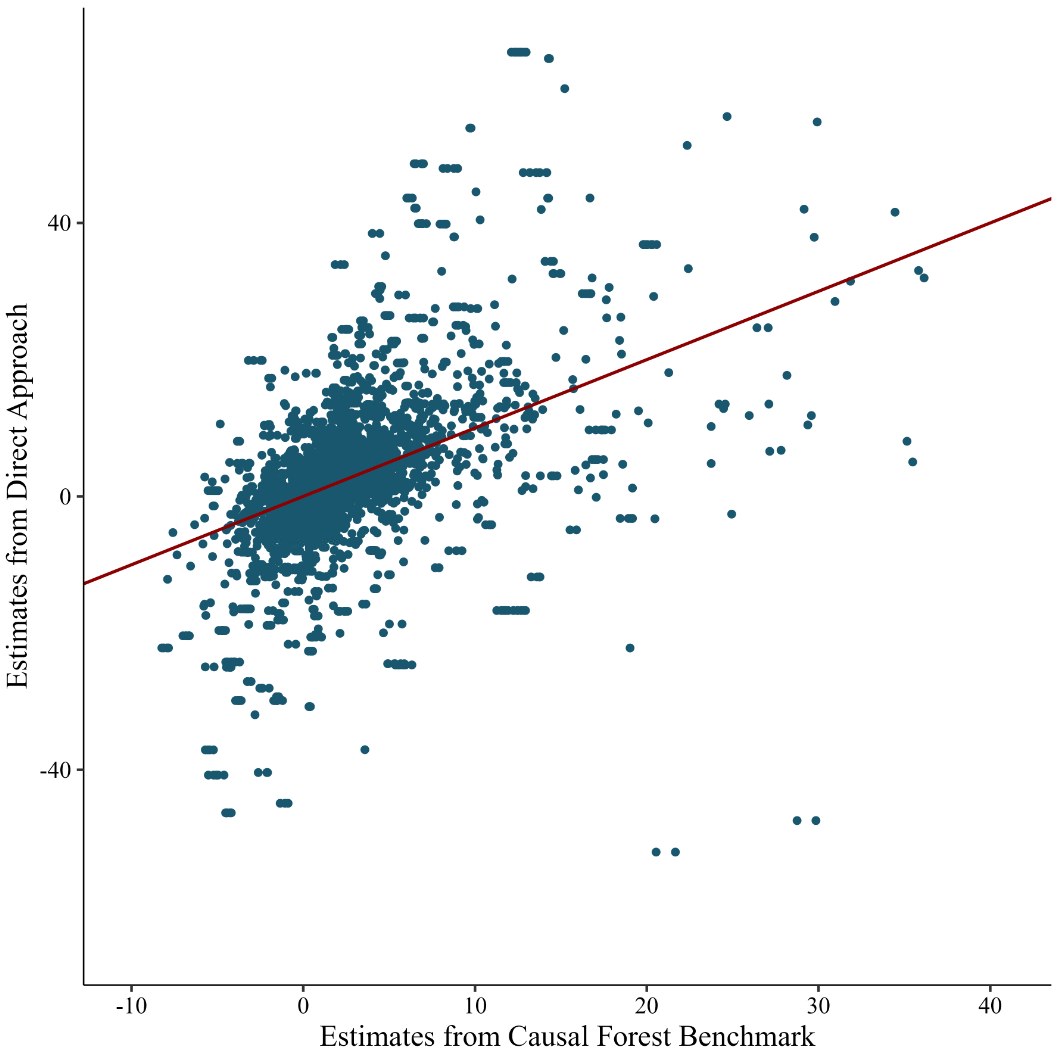}
    \caption{Predicted Treatment Effects from Alternative Approaches}
    \label{figA:sl_cf}
\figurenotes{This scatterplot compares the predicted conditional average treatment effects from the direct approach (vertical axis) against those from the causal forest benchmark (horizontal axis). The solid red line represents the 45-degree line. The clustering of estimates around this line demonstrates that the predicted treatment effects are consistent across both strategies.}
\end{myfigure}

\subsection{Accumulated Local Effects (ALEs)}
\label{sec:driverapp}

We estimate the marginal influence of individual characteristic on the predicted treatment effect using the Accumulated Local Effects (ALEs) framework of \citet{AZ2020}, applied here in a heterogeneous-treatment-effects setting. Let $\hat{\tau}(X)$ denote the bias-corrected estimate of the treatment effect function obtained by nonparametrically regressing the doubly robust score $\psi_i$ on covariates $\mathbf{X}_i$. For a given characteristic $X_k$, we compute the accumulated local effect at a specific value $x$ as:

\begin{equation}
\text{ALE}_k(x) = \sum_{b=1}^{b(x)} \frac{1}{|S_b|} \sum_{i: X_{ik} \in S_b} \left[ \hat{\tau}(z_b, \mathbf{X}_{i,-k}) - \hat{\tau}(z_{b-1}, \mathbf{X}_{i,-k}) \right],
\end{equation}
where the support of $X_k$ is partitioned into $B$ intervals, with $b(x)$ denoting the index of the interval containing $x$. Let $S_b$ represent the $b$-th interval and $|S_b|$ the number of observations within it. The inner summation aggregates over all observations $i$ for which the realized feature value, $X_{ik}$, falls within $S_b$, while $z_b$ and $z_{b-1}$ define the upper and lower boundaries of the $b$-th interval, respectively. The vector $\mathbf{X}_{i,-k}$ captures all covariates other than $k$ for observation $i$. Intuitively, ALE aggregates local changes in the predicted treatment effect as $X_k$ moves across adjacent intervals, while holding the remaining covariates at their observed values.

We compute ALEs using the \texttt{iml} package in \textit{R}, setting \texttt{grid.size = 25} so that the support of each characteristic is partitioned into 25 equal-frequency intervals based on empirical quantiles. Relative to Partial Dependence Plots (PDPs), ALE has an important advantage when covariates are correlated. PDPs average predictions over the unconditional distribution of the remaining covariates and therefore evaluate the model at feature combinations that may be rare or unsupported in the data. By contrast, ALE computes changes in the predicted treatment effect over small intervals of $X_k$, using only observations whose realized $X_k$ falls in the relevant interval $S_b$. As a result, the estimated marginal influence of $X_k$ is based on observed covariate combinations rather than extrapolated values.

\subsection{Drivers of Treatment Effect Heterogeneity}
\label{subsec:driverhte}

Figure \ref{fig:vip} presents variable importance scores from the causal forest estimation, measuring the weighted frequency with which each covariate is selected for splitting across trees. Consumption habits (spending per order and order frequency), wealth, and neighborhood consumption amenities emerge as the most important predictors, consistent with the BLP results in the main text.

\begin{myfigure}[H]
    \includegraphics[width = .6\linewidth]{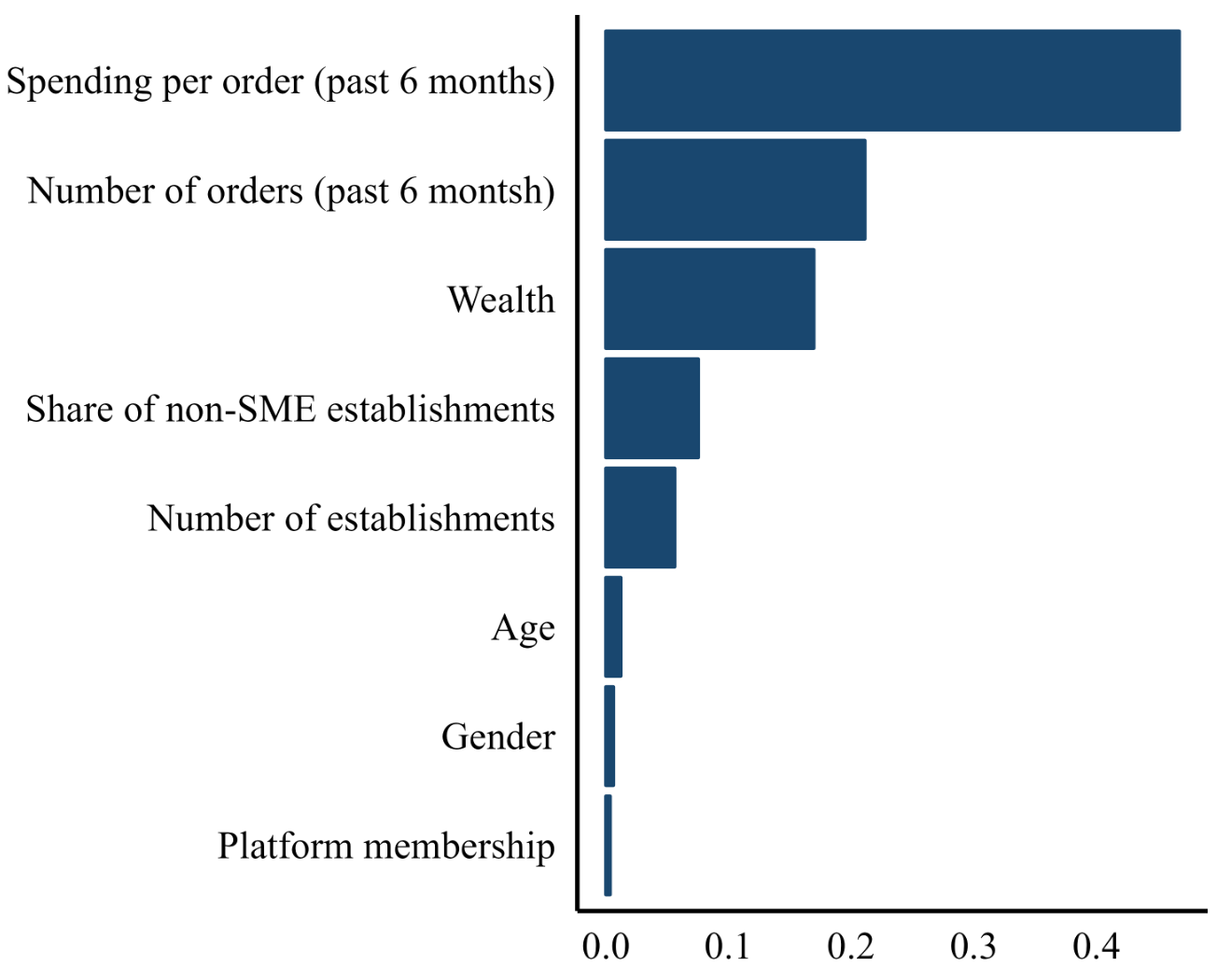}
\caption{Variable Importance Plot}
\label{fig:vip}
\figurenotes{This figure displays variable importance scores derived from causal forest estimation. These scores represent the weighted sum of the relative frequency with which each covariate is selected for splitting at each depth across all trees in the forest. The final scores are normalized to sum to one.}

\end{myfigure}

Table \ref{tab:cf_blp} reports the full BLP coefficient estimates. Column (1) presents the baseline specification corresponding to Figure \ref{fig:cf_blp} in the main text. Columns (2) and (3) assess the robustness of the non-SME establishment share, the key supply-side predictor, by controlling for the price composition of local restaurants. Because the program's two coupon thresholds are \yen 50 and \yen 100, restaurants in higher price tiers may facilitate redemption, raising the concern that the non-SME share proxies for local price composition rather than establishment size. Column (2) replaces the non-SME share with two price-tier variables capturing the share of nearby establishments whose average order price falls between \yen 50 and \yen 100, and the share above \yen 100. Neither is statistically significant, indicating that local price composition alone does not predict treatment effect heterogeneity. Column (3) includes both the non-SME share and the price-tier variables. The non-SME share remains strongly positive and significant, while the price-tier coefficients are positive but statistically insignificant. Establishment size, not price level, emerges as the primary supply-side driver of heterogeneous consumption responses in our setting.

\begin{mytable}[H]{2pt}
    \centering
    \caption{Best Linear Projection of Heterogeneous Treatment Effects}
    \label{tab:cf_blp}
\begin{tabular}{p{0.35\textwidth}T{0.2\textwidth}T{0.2\textwidth}T{0.2\textwidth}}
    \toprule
    \midrule
& \multicolumn{1}{c}{(1)} & \multicolumn{1}{c}{(2)} & \multicolumn{1}{c}{(3)} \\
    \midrule
    Spending per order (past 6 months) & 1.710{***} & 1.996{***} & 1.824{***} \\
                                       & (0.065) & (0.068) & (0.066) \\
    Number of orders (past 6 months)   & 0.418{***} & 0.697{***} & 0.480{***} \\
                                       & (0.063) & (0.067) & (0.064) \\
    Wealth                             & 0.585{***} & 0.514{***} & 0.461{***} \\
                                       & (0.066) & (0.070) & (0.067) \\
    Platform membership                & -0.380{***} & -0.331{***} & -0.371{***} \\
                                       & (0.064) & (0.067) & (0.065) \\
    Age                                & -0.119{*}   & -0.118{*}       & -0.109{*}   \\
                                       & (0.064) & (0.067) & (0.065) \\
    Female                             & -0.244{***} & -0.227{***} & -0.215{***} \\
                                       & (0.063) & (0.067) & (0.064) \\
    Number of establishments           & -0.740{***} & -0.550{***} & -0.705{***} \\
                                       & (0.065) & (0.071) & (0.070) \\
    Share of non-SME establishments    & 0.520{***}  &          & 0.503{***} \\
                                       & (0.065) &          & (0.066) \\
    Share of establishments (¥50--100) &          & 0.007       & 0.005 \\
                                       &          & (0.068) & (0.065) \\
    Share of establishments ($>$¥100)  &          & 0.011        & 0.013     \\
                                       &          & (0.068) & (0.065) \\
    Constant                           & 1.958{***}  & 2.179{***} & 2.047{***} \\
                                       & (0.061) & (0.065) & (0.062) \\

    \bottomrule
\end{tabular}
\tablenotes{This table reports the coefficient estimates from the best linear projection (BLP) of the doubly robust score on the set of observed characteristics. Column (1) includes the baseline characteristics, Column (2) replaces the share of non-SME establishments with shares of establishments whose average order price (in the six months prior to the coupon event) is between ¥50 and ¥100, or above ¥100; Column (3) includes both the share of non-SME establishments and the aforementioned price-tier shares. All continuous covariates are standardized to have a mean of zero and unit variance, allowing for direct comparison across coefficients as measures of variable importance. The constant term represents the average of bias-corrected conditional average treatment effects. Robust standard errors are reported in parentheses.  *, **, and *** indicate statistical significance at the 10\%, 5\%, and 1\% levels, respectively.}
\end{mytable}


\newpage
Figure~\ref{fig:inframarginal} reproduces the ALE curve for average spending per order from Figure~\ref{fig:ale_threshold}, overlaying the \yen 50 and \yen 100 coupon thresholds. Treatment effects begin to rise around the \yen 50 threshold and continue increasing above \yen 100. Section 4.4 discusses the implications of this pattern for models of consumption behavior.\footnote{Threshold-based incentives may generate responses not fully captured by standard utility maximization. In particular, the act of crossing the threshold itself may carry intrinsic utility.}

\begin{myfigure}[H]
    \includegraphics[width = .6\linewidth]{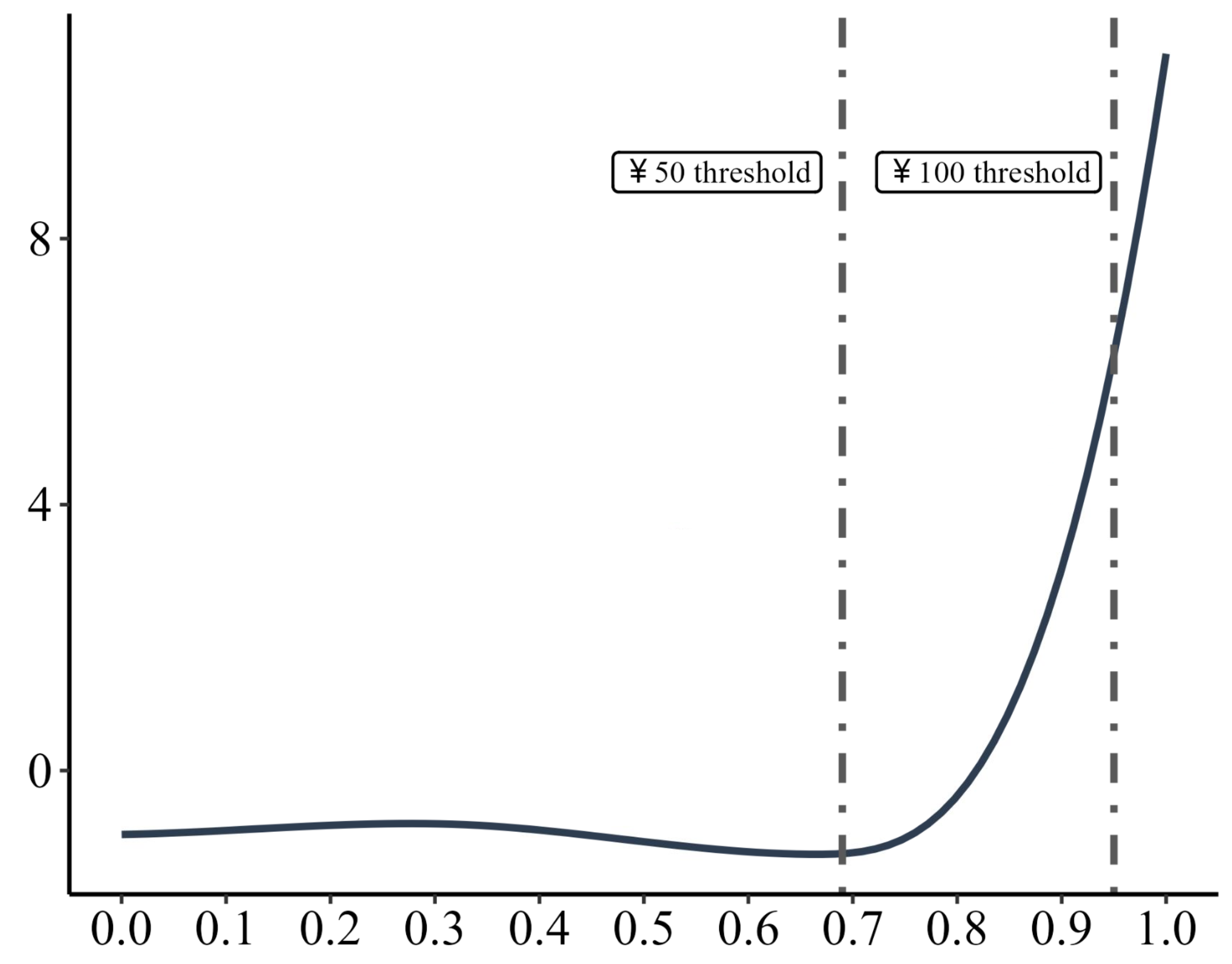}
    \caption{Treatment Effects by Spending Level}
    \label{fig:inframarginal}
    \figurenotes{This figure reproduces the ALE curve for average spending per order from Figure \ref{fig:ale_threshold}. The horizontal axis shows empirical quantiles of average spending per order during the six months prior to the coupon event, and the vertical axis shows the corresponding marginal effect on the estimated individual treatment effect, centered to have mean zero. The vertical lines mark the \yen 50 and \yen 100 minimum spending thresholds for the ``50--15'' and ``100--30'' coupons, respectively.}
\end{myfigure}

To complement the ALE evidence in Figure~\ref{fig:inframarginal}, Table~\ref{tab:interaction} reports subgroup difference-in-differences estimates by baseline spending per order. Specifically, we augment the baseline difference-in-differences specification in Equation~\eqref{eq:block_did} by allowing the treatment effect to vary across three groups defined by average spending per order in the six months before the coupon event: below \yen 50, \yen 50--100, and at least \yen 100. The estimates show that spending responses remain substantial among consumers whose baseline spending already exceeds one or both coupon thresholds, with the largest grouped effect appearing among those above \yen 100.

  \begin{mytable}[H]{6pt}
    \centering
   \caption{Treatment Effects by Baseline Spending Per Order}
    \label{tab:interaction}
\begin{tabular}{p{0.64\textwidth}C{0.3\textwidth}}
    \toprule
    \midrule
                            &     Out-of-pocket spending      \\ 
                            &     (1)      \\ \midrule
    Treat $\times$ Post $\times$ $\mathbf{1}(\bar{E}_i < 50)$      &     1.591*** \\
                            &   (0.619)    \\               
  Treat $\times$ Post $\times$ $\mathbf{1}(50 \leq \bar{E}_i < 100)$      &     1.735** \\
                            &   (0.861)    \\               
  Treat $\times$ Post $\times$ $\mathbf{1}(\bar{E}_i \geq 100)$      &     5.883** \\
                            &   (2.575)    \\

                            &              \\
    Observations            &  416,570    \\
    \bottomrule
    \end{tabular}

\tablenotes{This table reports subgroup difference-in-differences estimates based on Equation~\eqref{eq:block_did}, allowing the treatment effect to vary across three groups of consumers defined by baseline spending per order. The dependent variable is daily out-of-pocket spending. Consumers are grouped by their baseline spending per order ($\bar E_i$), defined as average spending per order in the six months before the coupon event, using the program's minimum-spending thresholds as cutoffs: below \yen 50, \yen 50--100, and at least \yen 100. Standard errors are clustered at the individual level. *, **, and *** indicate statistical significance at the 10\%, 5\%, and 1\% levels, respectively.}
\end{mytable}

\subsection{Spatial Heterogeneity in Digital Coupon Effects}

To further explore the geographic variation in consumer responsiveness to the digital coupon program, we aggregate the estimated individual-level treatment effects into spatial neighborhoods. Specifically, we divide the Beijing municipality into 3km $\times$ 3km grid cells and assign each treated consumer to a cell based on their location. We then calculate the average conditional treatment effects on the treated within each grid.

\begin{myfigure}[H]
    \includegraphics[width=.6\linewidth]{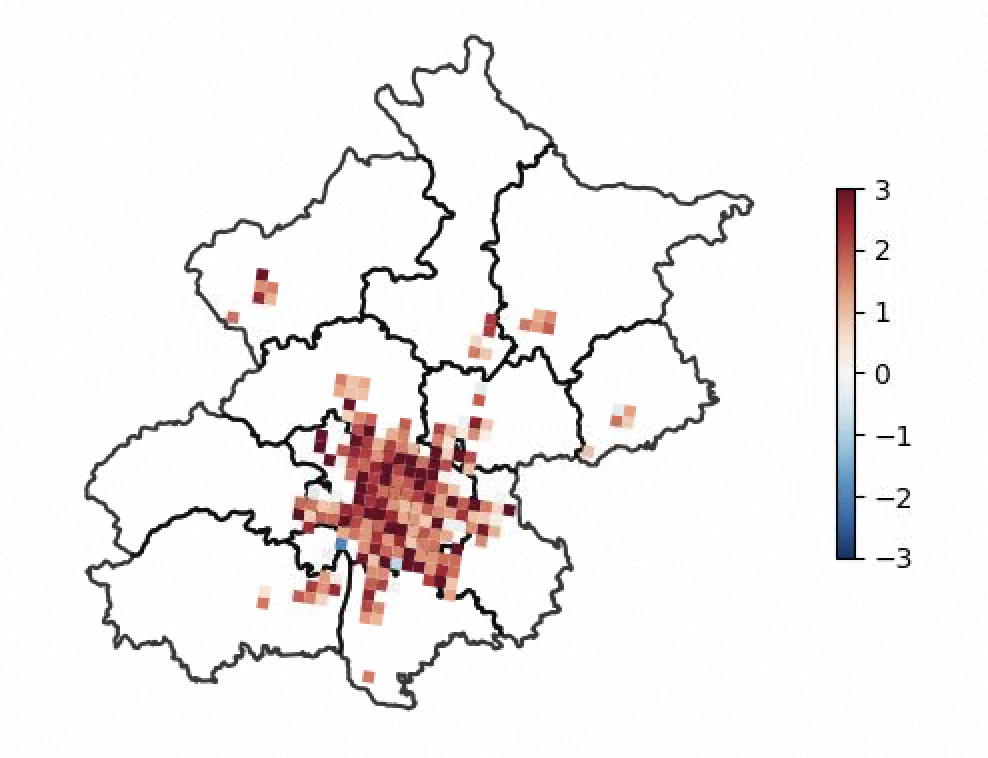}
    \caption{Spatial Distribution of Treatment Effects}
    \label{fig:nbhd_space}
  \figurenotes{This figure maps the spatial distribution of neighborhood-level average conditional treatment effects across the Beijing municipality using 3km $\times$ 3km grid cells. The value for each grid cell represents the mean of the estimated treatment effects in daily out-of-pocket spending for all treated individuals residing in the grid. Red cells indicate positive average effects. Blue cells indicate negative average effects. Grids containing only a single treated individual or those where the average pre-treatment spending was zero are excluded from the visualization. In total, the map displays 212 valid grids, 203 of which exhibit a positive average treatment effect. Over 6 percent of grids experienced spending increases exceeding 50 percent, while 33 percent saw gains of less than 10 percent.}
\end{myfigure}

\subsection{Shapley Variance Decomposition of Treatment Effect Heterogeneity}
\label{sec:shapley}

Let $X=(D,S,E)$, where $D$ denotes demand-side variables (e.g., wealth), $S$ denotes supply-side variables (e.g., local consumption amenities), and $E$ denotes variables that may be influenced by both demand-side and supply-side factors (e.g., consumption habits). As noted in the main text, consumption habits may reflect equilibrium outcomes shaped jointly by consumer preferences and local consumption opportunities. 

Let $V_{\mathrm{tot}} \equiv \operatorname{Var}(\hat{\alpha}_i)$ denote the variance of the estimated CATEs $\hat{\alpha}(X_i)$. For each subset $G \subseteq \{D,S,E\}$, define
\[
v(G) \equiv \operatorname{Var}\!\left(\mathbb{E}[\hat{\alpha}(X)\mid G]\right).
\]
We estimate $v(G)$ by projecting $\hat{\alpha}_i$ onto a fully expanded fourth-degree polynomial in the variables contained in $G$, including all interaction terms up to degree four, and then computing the variance of the fitted values. We do this for all seven nonempty subsets of $\{D,S,E\}$: $v(D)$, $v(S)$, $v(E)$, $v(D,S)$, $v(D,E)$, $v(S,E)$, and $v(D,S,E)$.

These quantities summarize how much of the variation in the estimated treatment effects can be explained by different combinations of variable groups. However, they do not by themselves tell us how to allocate $V_{\mathrm{tot}}$ across $D$, $S$, and $E$, because the incremental contribution of any one group depends on which other groups are already included. For example, the contribution of $D$ is $v(D)$ if it enters first, $v(D,S)-v(S)$ if it enters after $S$, and $v(D,S,E)-v(S,E)$ if it enters last. To avoid relying on any single ordering, we use the Shapley decomposition, which assigns to each group its average incremental contribution across all possible orderings of entry. With three groups, there are six possible orderings. Averaging over these orderings yields the Shapley contributions $\phi_D$, $\phi_S$, and $\phi_E$:
\begin{align*}
\phi_D &= \tfrac{1}{6}\left[2v(D) + \big(v(D,S)-v(S)\big) + \big(v(D,E)-v(E)\big) + 2\big(v(D,S,E)-v(S,E)\big)\right], \\
\phi_S &= \tfrac{1}{6}\left[2v(S) + \big(v(D,S)-v(D)\big) + \big(v(S,E)-v(E)\big) + 2\big(v(D,S,E)-v(D,E)\big)\right], \\
\phi_E &= \tfrac{1}{6}\left[2v(E) + \big(v(D,E)-v(D)\big) + \big(v(S,E)-v(S)\big) + 2\big(v(D,S,E)-v(D,S)\big)\right].
\end{align*}

By construction, these contributions sum to $V_{\mathrm{tot}}$. This decomposition is useful in our setting because demand-side variables, supply-side variables, and consumption habits are correlated, so the relative importance of any one group cannot be summarized by a single ordering alone. The Shapley approach provides a symmetric way to apportion the total variation in estimated treatment effects across the three groups.

Applying this decomposition yields $\phi_D = 4.09$, $\phi_S = 2.93$, and $\phi_E = 18.10$. To summarize the relative importance of demand-side and supply-side factors, we report two comparisons. First, we compare demand and supply after excluding consumption habits from both sides. Under this comparison, the demand share is
\[
\frac{\phi_D}{\phi_D+\phi_S}=58\%
\]
and the supply share is 42 percent.

Second, because consumption habits may reflect both demand-side and supply-side forces, we allocate their contribution equally between the two sides. Under this comparison, the demand share is
\[
\frac{\phi_D+\phi_E/2}{\phi_D+\phi_S+\phi_E}=52\%
\]
and the supply share is 48 percent.

Under both comparisons, demand-side and supply-side factors contribute in broadly comparable magnitudes, with the demand side playing a modestly larger role.

\section{Distributional Consequences and Policy Design}
\label{sec:oa-welfare}

\subsection{Cost Estimation}
\label{sec:cost}

To evaluate the fiscal implications of alternative targeting policies, we estimate the government subsidy (cost) associated with treating any given individual. We accomplish this by training a regression forest to predict individual-level coupon redemption costs based on observed characteristics. The estimation procedure proceeds in three steps.

First, for each individual $i$ in the treatment group, we define the realized cost, $C_i$, as the average daily coupon subsidy redeemed by individual $i$ during the treatment period. This variable represents the exact fiscal outlay incurred by the government for treating individual $i$. 

Second, we train a regression forest model on the treated users by regressing the realized cost, $C_i$, on our full set of observed characteristics, $X_i$. We implement this using the package \texttt{grf} from \textit{R}, retaining the package's default hyperparameter settings.

Finally, we apply the trained model to generate predicted costs, $\hat{C}_i(X_i)$, for all individuals in our full sample. These predictions represent the expected fiscal cost if a given individual were to receive a digital coupon bundle. For our policy counterfactuals, the total baseline government budget is subsequently calibrated by summing these predicted costs across all individuals who are treated during the program.

\subsection{Impact on Local Businesses}
\label{sec:mapping}

To map individual treatment effects to business earnings, we make two simplifying assumptions. First, we assume no coupon-induced consumption reallocation. Under this assumption, each consumer's estimated treatment effect is allocated across establishments in proportion to their observed spending shares during the treatment period. Second, we assume no supply-side adjustment. In particular, establishments do not (a) raise prices, (b) introduce additional promotions, or (c) change menus in response to the coupon event.

Suppose there are $N$ individuals and $K$ establishments. For each individual $i$, we define $r_{ik}$ as the spending placed by individual $i$ on establishment $k$ in the coupon event period and  $p_{ik}=\left.r_{ik}\right/\sum_{j}r_{ij}$ as the corresponding spending share. Let $\alpha_{i}$ be the treatment effect on buyer $i$'s total spending (including OOP expenditure and coupon subsidy) and $\tau_k$ be treatment effect on seller $k$'s total revenue during the coupon event period, and $\Psi$ be the establishment-level effects we want to recover:
$$
\Psi=\left[\begin{array}{c} \tau_{1}\\ \vdots\\ \tau_{K} \end{array}\right],\qquad\Phi=\left[\begin{array}{c} \alpha_{1}\\ \vdots\\ \alpha_{N} \end{array}\right],\qquad\mathbb{P}=\left[\begin{array}{ccc} p_{11} & \ldots & p_{1K}\\ \vdots & \ddots & \vdots\\ p_{N1} & \cdots & p_{NK} \end{array}\right].
$$
Given the two assumptions, we calculate the change in business revenue for all establishments, $\Psi$,  during the coupon period as $\Psi=\Phi'\cdot\mathbb{P}$.

\subsection{Optimal Policy Trees for Coupon Allocation}
\label{sec:oa-policytree}

This section provides technical details on the optimal policy tree method used in Section~\ref{sec:sme_support} to evaluate alternative coupon allocation rules. Consider a policymaker who must decide which individuals should receive a digital coupon. Let $\pi(\textbf{X}_i): \mathcal{X} \rightarrow \{0,1\}$ be a policy that maps an individual's observed characteristics $\textbf{X}_i$ to a treatment assignment, and let $\Gamma_i(d)$ denote the reward from assigning treatment $d$ to individual $i$. The policymaker's objective is to find the policy that maximizes total reward:
\begin{equation*}
	\label{eq:policy_obj}
	\pi^{*}=\underset{\pi\in\Pi}{\arg\max}\left[\sum_{i=1}^{N}\Gamma_i\left(\pi\left(\textbf{X}_{i}\right)\right)\right],
\end{equation*}
where $\Pi$ is the space of admissible policy functions.

Let $p_{i}$ denote consumer $i$'s spending shares on SMEs (see Section~\ref{sec:mapping}). We normalize the reward from non-treatment to zero, $\Gamma_i(0) = 0$, and define
\begin{equation*}
    \Gamma_{i}\left(1\right)=\left(\lambda\cdot p_{i}+\left(1-\lambda\right)\cdot\left(1-p_{i}\right)\right)\cdot\alpha\left(X_{i}\right),	
\end{equation*}
where $\alpha\left(X_{i}\right)$ is the individual treatment effect and $\lambda\in[0,1]$ governs the weight placed on SME revenue in the policy objective. This formulation treats coupon-induced spending at SMEs and non-SMEs as differentially valued by the policymaker. When $\lambda = 0.5$, the objective reduces to maximizing the total stimulus. As $\lambda$ increases from 0.5 to 1.0, the policy increasingly favors treating consumers whose spending patterns direct more revenue toward SMEs. When $\lambda = 1$, the policy focuses exclusively on SME revenue.

We search for the optimal policy within the class of depth-2 decision trees \citep{Athey2021}. In implementation, we replace $\alpha\left(X_{i}\right)$ with its estimated doubly robust score, $\psi_i$, and solve the policy problem using the \texttt{policy\_tree} function from the \texttt{policytree} package in \textit{R} \citep{Athey2021}, with the tree depth set to 2. We vary $\lambda$ from 0.5 to 1.0, solve for the optimal depth-2 tree at each value, and repeat the exercise under three alternative SME definitions based on the 20th, 50th, and 86th percentiles of establishment revenue.

\subsection{Policy Design}
\label{subsec:policy-design}
\begin{mytable}[H]{8pt}
\centering
\caption{Ranked Average Treatment Effect by Decile}
\label{tab:rate}
\footnotesize
\begin{tabular}{l*{10}{r}}
\toprule
\midrule
Variable for Ranking & 0.1 & 0.2 & 0.3 & 0.4 & 0.5 & 0.6 & 0.7 & 0.8 & 0.9 & 1.0 \\
\midrule
All characteristics & 12.39 & 8.22 & 6.35 & 5.25 & 4.50 & 3.93 & 3.46 & 3.04 & 2.58 & 1.80 \\
Spending per order & 7.43 & 5.29 & 3.84 & 3.08 & 2.69 & 2.39 & 2.19 & 2.01 & 1.87 & 1.80 \\
Number of orders & 4.68 & 2.54 & 1.95 & 2.04 & 1.94 & 1.88 & 1.83 & 1.84 & 1.80 & 1.80 \\
Wealth & 4.50 & 3.27 & 3.02 & 2.79 & 2.52 & 2.36 & 2.19 & 2.04 & 1.92 & 1.80 \\
Large establishments & 3.90 & 3.12 & 2.57 & 2.32 & 2.27 & 2.11 & 2.01 & 1.94 & 1.88 & 1.80 \\
Number of establishments & 2.47 & 2.54 & 2.33 & 2.16 & 2.09 & 2.10 & 2.16 & 2.10 & 1.99 & 1.80 \\
Female & 2.00 & 1.99 & 2.27 & 2.21 & 2.12 & 2.06 & 2.05 & 1.97 & 1.92 & 1.80 \\
Platform membership & 1.98 & 1.92 & 1.92 & 1.93 & 1.97 & 1.99 & 1.94 & 1.93 & 1.89 & 1.80 \\
Age & 1.93 & 2.33 & 2.18 & 2.05 & 2.10 & 2.12 & 2.01 & 1.95 & 1.84 & 1.80 \\
\bottomrule
\end{tabular}

\tablenotes{This table reports the cumulative Ranked Average Treatment Effect (RATE) to evaluate the efficacy of alternative targeting strategies. Each row denotes one specific strategy, where individuals are ranked in descending order based on their predicted treatment effect. For the ``All characteristics'' benchmark (first row), predictions are derived from the full causal forest model. For the single-characteristic strategies (subsequent rows), predictions are obtained by projecting the causal forest estimates onto the specified covariate using a one-dimensional Generalized Additive Model to account for potential nonlinearities. Columns 0.1 through 1.0 denote the cumulative fraction of the population targeted. The reported estimates represent the accumulated average treatment effect for the entire targeted subpopulation up to that specific quantile, rather than the marginal effect within an isolated decile bin. For instance, a value in the 0.1 column represents the average treatment effect achieved when allocating the digital coupons only to the top 10\% most responsive consumers according to that row's specific predicted ranking. Because targeting 100\% of the population is equivalent to a universal rollout, the estimates in the final column (1.0) naturally converge to 1.80 across all rows, which corresponds to the ATT.}
\end{mytable}

\begin{mytable}[H]{6pt}
\caption{Consumer Characteristics by Non-SME Expenditure Quintiles}
\label{tabA:btmean}
\begin{tabular}{p{0.33\textwidth}R{0.12\textwidth}R{0.12\textwidth}R{0.12\textwidth}R{0.19\textwidth}}
\toprule
\midrule
    &   \multicolumn{2}{c}{Mean} &  \text{Difference} &  \text{Standard Error}\\  \cmidrule{2-3}
Expenditure  & \text{0 - 20\%} & \text{80 - 100\%} &  & \text{of Difference}  \\
\midrule
Age                                & 31.68 & 32.45 & 0.77  & 1.39 \\
Female                             & 0.61  & 0.65  & 0.05  & 0.07 \\
Platform membership                & 0.11  & 0.41  & 0.30  & 0.04 \\
Wealth                             & -0.22 & 0.10  & 0.32  & 0.13 \\
Number of establishments           & 27.65 & 54.14 & 26.49 & 3.33 \\
Share of non-SME establishments    & 0.38  & 0.54  & 0.16  & 0.03 \\
Number of orders (past 6 months)   & 18.76 & 58.04 & 39.29 & 3.65 \\
Spending per order (past 6 months) & 33.25 & 46.28 & 13.03 & 3.15 \\
\bottomrule
\end{tabular}
\tablenotes{This table compares characteristics of consumers with different expenditure patterns across establishment types. The first column shows mean values for consumers in the lowest quintile of non-SME expenditure share (0-20\%), representing those who predominantly patronize SMEs. The second column shows consumers in the highest quintile (80-100\%), representing those who primarily patronize non-SME establishments. The third and fourth columns report the differences in means and the associated standard errors.}

\end{mytable}

\begin{mytable}[H]{1pt}
\centering
\caption{Tradeoff Between Overall Stimulus and SME Support}
\label{tab: sme_tradeoff}
\begin{tabular}{C{0.17\textwidth}R{0.13\textwidth}R{0.13\textwidth}R{0.13\textwidth}R{0.13\textwidth}R{0.13\textwidth}R{0.13\textwidth}}
\toprule
\midrule
\multirow{2}{*}{Weight on SME} & \multicolumn{2}{c}{20th Percentile} & \multicolumn{2}{c}{50th Percentile} & \multicolumn{2}{c}{86th Percentile} \\
\cmidrule(lr){2-3} \cmidrule(lr){4-5} \cmidrule(lr){6-7}
 & SME & non-SME & SME & non-SME & SME & non-SME \\
\midrule
0.50 & 48.34 & 11792.92 & 739.19 & 11102.07 & 4203.70 & 7638.05 \\
0.60 & 48.34 & 11792.92 & 739.19 & 11102.07 & 4203.20 & 7638.05 \\
0.70 & 48.34 & 11792.92 & 768.73 & 11055.87 & 4251.86 & 7541.98 \\
0.80 & 182.34 & 11454.89 & 1067.00 & 10120.84 & 4251.86 & 7541.98 \\
0.90 & 218.71 & 11219.01 & 1085.37 & 9986.10 & 4388.43 & 6954.43 \\
1.00 & 279.04 & 5706.24 & 1085.57 & 9956.80 & 4389.21 & 6809.37 \\
\bottomrule
\end{tabular}
\tablenotes{This table presents the numerical values (¥) of Figure \ref{fig:sme}, illustrating the tradeoff between maximizing total revenue and increasing the SME revenue share. This tradeoff is evaluated by varying the weight parameter ($\lambda$) placed on SME revenue in the targeting algorithm: $\lambda \cdot R_{\text{SME}} + (1-\lambda) \cdot R_{\text{large}}$, as detailed in Section \ref{sec:sme_support}. To demonstrate how this tradeoff depends on the size of the targeted group, the table reports outcomes under three alternative SME definitions. Specifically, an establishment is classified as an SME if its total sales revenue during the six months prior to the coupon program falls below a specified percentile threshold (the 20th, 50th, or 86th percentile) of the citywide establishment revenue distribution. Establishments above the given threshold are classified as non-SMEs. The weight parameter $\lambda$ ranges from 0.5 (equal weighting between SMEs and non-SMEs) to 1.0 (exclusive focus on maximizing SME revenue). As $\lambda$ increases, the revenue directed to SMEs mechanically rises, while the revenue directed to non-SMEs and consequently the overall total stimulus declines, highlighting the fundamental policy tension between aggregate efficiency and small business support. All reported monetary values represent daily total amounts. }

\end{mytable}

\clearpage 

\setstretch{1} 

\singlespacing 

\end{document}